\documentclass[]{article}
\usepackage{emulateapj}
\usepackage{epsfig}

\usepackage{ifthen}
\submitted{Accepted by The Astrophysical Journal}

\newboolean{apj}
\setboolean{apj}{true}
\def\gsim{\;\rlap{\lower 2.5pt
 \hbox{$\sim$}}\raise 1.5pt\hbox{$>$}\;}
\def\lsim{\;\rlap{\lower 2.5pt
   \hbox{$\sim$}}\raise 1.5pt\hbox{$<$}\;}
\def\msol{{\rm\,M_\odot}}

\def\kpc{{\rm\,kpc}}

\def\mic{{\,\mu{\rm m}}}

\def\kms{{\rm\,km\,s^{-1}}}

\def\spose#1{\hbox to 0pt{#1\hss}}
\def\lta{\mathrel{\spose{\lower 3pt\hbox{$\mathchar''218$}}
     \raise 2.0pt\hbox{$\mathchar''13C$}}}
\def\gta{\mathrel{\spose{\lower 3pt\hbox{$\mathchar''218$}}
     \raise 2.0pt\hbox{$\mathchar''13E$}}}

\def\zsol{{\,Z_\odot}}
\def\lsol{{\,L_\odot}}
\def\tc{{t_{\rm c}}}
\def\tp{{t_{\rm p}}}

\begin{document}
	
\title{Metal Enrichment of the Intergalactic Medium in Cosmological
Simulations}

\author{Anthony Aguirre,\footnote{Institute for Advanced Study, School of Natural Sciences, Princeton NJ 08540}$^{,b}$
Lars Hernquist,\footnote{Department of Astronomy, Harvard University
60 Garden Street, Cambridge, MA 02138}
Joop Schaye,$^{a}$
Neal Katz,\footnote{Department of Astronomy, University of Massachusetts, Amherst, MA 01003}
David H. Weinberg,\footnote{Department of Astronomy, Ohio State University, Columbus, OH 43210}
\& Jeffrey Gardner\footnote{Department of Astronomy, University of Washington, Seattle, WA 98195}
}
\setcounter{footnote}{0}

\begin{abstract}
  
	Observations have established that the diffuse
	intergalactic medium (IGM) at $z \sim 3$ is enriched to $\sim
	10^{-2.5}$ solar metallicity and that the hot gas in large
	clusters of galaxies (ICM) is enriched to $1/3-1/2\,Z_\odot$ at
	$z=0$.  Metals in the IGM may have been removed from galaxies
	(in which they presumably form) during dynamical encounters
	between galaxies, by ram-pressure stripping, by
	supernova-driven winds, or as radiation-pressure driven dust
	efflux.  This study develops a method of investigating the
	chemical enrichment of the IGM and of galaxies, using already
	completed cosmological simulations.  To these simulations, we
	add dust and (gaseous) metals assuming instantaneous
	recycling, and distributing the dust and metals in the gas
	according to three simple parameterized prescriptions, one for
	each enrichment mechanism.  These prescriptions are formulated
	to capture the basic ejection physics, and calibrated when
	possible with empirical data.  Our method allows exploration of
	a large number of models, yet for each model yields a specific
	(not statistical) realization of the cosmic metal distribution
	that can be compared in detail to observations.  Our results
	indicate that dynamical removal of metals from $\gsim
	10^{8.5}\msol$ galaxies cannot account for the observed
	metallicity of low-column density Ly$\alpha$ absorbers, and
	that dynamical removal from $\gsim 10^{10.5}\msol$ galaxies
	cannot account for the ICM metallicities.  Dynamical removal
	also fails to produce a strong enough mass-metallicity
	relation in galaxies. In contrast, either wind or
	radiation-pressure ejection of metals from relatively large
	galaxies can plausibly account for all three sets of
	observations (though it is unclear whether metals can be
	distributed uniformly enough in the low-density regions
	without overly disturbing the IGM, and whether clusters can be
	enriched quite as much as observed).  We investigate in detail
	how our results change with variations in our assumed
	parameters, and how results for the different ejection
	processes compare.  

\end{abstract}

\section{Introduction}

	In the standard hot big-bang cosmological model, elements of
atomic number $Z > 2$ cannot form in appreciable quantities until the
first stars form.  Thereafter, the universe becomes progressively
enriched with heavy elements (metals) as stars release these fusion
products in stellar winds or supernova ejecta.  Stars only form
efficiently in dense gas, almost all of which is bound in galaxies (or
protogalaxies at high $z$) with deep gravitational wells.  Despite
this, the observed intergalactic medium (IGM) shows substantial metal
enrichment at all redshifts and at all densities yet measured, from
the lowest-density Ly$\alpha$ absorbers at $z > 3$ with $\gsim 1/1000$
solar metallicity, to the hot gas with 1/3-1/2 $Z_\odot$ bound in
present-day galaxy clusters.  A significant fraction, perhaps even a
majority, of cosmic metal appears to lie in the IGM.  At very high
redshift ($z \gg 5$) population III stars could enrich the IGM to a
low level (e.g., Carr, Bond \& Arnett 1984; Ostriker \& Gnedin 1996;
Haiman \& Loeb 1997; Tegmark et al. 1997; Abel et al. 1998), but since
most cosmic metals presumably form in stars in galaxies, a
fundamental question arises as to how these metals escape their
progenitor galaxies and spread throughout the IGM.

	Efforts to understand the enrichment of the IGM by galactic
stars (after any Pop. III epoch) have focused on three mechanisms
whereby metals could be removed from a galaxy.  First, metal-enriched
gas (or stars that later explode as Type Ia supernovae) might be
unbound during a merger or tidal interaction with another galaxy, or
by the ram pressure of the IGM through which the galaxy moves.  We
shall combine these processes under the name of `dynamical removal'.
Second, the energy input from supernovae may impart sufficient kinetic
and thermal energy to galactic gas for it to escape the gravitational
well of the galaxy. We shall denote this process as the `galactic
wind' mechanism.  Third, the radiation pressure on dust grains due to
stellar light may exceed the gravitational force of the matter,
leading (if the dust can decouple from the gas) to an outflow of dust.
We will denote this possibility by `radiation-pressure ejection' or
`dust ejection'.\footnote{Dust will, however, also be ejected in
dynamical removal or wind ejection along with the gas.} (The ejected
dust adds metals to the intergalactic {\em gas} when it is destroyed
by thermal sputtering.)

	Investigation of the metal enrichment of the IGM requires not
only some understanding of how these mechanisms function in a given
galaxy, but also calls for knowledge of the properties,
distribution, and evolution of galaxies in a cosmological context.
Studies of these matters have generally adopted one of two rather
different, yet complementary, approaches: the `numerical' and the
`semi-analytic' methods.

In the numerical approach, an attempt is made to numerically evolve
the state of a single sample of the universe from some set of initial
conditions, by solving equations of motion encapsulating the most
relevant (gravitational, hydrodynamical, etc.) physics.  Among the
advantages of this approach are that it captures the incorporated
physics very well, and that it generates a specific realization of a
possible cosmological volume.  On the other hand, the available
computing power limits the amount of detailed physics and the dynamic
range that can be handled. Also, some physical processes must be
`parameterized', and the large computational time prohibits the
investigation of large regions of parameter space. The numerical
method has been used by Gnedin (1998) to study the enrichment of the
IGM by winds and dynamics and employed by Cen \& Ostriker (1999) and
Gnedin \& Ostriker (1997) to study IGM enrichment by dynamics or other
(unspecified) processes.  Cluster enrichment by galactic winds has
been studied using simulations by Metzler \& Evrard (1994; 1997) and
Murakami \& Babul (1999), and the enrichment of the intracluster
medium (ICM) through dynamics has been numerically studied by Abadi,
Moore \& Bower (1999), by Balsara, Livio \& O'Dea (1994), and by
Quilis, Moore \& Bower (2000).

In the semi-analytic approach, predictions are made on a statistical
basis, by layering together a number of prescriptions that are
individually derived either from theory or observation.  This approach
allows the investigation of large regions of parameter space, and
employs somewhat more complicated physical prescriptions than the
numerical method.  On the other hand, the flexibility in the input
parameters and physical prescriptions leads to a corresponding range
in actual predictions, and it is not always clear which physical
processes are accurately captured and which are not; nor is it clear
that the parameters chosen to best fit the observations are unique.
The semi-analytic method has been used by a number of investigators to
address IGM enrichment in greatly varying levels of complexity.  Nath
\& Trentham (1997), Ferrara, Pettini \& Shchekinov (2000), and Madau,
Ferrara \& Rees (2000) have studied IGM enrichment by winds in this
way, attempting also to calculate the statistical properties of the
metal distribution in the IGM. Cluster enrichment by winds has been
studied by Dekel \& Silk (1986), Nath \& Chiba (1995), and David,
Forman \& Jones (1990), among others.  The dynamical enrichment of
clusters has been similarly examined a number of times (e.g., Gunn \& Gott
1972; Renzini et al. 1993).  In both cases the overall degree of
enrichment has been assessed, but not the distribution of the metals.
Finally, ejection of dust by radiation pressure has been calculated
for individual sample galaxies (Chiao \& Wickramasinghe 1972; Ferrara
et al. 1990; Shustov \& Vibe 1995; Davies et al. 1998; Simonsen \&
Hannestad 1999), but only rough estimates of the overall ensuing IGM
enrichment are offered.

The method used in the present work combines aspects of both
approaches.  The time-dependent distribution of dark matter, gas and
stars is taken from already completed cosmological smoothed-particle
hydrodynamics (SPH) simulations computed using the method described by
Katz, Weinberg \& Hernquist (1996).  To the gas and star particles,
metals are added using fairly simple prescriptions formulated to
capture the basic physics of the various processes that
can transport metals from galaxies into the IGM.  This method yields a
numerical realization of the actual distribution of metals in the IGM
(as well as in galaxies), yet uses relatively little computing time so
that a large number of models can be tested, and the effects of
changing both physical prescriptions and parameters can be explored.
This method has a number of limitations that will be discussed in
detail in the following sections, but it can nevertheless yield important
insight into metal ejection and distribution unobtainable by either
`purely' numerical computations (using present technology), or
semi-analytical methods.


This paper presents an explication and investigation of the method we
have developed.  We do not attempt to `fit' a set of observations
using our results, but rather make specific assumptions about our
input parameters based on independent considerations, and compare the
predictions to observations where available.  We also investigate
directly the variations in predictions resulting from different
parameter choices, and address various methodological and numerical
considerations.  Studies addressing specific astrophysical questions using the
method presented here will be published separately (e.g., Aguirre et al.
2001a,b).

We have organized this paper as follows.  Section~\ref{sec-method}
describes our calculation method, with
subsections~\ref{sec-locmeth}, ~\ref{sec-windmeth}, and~\ref{sec-radmeth} 
detailing prescription for
dynamical, wind, and radiation pressure ejection of metals,
respectively.  These sections also describe the parameters used in the
calculation and the fiducial values for these parameters (summarized
in Tables~\ref{tab-fidparwind}, \ref{tab-fidpardust},
\ref{tab-fidpargen} and~\ref{tab-fidpargrain}).  In
\S~\ref{sec-obssumm} we review observations of cluster, galaxy, and
Ly$\alpha$ forest metallicity with which we will compare our
results. Section~\ref{sec-simsvsobs} briefly compares the SPH
simulations (to which we apply our method in this paper) to
observations.  Sections~\ref{sec-resdyn},~\ref{sec-reswind}
and~\ref{sec-resrad} present and discuss the trials we have run and
the results obtained.  We summarize our findings and draw general
conclusions in~\S~\ref{sec-conclusions}.

\section{Method}
\label{sec-method}

The procedure begins with a set of coarsely-spaced (every
$10^8-10^9\,$yr) snapshots from an SPH cosmological simulation, each
containing the states of the dark matter, star, and gas particles at a
given redshift.  We shall index these particles by `i', and
superscript them by particle type as `d', `s' or `g' respectively. We
will also denote by $\tc$ the time of the output being processed, and by
$\tp$ the time at the previous output.

Beginning with the first output time where stars exist, the
procedure for each time step is as follows.  For a star particle $i$,
the stellar mass created since the last step is $\Delta
m^s_i(\tc)\equiv m^s_i(\tc)-m^s_i(\tp)$.  This mass was taken from a
gas particle also indexed by $i$, so the first task is to transfer the
metals associated with this gas mass from the gas particle to the star
particle:
\begin{eqnarray}
w^s_i(\tc)&=&w^s_i(\tp)+\Delta
m^s_i(\tc)\times[w^g_i(\tp)/m^g_i(\tp)] \\ \nonumber
w^g_i(\tc)&=&w^g_i(\tp)-\Delta
m^s_i(\tc)\times[w^g_i(\tp)/m^g_i(\tp)],
\end{eqnarray}
where $w_i$ is the metal mass of the $i$th particle.

We next assume that each unit of forming stellar mass instantaneously
ejects $y_*$ units of metal mass into the gas, so that the star
particle $i$ adds a metal mass of $\Delta m^s_i(\tc) \times y_*$ to
the gas particles. The metals are distributed over the gas particles
in a manner appropriate to the particular
process, as described in the next few sections.  
After all metal is deposited, the process repeats for the
next simulation output time.  

\subsection{Prescription 1: Local Metal Distribution}
\label{sec-locmeth}

The first prescription for metal distribution is the simplest.  The
metal mass $\Delta m^s_i(\tc) \times y_*$ is `scattered' over the 32
gas particles nearest to the star particle $i$, weighted by the SPH
smoothing kernel $W(r,h)$, where $2h$ is the radius of the sphere
(about particle $i$) containing exactly 32 gas particles, and $r$ is
the distance from particle $i$ to the gas particle in question (see
Hernquist \& Katz 1989).  Since stars form in these simulations only
in dense, cool regions with a converging flow, this prescription
places metals only in bound regions such as galaxies.\footnote{If stars form in bound groups with $< 32$ gas particles, some metals will be placed in particles outside the bound group.}
  Thus with this
prescription intergalactic metal must leave galaxies by being carried
by a gas particle that is dynamically removed from a bound region.

\subsection{Prescription 2: Ejection by Galactic Winds}
\label{sec-windmeth}

Galactic winds can distribute metals `non-locally', i.e.\ disperse them
into gas far away from where they are formed.  The prescription used
to model this physical effect is as follows.  First, we divide the gas
and star particles into bound groups (i.e.\ galaxies) using the SKID
algorithm.\footnote{SKID is publicly available at
http://www-hpcc.astro.washington.edu/tools. We require also that these
groups have at least 4 (gas+star) particles and a minimal overdensity
of 50; groups of between 1 and 4 particles are treated as `ungrouped',
but essentially no star formation takes place in these areas.}  At
each simulation step, each bound group is considered in turn. As
described above, the new metal mass $\Delta m^s_i(\tc) \times y_*$ in
each star particle $i$ within the group is computed.  A fraction
$(1-Y_{\rm ej})$ of this metal is distributed among the 32 gas
particles nearest to star particle $i$ as per prescription 1.  The
remaining fraction is added to the tally for the new metal in the
group, $\Delta w^{\rm grp}$:
\begin{equation}
\Delta w^{\rm grp} = y_*\times Y_{\rm ej} \times \sum_{i \in {\rm grp}}\Delta m^s_i(\tc).
\end{equation}
We then distribute the metal mass $w^{\rm grp}$
within a radius $h^{\rm wind}(\theta,\phi)$ about the `center of star formation'
\begin{equation}
\vec r_c \equiv \left[\sum_{i \in {\rm grp}} \vec x^s_i \Delta m^s_i\right]/\left[
\sum_{i \in {\rm grp}}  \Delta m^s_i\right],
\label{eq-cosf}
\end{equation}
where $\vec x_i^s$ is the position of star $i$.  
This simulates the non-local dispersal
of metal into the regions where winds carry and deposit them.
The metal is distributed so that
within the angular ranges $[\cos\theta,\cos\theta+\Delta_{\cos\theta}]$ and $[\phi,\phi+\Delta_{\phi}]$,
the metal mass
within a shell of width $dr$ at radius $r$
is proportional to $W^{\rm
grp}(r,h^{\rm wind}(\theta,\phi))dr$; uniform distribution of metals within the
angular region thus corresponds to $W^{\rm grp}(r,h^{\rm wind}(\theta,\phi)) = r^2$. 

Choosing values of $h^{\rm wind}(\theta,\phi)$, $Y_{\rm ej}$ and $W^{\rm
grp}(r,h^{\rm wind})$ for each galaxy requires some understanding of
the physics of galactic winds.  The idea that galaxies might drive
outflowing winds has a fairly long history. Galactic-scale winds, in
which supernova bubbles can overlap and drive a coherent wind across
the galaxy before they can cool, have been theoretically investigated in
dwarf (e.g., Mac Low \& Ferrara 1999), elliptical (e.g., David, Forman
\& Jones 1990; 1991 and references therein), and starburst spiral
(e.g., Heckman et al. 2000, hereafter HLSA; Lehnert \& Heckman 1996)
galaxies.  In the last case, such winds can be observed in some detail
as high velocity bipolar outflows seen in many starburst galaxies such as M82
(e.g., Lehnert, Heckman \& Weaver 1999).  Even if the conditions to
drive a coherent galactic-scale wind do not exist, winds capable of
ejecting matter into the IGM may still develop: a two-phase ISM with a
hot phase fed by supernova remnants can lead to a stochastic
`steady-state' in which some fraction of matter has high enough
kinetic energy -- perhaps aided by cosmic-ray pressure
(e.g., Breitschwerdt et al. 1991) -- to escape the galaxy (especially
near concentrations of supernovae).  Winds of this sort have been
investigated theoretically by Efstathiou (2000), Ferrara \& Tolstoy (2000),
Breitschwerdt \& Schmutzler (1999), and Ferrara, Pettini \& Shchekinov
(2000).  Thus we see that winds may be starburst-driven, driven by
`quiescent' star formation, or cosmic ray-driven, and that they may be
`global' or `local'.  These distinctions can readily break down,
however: in dwarfs or in galaxies with very rapid star-formation, the
local/global distinction breaks down, and at high-$z$ (where star
formation is vigorous and mergers common) there is probably no clear
line between starbursts and quiescent star formation.

Despite their differences, all of the wind types we have described
share some common features:
\begin{enumerate}
\item{The energy release in supernovae is the ultimate source of the
wind energy.  Some critical supernova rate is necessary for a
galactic-scale wind to form and blow out of the disk.}
\item{The wind speed may exceed the escape velocity of the
progenitor galaxy, but the wind and the swept-up material must stall
or become pressure-confined at some radius.}
\item{It is physically reasonable for the wind's energy to be tied
to the star formation rate.}
\end{enumerate}

The method used in our investigation is based primarily on
observations of galactic-scale `superwinds' and has been formulated to
capture these three key physical features in a simple and general way,
so that it can be reasonably applied to galactic winds of all types
(with varying degrees of confidence).  First, we assume some critical
SFR/(area), SFR$_{\rm crit}$, below which wind development is
suppressed.  For thermal winds, this should physically correspond to a
rate above which supernova remnants can overlap before cooling (David,
Forman \& Jones 1990; Efstathiou 2000; Heckman, Armus \& Miley 1990).
This assumption is supported by observations indicating that
superwinds in spirals develop when the SFR/unit area (averaged over
the disk) reaches a critical value (Martin 1999; Heckman 2000).
Since the `areas' of spiral galaxies are not robustly determined in
the simulations, we compute galaxy areas from their masses, using the
empirical mass-radius relation\footnote{The masses derived by Gavazzi
et al. from rotation curves will include some dark matter contribution
not included in the corresponding simulation galaxy mass.  This should
not affect large galaxies too much but may be important in dwarfs. See
\S~\ref{sec-windmeth}.} of Gavazzi, Pierini \& Boselli (1996), who
find (area)$\propto M^{0.78}.$

We allow all galaxies to drive winds, with an initial wind velocity
$v_{\rm out}$ and mass outflow rate $\dot m_{\rm out}$.  Guided by
observations of starburst-driven superwinds (Heckman et al. 2000;
Martin 1999) indicating mass outflow rates similar to the galaxies'
SFRs (for a wide range of SFRs), we express $\dot m_{\rm out}$ in
units of the SFR.\footnote{This mass outflow rate, using the current
methodology, does not really determine how {\rm much} metal leaves the
galaxy, merely the `strength' of the wind.}  Assuming that one
supernova forms per $M_{100}100\,\msol$ of star formation and releases
$10^{51}E_{51}{\rm\,ergs}$ of energy, 
the outflow rate $\dot
m\times({\rm SFR})$ (in $\msol/$yr) can be related to the fraction
$\chi$ of the supernova energy that is incorporated into the wind's
kinetic energy $\dot mv_{\rm out}^2\times({\rm SFR})$, by
\begin{equation}
\chi \simeq 0.75\,\dot m_{\rm out}\left({v_{\rm out}\over 600{\rm\,km\,s^{-1}}}\right)^2
E_{51}^{-1}M_{100},
\label{eq-mdotchi}
\end{equation}
where $v_{\rm out}$ is the wind velocity.

We use fixed values of SFR$_{\rm crit}$ and $\chi$ that are based
on the data available for superwinds (see \S~\ref{sec-fidparwind}
below). For $v_{\rm out}$, we assume a uniform distribution of values
with mean $v_{\rm out}^{\rm fid}$ and width $\sigma_{\rm out}$.  Given
$\chi$ and $v_{\rm out}$ for a given galaxy, we compute $\dot m_{\rm out}$ using
Eq.~\ref{eq-mdotchi}.  For galaxies with SFR/(area) $<$ SFR$_{\rm
crit}$, we attenuate $v_{\rm out}$ by a factor $({\rm SFR}/{\rm
SFR}_{\rm crit})^\beta$ and attenuate $\chi$ by a factor $({\rm
SFR}/{\rm SFR}_{\rm crit})^{2\beta}$.  This lowers both the energy of
the wind and the energy of the initial shell by a factor $({\rm
SFR}/{\rm SFR}_{\rm crit})^{2\beta}$, and therefore suppresses winds
in low-SFR galaxies as desired.

We assume that in each direction $(\theta,\phi)$ the wind flows to
some maximum radius $h^{\rm wind}(\theta,\phi)$, then joins the
dynamics of the ambient gas.  This radius, within which the galaxy's
metals are distributed, is again determined from the physics of winds.
When a bubble of supernova-heated gas forms and begins to expand,
several things can stop its growth.  First, the ISM of the host galaxy
will be swept into a shell which, if massive enough, can confine the
wind to the galaxy, as happens in an ordinary single supernova
explosion. If the bubble has enough energy, the shell can be blown out
of the disk, whereupon it may partially fragment due to the
Raleigh-Taylor instability or because of density inhomogeneities in
the ambient medium; the hot wind can then stream past the shell,
entraining it (and perhaps other portions of the ISM) into a
mass-loaded outflow.  The process may then repeat.

	For each of $N_a$ directions $(\theta,\phi)$, we compute the
maximum radius to which the wind could expand by following the
dynamics of a `test' shell of physical radius $r$ about $\vec r_c$,
with mass $m(r)$ and outward radial velocity $v(r)$.  This shell
represents the initial wave of matter swept up by the developing wind,
and feels four forces: the wind's ram pressure, gravity, the thermal
pressure of the ambient gas, and the ram pressure of any infalling
ambient gas\footnote{This purely momentum-based method is similar to
that of Theuns, Mo, \& Schaye 2001; other studies such as Tegmark,
Silk, \& Evrard 1993, Nath \& Trentham (1997) and Scannepieco \&
Broadhurst 2001 use slightly different approach in which the shell is
driven by thermal (not kinetic) energy so that the internal energy of
the `bubble' must be evolved.}  Under these four forces the shell
momentum evolves according to
\begin{eqnarray}
\label{eq-shellev}
{d\over dt}(mv)&=&{\dot m_{\rm out}\over N_a}\left(v_w-v\right)-m{d\bar\phi \over dr}-\left({4\pi\over N_a}\right)r^2 {\bar p}\\ \nonumber
&+&\epsilon_{\rm ent}\left({4\pi\over N_a}\right)r^2\bar\rho\left(
{\bar v_{\rm rad} + Hr}\right)(v-\bar v_{\rm rad}-Hr),
\end{eqnarray}
and the mass evolves as
\begin{equation}
{d\over dt}(m)={\dot m_{\rm out}\over N_a}\left(1-{v\over v_w}\right)+
\epsilon_{\rm ent}
\left({4\pi\over N_a}\right)
r^2\bar \rho\left(v-\bar v_{\rm rad}-Hr\right).
\label{eq-mev}
\end{equation}
Here, $\bar\rho(r)$, $\bar p(r)$, $\bar\phi(r)$ and $\bar v_{\rm rad}(r)$ are
the average density, thermal pressure, gravitational
potential and outward radial peculiar velocity
 of the ambient medium
at radius $r$, $v_w(r)$ is the
wind velocity, and $H(z)$ is the Hubble constant at redshift $z$:
$$
H^2(z)=H_0^2\left[(1+z)^3\Omega+\Omega_\Lambda+(1+z)^2(1-\Omega-\Omega_\Lambda)\right].
$$
The averages are done over all particles within the angular ranges
$[\cos\theta,\cos\theta+\Delta_{\cos\theta}]$ and
$[\phi,\phi+\Delta_{\phi}]$ and the radial range\footnote{Other
prescriptions for the radius to average over could be used, but
experimentation shows that the integration is not sensitive to
the averaging scheme so we employ this simple one.} $[0.95r,1.05r]$.
The angles are spaced in $N_\theta$ segments of
$\Delta_{\cos\theta}\equiv 2/N_\theta$ in $\cos\theta$ and
$N_\phi=2N_\theta$ segments of $\Delta_{\phi}\equiv 2\pi/N_\phi$ in
$\phi$, giving $N_a=N_\theta N_\phi$ portions of equal solid angle.
We choose $N_a$ for each galaxy to be 1/16th of the number of gas+star
particles in the galaxy; tests show that this gives a small enough
number of angles for the radial integrations to be accurate (see also
\S~\ref{sec-reswind}).

The entrainment parameter $\epsilon_{\rm ent}$ is the fraction of ambient material
we assume to be swept up by the wind; a perfectly homogeneous shell
expanding into a homogeneous medium would give $\epsilon_{\rm ent}=1$,
whereas $\epsilon_{\rm ent}\ll 1$ might describe a very clumpy ambient
medium with a very small filling factor that the wind can easily
stream past (while filling in the `holes' after passing), or a shell
that repeatedly fragments, leaving behind a fraction $(1-\epsilon_{\rm
ent})$ of its mass.  (Note that we have not included $\epsilon$ in the
terms accounting for the wind on the shell, as would be appropriate
if the shell itself had a small filling factor.)

The wind velocity
$v_w(r)$ is $v_{\rm out}$, attenuated by gravity:
\begin{equation}
v_w(r)=\sqrt{(v_{\rm out})^2-2[\bar\phi(r)-\bar\phi(r_0)]},
\label{eq-vworf}
\end{equation}
where $r_0$ is the initial radius of the test shell,
defined below.

The physical radial velocity of the shell with respect to the ambient
gas is $v-\bar v_{\rm rad}-Hr$, so evolving $(mv)$ and $m$ until
$v(r)-v_{\rm rad}-Hr < \delta \times Hr$ ($\delta \ll 1)$ gives the
radius $r_{\rm stall}$ and time $\tau_{\rm stall}$ at which the shell
stalls and mixes with the ambient gas (i.e.\ moves a distance $\delta
\times r$ in another Hubble time; we choose $\delta =0.05$).  We set
initial conditions for $m$, $v$ and $r$ by choosing a radius $r_0$ to
include a fixed fraction $\xi$ of the galaxy's mass $M_{\rm gal}$. The
initial shell mass is then $m(r_0)=\epsilon_{\rm ent}\xi M_{\rm
gal}/N_a$, and $v(r_0)=v_{\rm out}^0$.  (The fraction $\xi$ is
calibrated to give values of $r_0$ similar to the radii at which winds
from starburst galaxies are observed; see \S~\ref{sec-fidparwind}
below).

	The stalling radius must be modified by two effects resulting
from our general method, in which we deposit metal generated between
times $\tp$ and $\tc$ in the IGM at time $\tc$ rather than at the more
appropriate average time $(\tp+\tc)/2+\tau_{\rm stall}$.  First, if
$(\tp+\tc)/2+\tau_{\rm stall}$ exceeds the time $t_{\rm obs}$ at which
the results for metal enrichment are desired, we would
allow metals to move to erroneously large radii; thus we limit the
shell to have $\tau_{\rm stall} < t_{\rm obs}-(\tp+\tc)/2$.  Second, if
$\tau_{\rm stall}+(\tp+\tc)/2 > \tc$, then metal deposited at $\tc$
will effectively be artificially carried by the movement of the gas
particles between $\tc$ and $\tau_{\rm stall}+(\tp+\tc)/2$.  For
example, if the shell stalls in a region with radial velocity $v_{\rm
rad}$ and its metal were deposited at $\tc$, then at the appropriate
distribution time $\tau_{\rm stall}+(\tp+\tc)/2$ the metal would be in
particles at radius (roughly) $r_{\rm stall}+v_{\rm rad}[\tau_{\rm
stall}+(\tp-\tc)/2]$, rather than at $r_{\rm stall}$.  To compensate
for this, after the shell reaches $r_{\rm stall}$ we continue to
integrate in radius, but using $ dr/dt=-v_{\rm rad}(r)$ for a time
$\tau_{\rm stall}+(\tp-\tc)/2$ (or $t_{\rm obs}-\tc$ if this is
smaller), to reach a final radius $h_{\rm wind}.$ Then $h_{\rm wind}$
will be such that metal deposited at $h_{\rm wind}$ will end up at
$r_{\rm stall}$ after being carried by the gas particle movement for
time $\tau_{\rm stall}+(\tp-\tc)/2$.  This assumes that the radial peculiar
velocity field in the galaxy's neighborhood changes on a timescale
somewhat longer than $\tau_{\rm stall}+(\tp-\tc)/2$.  A similar
(small) error of the opposite sign arises if $\tau_{\rm
stall}+(\tp+\tc)/2 < \tc$.  In this case, we compensate for this error
by integrating $dr/dt=v_{\rm rad}(r)$ after stalling, for the time
$(\tc-\tp)/2-\tau_{\rm stall}$ to obtain $h_{\rm wind}$.

The metal mass is then distributed within the radius $h^{\rm
 wind}=\min(r_{\rm stall},r_{\rm max})$, if $r_{\rm stall} > 2r_0$.  If
 $r^{\rm stall} < 2r_0$, the wind has been efficiently confined, and we
 instead distribute the metal `locally' as per prescription 1.

The outlined prescription is based on empirical data concerning
galactic-scale superwinds driven by starburst nuclei.  For other types
of winds (cosmic ray driven winds, `quiescent' winds, `local' winds,
etc.) there is far less useful observational data; most of our
understanding of such winds derives from theoretical work (e.g.,
Breitschwerdt et al. 1991; Efstathiou 2000, Ferrara \& Tolstoy 2000;
Breitschwerdt \& Schmutzler 1999, and Ferrara, Pettini \& Shchekinov
2000).  Rather than invent a new prescription for these winds (based
on the theoretical work), we have chosen to model them with the same
prescription as for starburst-driven superwinds, but with different
parameters.  We shall lump these other types of winds into the
category of `quiescent winds', which will be characterized by a lower
critical SFR (so that essentially all galaxies drive winds), but also
a lower $\chi$ since the winds would not be as efficient.

The general wind prescription captures the essential physics of winds
well, and and should provide a good prediction of the dispersal of
metals by winds with a given assumed set of physical parameters. Since
we do not model the winds in full physical detail, however, a number
of methodological details merit attention.

\begin{enumerate}
\item{Adding metals after the completion of the simulation precludes
any self-consistent treatment of the effects of the winds on the gas
near the galaxy. This means that we cannot assess the importance of
earlier outflows in the escape of later outflows and motivates our
approach of assuming a `steady state' wind running into an undisturbed
IGM.  Also, we do not treat the interaction of winds from a given
galaxy with those from nearby galaxies.  Thus galaxies can
(unrealistically) pollute wind-driving neighbors, but only at a small
level.}
\item{The method also glosses over the detailed structure of the
outflowing wind; we have lumped the details of the wind's interactions
with the ambient medium into the parameter $\epsilon_{\rm ent}$,
assumed not to vary with radius or among different galaxies.}
\item{We neglect the effect of the matter within the test shell radius
on the wind that drives the shell.  This is correct for
$\epsilon_{\rm ent}=1$ (since then there {\em is} no matter within the
shell) and for $\epsilon_{\rm ent}=0$ (since then the wind passes
through the matter). For intermediate values, the wind will have
a somewhat lower mass $\dot m$ and higher velocity $v_w$ than the mass
and velocity $\dot m(\epsilon_{\rm ent}), v_w(\epsilon_{\rm ent})$ that it
should have had.  Fortunately these effects largely cancel out since (in
the limit where gravity is unimportant), $\dot m(\epsilon_{\rm
ent})v_w(\epsilon_{\rm ent})=\dot m v_w$ by momentum conservation.
}
\item{We assume a purely kinetic wind impinging upon the shell at all
stages.  This leads to the neglect of the pressure inside the shell in
Equation~\ref{eq-shellev}.  It also allows us to disregard shocks and
cooling, since the internal energy of the shell does not affect its
propagation.  We retain the external pressure term so that
environments of very high pressure (i.e.\ clusters) are not treated
incorrectly.}
\item{Although we compute the {\rm maximum} wind radius with good
accuracy, the deposition profile within this radius, $W^{\rm
grp}(r,h)$, must simply be assumed. (But we will demonstrate that the
details of the assumed form are not too important).}
\item{Our method assumes that the properties of galaxies driving winds
changes on a timescale that is long compared to the outflow time.
In particular, we assume that the SFR is constant while the wind propagates.
We check this assumption in \S~\ref{sec-reswind}.}
\item{
The SFRs we use are averaged over the interval between simulation output times
(typically $10^8-10^9$\,yr).  Weinberg et al. (2000) find that such
averages are a good approximation to the computed `instantaneous'
SFRs.  However, the simulations cannot resolve the small (few hundred
pc) scale of starburst nuclei, so real starbursts may be shorter and
more intense than the smoothly-varying SFR the simulations would
suggest, and lead to a larger scatter in the SFR/area than occurs in
the simulated galaxies (see Dav\'e et al. 1999b).  This is not a
problem for more `quiescent' winds but complicates the use and choice
of a critical wind-driving SFR if winds are assumed to be primarily
starburst-driven.}
\item{Our method concentrates all star formation at a point, from
which the wind emanates.  This maximizes the effect of the wind and is
appropriate in modeling starbursts, but overestimates the wind
effectiveness if star formation is distributed over the galaxy.  On
the other hand, we assume that the wind energy is distributed
isotropically.  Realistically, winds in their early
thermally-dominated phase will be funneled into a bi-conical outflow
(e.g., Mac Low \& Ferrara 1999) 
observations show bi-conical outflows with solid angles of $\sim
0.8\pi-2.4\pi$ radians at small (few kpc) radii (HLSA; Lehnert \&
Heckman 1996).  Thus realistic winds would have a higher energy per unit
solid angle and a more effective outflow in those directions.}
\item{If $h^{\rm wind}$ is small compared to the size of the galaxy
but $r_{\rm stall} > 2r_0$, the given prescription will
artificially concentrate metals in the radius $h^{\rm wind}$ about
$\vec r_c$. This indicates that metallicity gradients in the objects
should not be trusted, and that the efficiency of dynamical removal
(which cannot easily remove metals from galaxy cores) may be
suppressed.}

\end{enumerate}

\begin{deluxetable}{cccc}
\singlespace
\footnotesize
\tablecaption{Wind Parameters}
\tablehead{\colhead{Parameter} & \colhead{Description} & \colhead{Default value} & \colhead{Range}}
\startdata
SFR$_{\rm crit}$ & critical SFR/(disk area) for thermal wind & $0.1{\rm M_\odot/yr/kpc^{2}}$ & 0.001-0.2 \nl
$v^{\rm fid}_{\rm out}$ & outflow velocity at initial radius $r_0$ & 600 km/s &
300-1000\nl
$\sigma_{\rm out}$ & width of outflow velocity distribution & 200 km/s & - \nl
$\chi$ & `fraction' of supernova energy in wind (c.f.~\S~\ref{sec-windmeth}) & 1.0 & 0.5-2 \nl
$\epsilon_{\rm ent}$ & fraction of ambient material entrained in wind & 0.1 & 0.01-1 \nl
$\xi$ & fraction of galaxy mass enclosed within initial radius & 0.1 & 0.05-0.2 \nl
$Y_{\rm ej}$ & fraction of metals distributed non-locally & 1.0 & 0.5-1.0\nl
$f^{\rm ej}_{\rm dust}$ & portion of ejected metals in dust & 0.5 & 0.0-0.5 \nl
$\alpha$ & $W^{\rm grp} \propto (r/h)^\alpha$ & 3 & 1-4 \nl
$\beta$ & Wind and initial shell energy attenuated by $({\rm
SFR}/{\rm SFR}_{\rm crit})^{2\beta}$. &
2 & $1,2,4,\infty$ \nl\hline
\enddata
\label{tab-fidparwind}
\doublespace
\end{deluxetable}

\subsubsection{Fiducial Parameter Values for Wind Ejection}
\label{sec-fidparwind}

	The key parameters we need to determine are the mean outflow
velocity $v_{\rm out}^{\rm fid}$, the width $\sigma_{\rm out}$ of
the velocity distribution, the fraction of supernova energy in the
wind $\chi$, the critical SFR/kpc$^2$ (SFR$_{\rm crit}$), the
enclosed mass fraction $\xi$ determining the initial shell radius, and
the entrainment fraction $\epsilon_{\rm ent}$.

	We base these values as much as possible on observations
of galaxies with supernova-driven winds, as recently compiled by HLSA
and by Martin (1999).  HLSA measure both the width $W \sim
300-600\kms$ and velocity offset $v_{\rm out} \sim 100-300\kms$ (from
the galaxy's inferred center-of-mass velocity) of NaI absorption
lines.  Their interpretation of these values is that cool material is
being accelerated by a hot outflowing wind. $v_{\rm out}$ is the
characteristic velocity of this outflowing material at small radii and
$v_{\rm term} \equiv v_{\rm out}+W/2 \sim 600\kms$ is the
inferred `terminal velocity' to which the dense gas is accelerated, i.e.
the velocity where it is roughly comoving with the hot gas (see HLSA).  This
view is supported by the rough agreement of the inferred $v_{\rm
term}$ values with those inferred from X-ray data (HLSA; Martin 1999).
>From the combined data set of HLSA we take $v^{\rm fid}_{\rm out} = 600\kms$,
$\sigma_{\rm out} = 200\kms.$ Very similar values of $v_{\rm out}$ and
$W$ are found for Lyman-break galaxies at $z\sim3$ (Pettini et al. 2000).
Somewhat unexpectedly, according to current data the outflow velocities do
{\em not} seem to be correlated with either the SFR or the mass of the
host galaxies (HLSA; Martin 1999; Heckman, private communication).
Hence we use a fixed value, though a different prescription may be
called for if future observations reveal some dependence on galaxy
properties.

In using the observed outflow velocities, we are making an important
assumption: a single velocity characterizes the outflowing cool gas
and the hotter wind. It is likely, however, that the multiple
components of the outflow have rather different
velocities.\footnote{Indeed, as shown in Fig.~\ref{fig-intsampz1.5}
the wind velocity substantially exceeds the shell velocity.}
Two-component plasma fits to the X-ray data tend to give temperatures
corresponding to velocities that bracket the $v_{\rm term}$ inferred
from absorption lines, but the X-ray observations only measure the
thermal energy of the gas; numerical simulations suggest that it may
have a kinetic energy $\sim 2-3$ times higher, i.e.\ velocities of
$\sim 800-1000\kms$ (HLSA; see also Strickland \& Stevens 2000). If
this hotter gas contains a mass comparable to the cool material, it
may dominate the pressure of the outflowing gas.  Another possibility
is that the outflowing cool gas may be better characterized by the
observed velocity offset $v_{\rm out}\sim 200-300\kms$.  This would
follow if the absorption occurred in a thin shell that might break up
but is inefficiently accelerated by the hotter wind. In this case, the
ram pressure of the outflow might be dominated by the high-velocity
but low-density wind, or alternatively by the higher-density but
lower-velocity entrained clumps.  This ambiguity is a significant
source of uncertainty, but the results should be bracketed by models
with velocities of $\sim300\kms$ and $\sim 1000\kms$, and we will test
such values in our calculations.

	The mass outflow rate from a wind-producing galaxy can be
roughly estimated by measuring the column density of the wind
material, then assuming either a thin shell (Pettini et al. 2000;
Martin 1999) or a spherical mass-conserving wind (e.g., HLSA).  Either
way, estimated mass outflow rates are $\sim 1-4$ times the galaxy's
estimated SFR.  Using Equation~\ref{eq-mdotchi}, this can be converted
into a fiducial value of $\chi \approx 1$. (This does not mean that we
assume supernovae drive winds with perfect efficiency, since probably
$E_{\rm 51} \neq 1$ or $M_{\rm 100} \neq 1$ in Eq.~\ref{eq-mdotchi};
$\chi$ is {\em proportional} to the true wind-driving efficiency,
calibrated by the observed velocities and mass outflow rates for
winds.)  
As noted above, for SFR/(area) $<$
SFR$_{\rm crit}$ we attenuate the wind energy by $({\rm SFR}/{\rm
SFR_{crit}})^{2\beta}$.  We use $\beta=2$ as a default, but investigate
$\beta=1,2,4,\infty$ to check the dependence on the abruptness of the
cutoff.

	The entrainment fraction for the winds depends on the
clumpiness of the IGM, and on instabilities in the outer shell of an
expanding wind.  There is currently no good basis -- theoretical or
observational -- for a particular assumption of $\epsilon_{\rm ent}$,
so we will try values between 1\% and 100\% with a fiducial value of 10\%.

	The critical SFR/area to drive a wind can be roughly addressed
using observations.  Observational samples of `normal'
star-forming disks versus starbursts indicate that normal disks tend
to have SFR/(area) of $0.001 - 0.1\msol{\rm\,yr^{-1}}\kpc^{-2}$,
whereas starburst regions typically have
$0.1-1000\msol{\rm\,yr^{-1}}\kpc^{-2}$ (Kennicutt 1998).  Heckman
(2000) finds a similar threshold for starburst-driven superwinds of
$0.1\msol{\rm\,yr^{-1}}\kpc^{-2}$.\footnote{Martin (1999) finds a much
smaller threshold of (few)$\times
10^{-4}\msol{\rm\,yr^{-1}}\kpc^{-2}$.  The reason for this large
disparity is unclear.}  We adopt this value as our fiducial SFR$_{\rm
crit}$.  This choice, as will be discussed in \S~\ref{sec-reswind},
leads to winds in most (but not all) galaxies at $z>2$, and about half
of the galaxies at $z=1$.  However, as noted above, the simulated galaxies
do not attain the {\em very} high areal SFRs seen in starburst galaxies
because we average the SFR over $\gsim 10^8$\,yr (longer than a typical
starburst `event'), and because the simulations cannot resolve the
scale of a typical starburst nucleus. Starbursts that {\em should}
occur in simulated galaxies are therefore washed out in both time and
space when we compute our areal SFRs.  Thus we are essentially
assuming that simulated galaxies exceeding the critical SFR would
contain the same sorts of starburst regions as observed galaxies, if
only such regions could be resolved.  We also investigate `quiescent'
models in which we set a smaller SFR/(area) of
$0.001\msol{\rm\,yr^{-1}}\kpc^{-2}$ so that essentially all galaxies
in the simulation drive weak winds.

	We start the shells at a radius enclosing a fraction $\xi$ of
the galaxy's baryonic mass, and choose $\xi$ so that the initial radii
obtained are comparable in scale to the areas observed in the studies
upon which we base the wind mass outflow rates and initial velocities.
These observed regions are typically 10-100 kpc$^2$; taking $\xi\sim
0.1$ (our fiducial value) gives $r_0 \sim 1-100\,$kpc at $z=0$ for
simulation galaxies with circular velocities similar to those of the
observed galaxies.  We check the
importance of $\xi$ in \S~\ref{sec-reswind}.

	Finally, we must choose a form for the distribution function
$W(r,h)$.  The final distribution of the wind material should
presumably be `piled up' where it is stopped by the IGM (or gravity),
as in the bow shocks often observed at the end of jets.  But any more
detailed assumption of $W(r,h)$ seems difficult to justify.  For
simplicity and generality, we assume $W(r,h) \propto (r/h)^\alpha$,
for $1 \le \alpha \le 4$ (with a default of $\alpha=3$), which should
span a range of reasonable cases.

\subsection{Prescription 3: Radiation Pressure Ejection}
\label{sec-radmeth}

For bright galaxies, the outward radiation pressure felt by a dust
grain in the interstellar medium (ISM) can exceed the inward
gravitational pull, suggesting the possibility that dust grains can be
expelled from galactic disks into halos or even into the IGM (see
Chiao \& Wickramasinghe 1972; Ferrara et al. 1990; Shustov \& Vibe
1995; Davies et al. 1998; Simonsen \& Hannestad 1999).  For a
spherical model galaxy with a radially increasing mass/luminosity
ratio $M/L$, this outward efflux would inevitably occur within some
critical radius where the radiation pressure and gravitational
forces on the grain balance.

In our prescription to model this physical effect, we distribute dust
near its progenitor galaxy in a way that reflects the equilibrium
distribution of dust `levitating' at the force-balance radius.  The
maximum radius at which dust can be in force equilibrium, $h^{\rm
dust}$, depends on the galaxy luminosity (which we compute using
spectral synthesis) and on the distribution of mass (given directly by
the simulation). The radial density profile of the levitating dust
depends mostly on the dust properties, described below in
\S~\ref{sec-dustmeth}.  We do not calculate the destruction of grains
before they reach this radius, but we do (as described in
\S~\ref{sec-dustmeth}) calculate thermal sputtering of dust
after it is deposited.

To implement this prescription numerically, we first divide the gas
and star particles into bound groups using the SKID algorithm as in the
wind prescription, and determine the center of star formation
$\vec r_c$ using Eq.~\ref{eq-cosf}.
Then about this center we determine a radius $h^{\rm dust}$ such that the
gravitational force due to the matter within $h^{\rm dust}$ balances
the outward radiation pressure due to the group's luminosity $L_\nu$,
i.e.\  
\begin{equation} 
{1\over 4\pi c h^2}\int d\nu\,L_\nu \kappa_\nu = {G M_{\rm
tot}(h) \over h^2}, 
\end{equation} 
where $M_{\rm tot}(h)$ is the total group mass within $h$, $L_\nu$ is
the group luminosity, $\kappa_\nu$ is the dust cross section/unit
mass, and we assume spherical symmetry.  Assuming a bolometric
luminosity $L_{\rm bol}$ and a radiation-pressure efficiency $Q_{\rm
pr}(T)$ (defined as the ratio of $\kappa_\nu$
to geometrical cross section per unit mass, $\pi a^2/(4\pi a^3\rho/3)$,
averaged over a blackbody of a temperature $T$ that reasonably
approximates the spectrum of a galaxy), we have
\begin{equation}
{L_{\rm bol} \over M_{\rm tot}(h^{\rm dust})} = {16\pi G c a \rho_{\rm dust}
\over 3 Q_{\rm pr}},
\label{eq-forcebal}
\end{equation}
where $a$ is the grain radius, giving
\begin{equation}
{L_{\rm bol}/L_\odot \over {M}_{\rm tot}/M_\odot(h^{\rm dust})} = 1.76
\left({Q_{\rm pr}(T) \over a[\mic]}\right)^{-1} \left({\rho_{\rm
dust}\over{\rm g\,cm^{-3}}}\right).
\label{eq-qpr}
\end{equation}

To compute $L_{\rm bol}$, we track the `effective' age $\tau_i$ of
each star particle, then take $L_{\rm bol}(\tau)$ (in units of
$L_\odot/M_\odot$) for a stellar population of that age from the
stellar synthesis models of Bruzual \& Charlot (1993),\footnote {The
models are available via anonymous FTP from ftp.noao.edu.} for an
assumed IMF.  We then sum this luminosity over all star particles in
the group:
\begin{equation}
L_{\rm bol} = \sum_{i \in {\rm grp}}m_i^s L_{\rm bol}(\tau_i).
\label{eq-lbol}
\end{equation}
  To compute $\tau$ for a newly formed star particle, we assume that
the SFR is constant between simulation outputs, so that the luminosity
of stars formed between time steps $\tp$ and $\tc$ can be
computed as
\begin{equation}
L_{\rm bol}(\tau_{\rm eff}) \equiv {1\over \tc-\tp}\int_{\tp}^{\tc}dt\,
L_{\rm bol}(t-\tp).
\end{equation}
We then invert $L_{\rm bol}(\tau_{\rm eff})$ to
obtain $\tau_{\rm eff}$.  When a new stellar mass
$\Delta m_i$ is added to an existing star particle of age $\tau_i(\tp)$ and
mass $m_i(\tp)$, the new effective age $\tau_i(\tc)$ will be given by
\begin{equation}
L_{\rm bol}(\tau_i(\tc)) = {L_{\rm bol}(\tau_{\rm eff})\Delta m_i+
L_{\rm bol}(\tau_i(\tp)+\tc-\tp)m_i \over m_i+\Delta m}.
\end{equation}

	As pointed out by Davies et al. (1998), the luminosity
calculated using Eq.~\ref{eq-lbol} must be corrected for extinction by
dust in the galaxy, because radiation re-emitted in the far-infrared
(FIR) provides negligible radiation pressure.  Large dust corrections
have been deduced for high luminosity local spirals (Wang \& Heckman
1996) and starbursts (Heckman et al. 1998), and in high redshift
star-forming galaxies (e.g., Calzetti \& Heckman 1999; Granato et
al. 2000).  As there is significant disagreement as to what dust
correction is appropriate in any given context, the topic requires
some discussion.  In both starburst nuclei and in disk galaxies, the
dust correction (i.e.\ the ratio $L_{\rm IR}/L_{\rm UV}$) appears to be
correlated with bolometric luminosity.  Heckman et al. (1998),
however, find that the correlation with metallicity is stronger, so
the correlation with luminosity (or mass) may be partially the
metallicity correlation combined with a luminosity-metallicity
relation.  For example, Heckman et al. (1998, Figure 2b) give a
metallicity-extinction relation of
\begin{equation}
\log(L_{\rm IR}/L_{\rm UV}) = 1.45+1.65\log(Z/Z_\odot),
\label{eq-zcorr}
\end{equation}
where $Z$ is the oxygen abundance relative to solar, $L_{\rm UV}$ is
luminosity at $0.19\mic$, and $L_{\rm IR}$ is the integrated
$\sim 40-120\mic$ luminosity.
  Combining this with
an (elliptical galaxy) luminosity-metallicity relation
\begin{equation}
\log(Z/Z_\odot)=0.4\log(L_B/\lsol)-4.4
\label{eq-massmet}
\end{equation}
(from Zaritsky, Kennicutt \& Huchra 1994),
we obtain
\begin{equation}
\log(L_{\rm IR}/L_{\rm UV}) = 0.66\log(L_B/\lsol)-5.8,
\end{equation}
that reproduces the slope of the 
disk galaxy extinction-luminosity relation 
\begin{equation}
\log(L_{\rm IR}/L_{\rm UV}) = (0.5\pm0.1)\log(L_B/\lsol)-(3.7\pm0.7)
\label{eq-lcorr1}
\end{equation}
of Wang \& Heckman (1996) tolerably well, even though
Eq.~\ref{eq-zcorr} was derived for starbursts, Eq.~\ref{eq-massmet} for
ellipticals, and Eq.~\ref{eq-lcorr1} was derived for spiral galaxies.
Neither Eq.~\ref{eq-zcorr} nor Eq.~\ref{eq-lcorr1} can
straightforwardly be applied as a dust correction: they relate the FIR
flux to the UV ($\approx 0.19\mic$) radiation rather than to the
UV-optical-NIR radiation (denoted henceforth by the subscript `opt')
that drives dust.  We assume a constant relation between UV and
UV-optical-NIR attenuation, given by $C_f\equiv \log(L_{IR}/L_{\rm
opt})$.\footnote{Realistically $C_f$ should be constant only for a
fixed effective galaxy temperature; see \S~\ref{sec-resrad}. Also,
since Eq.~\ref{eq-zcorr} applies to starburst regions, $C_f$ also
absorbs geometrical differences in extinction between starburst
regions and the dust corrected region.}
This constant is not `free', however, as there is the observational
constraint that the total integrated (over the cosmic history and all
galaxies) UV radiation as observed in the UV/optical/NIR background is
comparable to the total far infrared emission in the FIR background
radiation. Madau \& Pozzetti (2000) find that the energy in these two
radiation backgrounds have a ratio of $1/3 \lsim F_{\rm IR}/F_{\rm
opt} \lsim 2$, with a probable value of $\approx 1$.  Because
Eq.~\ref{eq-zcorr} along with a mass-metallicity relation can yield an
extinction-mass relation, and because the simulations should by
themselves give a mass-metallicity relation, we adopt
\begin{equation}
\log(L_{\rm IR}/L_{\rm opt}) = 1.45+1.65\log(Z/Z_\odot)+C_f
\label{eq-zcorr2}
\end{equation}
 as a fiducial dust correction, with $C_f$ set by requiring that the
total emitted FIR/UV ratio (an output of the calculations) is near
unity.  This typically leads to values of $C_f \sim -1$.
We also allow for the possibility of a dust correction depending
purely on luminosity (from Wang \& Heckman 1996):
\begin{equation}
\log(L_{\rm IR}/L_{\rm opt}) = 0.5\log(L_{\rm bol})-4.7+C_F.
\label{eq-lcorr}
\end{equation}
Here, we use the galaxy bolometric luminosities $L_{\rm bol}$
in place of Wang \& Heckman's $L_{\rm UV}+L_{\rm FIR}$.  Finally,
we also try a constant dust correction $\log(L_{\rm
IR}/L_{\rm opt}) = C_f$, in which case $C_f\sim 0$ from the background
density constraint.

The dust-corrected luminosity can be used with Eq.~\ref{eq-qpr} to
find $h^{\rm dust}$, the metal mass $\Delta w^{\rm grp}$ is then
distributed among the gas particles within the radius $h^{\rm dust}$
with the mass profile $W^{\rm grp}(r,h^{\rm dust})$ as in the wind
prescription.

For $W^{\rm grp} = \delta(r-h^{\rm dust})$, the prescription accurately
captures the physics of dust of a single grain size being ejected from
a spherical region with sufficient gas drag on the dust that the
grains do not attain escape velocity.  In general, the prescription
captures the essential physics of dust ejection fairly well, but
nevertheless has a number of important shortcomings:
\begin{enumerate}
\item{Charged grains can be confined to galaxies by magnetic fields.
Dust might escape if magnetic fields have a component perpendicular to
the galaxy; winds or Parker instabilities exacerbated by the radiation
pressure may enhance such a component (Chiao \& Wickramasinghe 1972;
Ferrara et al. 1991; Shustov \& Vibe 1995).  Ferrara et al. (1991) and
Davies et al. (1998) have also noted that grains are charged only
sporadically.  Magnetic fields may also be much weaker at high
redshifts if
they have been amplified by a dynamo.  Determining the effects of
magnetic fields would require information about the magnetic
structure of galaxies, their halos, and the IGM that is presently
unavailable, so we will {\em assume} in this study that dust escapes
magnetic confinement (though we emphasize that it also possible that
it does not).}
\item{After its deposition, dust is coupled to gas (i.e.\ follows the gas particles).
Thus dust distributed in the halo can simply fall back into the
galaxy, rather than maintaining its distance from the galaxy.}
\item{More detailed calculations show that grains often attain a high
velocity at small radii, carrying them past $h^{\rm dust}$ and perhaps
to `infinity'.  Thus our estimates of the radius to which grains could
escape is probably an underestimate, because the grains would inevitably
reach that radius with some outward velocity that would carry them
past it.  However, if the gas drag is high (or if magnetic
fields are important), the time to escape the galactic disk could
exceed the dust destruction time in that environment. Thus only in
certain situations would the dust escape the inner galactic region and
`levitate' in the halo as the prescription describes. See Aguirre et al.
(2001b) for some discussion.}
\item{As in the wind case, the method is only fully consistent if the
dust ejection time (i.e.\ the transport time from the disk into its
final equilibrium position) is shorter than the interval between
simulation outputs or between the ejection time and the time at which
results are desired, and shorter than the timescale over which the
galaxy properties change.  The errors at late times are unlikely to be
important (see \S\S~\ref{sec-windmeth},~\ref{sec-reswind}) unless the
transport time substantially exceeds the time between ejection and
`observation'. Dust moving with the high velocities (up to $\sim
1000\kms$) found by Shustov \& Vibe (1995) would not encounter this
problem but slower-moving dust might.}
\item{The distribution of metals occurs without regard to the
existence of nearby galaxies whose luminosity and gravitation would affect
the force balance.  Also, as in the wind prescription galaxies can
(unrealistically) enrich each other with their metallic outflows,
though this is a small effect with the distribution method we employ.}
\item{We ignore the radiation-related forces on grains due to
photodesorption and the photoelectric effect (see Weingartner \& Draine
2000).  Both of these effects would increase the radial force on the
grain by a small factor.  Also, we neglect the luminosity due to
accretion onto black holes.  
}
\item{Galaxies are not spherical, so the spatial distribution of
`levitating' grains will only be approximately spherical if $h^{\rm dust}$
is much larger than the characteristic size of the galaxy.}
\item{The dust correction we apply is rather uncertain.  We have
chosen a dust correction to the UV-optical-NIR luminosity based on
metallicity, and also tried a luminosity-dependent correction.  The
metallicity correction is reasonable and in accord with observations,
but it neglects the {\em distribution} of dust within galaxies, the
gas fraction, and also the effective temperature of the radiation.  At
higher redshift, galaxies are probably more compact, have higher gas
fractions, and emit more light at short wavelengths where dust
attenuation is more effective.  All of these effects may increase the
importance of dust at high $z$ more than the decrease in metallicity
suppresses extinction.  But the luminosity-dependent correction we try
{\em is} stronger at high $z$ and can be used as a check on our
assumptions.}
\end{enumerate}

While our prescription for modeling the ejection of dust could be
significantly improved given a better understanding of the dust
ejection process and of the correct dust correction, it should 
give a reasonable estimate of which galaxies could eject dust, and of the ejection
radius.

\begin{deluxetable}{cccc}
\singlespace
\footnotesize
\tablecaption{Parameters for Dust Ejection}
\tablehead{\colhead{Parameter} & \colhead{Description} & \colhead{Default value} & \colhead{Range}}
\startdata
$Y_{\rm ej}$ & fraction of metals distributed non-locally & 0.5 & 0.0-1.0 \nl
$f^{\rm ej}_{\rm dust}$ & portion of ejected metals in dust & 1.0 & 0.0-1.0 \nl
$Q_{\rm pr}/a\rho$ & absorption efficiency for dust & $19 (G), 4 (S)$ & - \nl
$C_f$ & free constant in dust correction & fixed by $L_{\rm IR}/L_{\rm opt}$ & - \nl
correction & dust correction type & $Z$ & $Z$/$L_{\rm bol}$/constant \nl
IMF & IMF for $L/M$ determination & Scalo & Scalo/Salpeter \nl\hline
\enddata
\label{tab-fidpardust}
\doublespace
\end{deluxetable}

\subsubsection{Fiducial Parameter Values for Dust Ejection}

	The key physical parameters in the dust ejection model are the
dust absorption efficiency, the dust distribution function $W^{\rm
grp}$, the adjustment factor $C_f$ in the dust correction to the
galaxy luminosity, and the IMF used in the spectral synthesis.

	For the dust absorption efficiency, we adopt the values
calculated by Laor \& Draine (1993) for silicate and graphite dust,
averaged over a Planck spectrum of $T=12000\,$K.\footnote{We do not
account for the effect of extinction on the effective temperature of
the escaping radiation.}
 We also assume a
specific gravity of $2.2{\rm \,g\,cm^{-3}}$ for graphite and $3.3{\rm
\,g\,cm^{-3}}$ for silicates.  These assumptions yield maximum values
of $Q_{\rm pr}/a\rho \lsim 19(4)$ for graphite (silicate) grains.  For
$T \approx 10000\,$K and $T\approx 8000\,$K these values are 16(3) and
13(2), respectively.  These values determine the {\em maximum} radius
to which grains could levitate; but grains of different sizes will
levitate to different radii, owing to their range in $Q_{\rm pr}$.

  For a general grain mass-size distribution in mass $dm(a)/da$ and
absorption efficiency $Q_{\rm pr}(a)/a$, we can derive an approximate
form of $W^{\rm grp}(r,h^{\rm dust})dr$ by assuming a flat galactic
rotation curve, i.e.\ $M_{\rm tot}(r) \propto r$.  Then the force
balance equation (Eq.~\ref{eq-forcebal}) gives $r = K Q_{\rm pr}(a)/a$ for some
proportionality constant $K$, and we can take
\begin{equation}
W^{\rm grp}(r) = \int_{a_{\rm min}}^{a_{\rm max}} da {dm(a)\over da}
\delta (r-K Q_{\rm pr}(a)/a),
\end{equation}
that should accurately capture the mass distribution of levitating
grains with a given grain-size distribution. For the $Q_{\rm pr}$
values of Laor \& Draine with $T=12000\,$K, and the `PED' grain size
distribution of Kim, Martin \& Hendry (1994; see
\S~\ref{sec-dustmeth}), the derived $W(r,h)$ can be reasonably fit by
a second-degree polynomial, which is used in the actual algorithm.

	The default value of $C_f$ will be taken as the value
necessary to give $F_{\rm IR}/F_{\rm opt} \simeq 1 $ in the background
radiation, and the default IMF is Scalo (1986), from $0.1\,M_\odot$ to
$125\,M_\odot$.

The fiducial parameter values just described are summarized in
Table~\ref{tab-fidpardust} and are used in the fiducial models listed in
Tables~\ref{tab-dynmodrun}, \ref{tab-windmodrun}, and \ref{tab-dustmodrun}.

\subsection{Other Parameters and Considerations\label{sec-otherpar}
}

A few more general considerations (and their associated parameters)
are common to two or more of our prescriptions and we discuss them
here.  The total normalization of the metal mass in our calculations
is given by the `effective yield' $y_*$, defined as the ratio of metal
mass returned to the ISM to the stellar mass formed, in solar units.
We take $y_*$=1 (i.e.\ solar yield) for all runs both because this is
conventional and because all results can be simply scaled to a
different effective yield.\footnote{The metallicity-dependent dust
correction in the radiation-pressure prescription does depend on
$y_*$, but the effect is largely counteracted by changing $C_f$ to
ensure the proper $F_{\rm FIR}/F_{\rm opt}$ in the derived optical and
FIR backgrounds.}  As discussed later, a higher overall effective
yield may in fact be called for by the observations (see also Renzini
1997; Pagel 1999; Aguirre 1999).

	The yield $y_*$ is split into gaseous metals and metals locked
in dust.  For metals distributed in the local gas, we take the ratio
$f_{\rm dust} \equiv $(dust mass)/(total metal mass)$ = 0.5$, as in
local galaxies, and as suggested by observations of damped Ly-$\alpha$
absorbers (Pei, Fall \& Hauser 1999; but see Pettini et al. 1997 for a
lower estimate).  For ejected metals, we allow a different ratio
$f^{\rm ej}_{\rm dust}$. For wind ejection, we should have  $f^{\rm ej}_{\rm
dust} \lsim f_{\rm dust}$, which is sensible if most of the ejected
metal mass is in the ISM of the galaxy that has been entrained by the
wind (it is an upper limit because some dust would be destroyed during
ejection). The value may be somewhat different -- in a quite unknown
way -- if most metals are contained in the hot wind itself. For
radiation pressure ejection, $f^{\rm ej}_{\rm dust} \lsim 1$ applies;
the fraction then represents the survival fraction of dust as it
traverses the halo during its ejection.

	The fraction $Y_{\rm ej}$ of metals that are ejected (versus
locally distributed) could vary anywhere from near zero, for radiation
pressure ejection where gas drag or magnetic confinement is very
strong, to near one or more for galactic-scale winds where the
metal-rich supernova ejecta escapes along with some entrained gas.  We
use a fiducial value of $Y_{\rm ej} = 1$ for `superwinds', and $Y_{\rm
ej}=0.5$ for quiescent winds.  For dust, $Y_{\rm ej} > 0.5$ could not
be maintained for much of a galaxy's lifetime since only $\sim 1/2$ of
a typical galaxy's metals are in dust at any given time, and even
values of $Y_{\rm ej} \sim 0.5$ would severely change the abundance
ratios of refractory vs. non-refractory elements in the galaxies.
Hence, this is probably an upper limit.

	A final parameter $\epsilon_*$ is introduced in an effort to
correct for a possible disparity between the simulations and reality:
whereas the simulations tend to find $\Omega_*^{\rm sim}(z=0) \gsim
0.011$, values of $0.002 \lsim \Omega_*^{\rm obs}(z=0) \lsim 0.006$ are
estimated from observations (e.g., Fukugita, Hogan \& Peebles 1998).  Three
possible reasons for this disparity (discussed in
\S\ref{sec-simsvsobs}) are: 1) The simulation $\Omega_*$ is correct but
most of the `stellar' mass is unobservable, e.g., brown dwarfs, 2)
the simulations over-estimate the efficiency of star formation, or 3)
the observations underestimate $\Omega_*$.  In cases 1) and 2), the
simplest reasonable correction to make is to multiply the SFR and the
luminosity of the simulation galaxies by $\epsilon_* = \Omega_*^{\rm
obs}(z=0) / \Omega_*^{\rm sim}(z=0) \gsim 0.011 \approx 0.36$ wherever
they are used.  The stellar yield $y_*$ should also be multiplied by
$\epsilon_*$, although if case 2) holds the resulting stellar
metallicities will be too low.

\begin{deluxetable}{cccc}
\singlespace
\footnotesize
\tablecaption{General Parameters}
\tablehead{\colhead{Parameter} & \colhead{Description} & \colhead{Default value} & \colhead{Range}}
\startdata
$y_*$ & mean stellar yield (solar units) & 1 & - \nl
$\epsilon_\star$ & factor multiplying SFR, $\dot m_{\rm out}$, $L_{\rm bol}$ & 1.0 & 
0.355-1 \nl
$f_{\rm dust}$ & portion of locally-distributed metals in dust & 0.5 & - \nl \hline
\enddata
\label{tab-fidpargen}
\doublespace
\end{deluxetable}

\subsection{Treatment of Dust and Dust Destruction}
\label{sec-dustmeth}

Our method incorporates a fairly detailed treatment of IG dust.  We
track both the dust mass and grain-size distribution for each gas
particle, and treat the conversion of dust to gaseous metals by thermal
sputtering.  Dust is added to pristine gas with a set grain-size
distribution, but when adding dust to a gas
particle that has dust, we average the grain size distributions.

The
extinction properties of dust in galaxies appear to be well fit by
models employing spherical grains with a power-law grain-size
distribution.  The grain sizes necessary to account for the extinction
data range from $a_{\rm min} \sim 0.001\mic$ up to some cutoff
(generally either sharp or exponential) above $\sim 0.2\mic.$ We
represent the grain-size distribution by a set of power laws, i.e.
\begin{equation}
{dN(a)\over da} \propto a^{\alpha_k-3}\ \ {\rm for\ } a_k \le a <
a_{k+1},\, k=1..N,
\end{equation}
for some set of $\alpha_k$, with $dN(a)/da$ a continuous function.  We
implement this by tracking the dust mass $d_k(i)\equiv dm_i^{\rm
dust}(a_k)/da$, where $i$ is the particle number.  The differential dust
mass values at the selected set of $a_k$ then determine
the shape of the grain size distribution; clearly for $N\rightarrow
\infty$ this allows for an arbitrary grain-size distribution; but in
practice the method is `efficient' enough that a small number ($N\sim
9$) provides good accuracy (see the upper panel of Figure~\ref{fig-dusttest} for
a demonstration).

	The total dust mass in a particle is determined by a piecewise
power-law integration over $d_k$, i.e.\ 
\begin{equation}
m_i^{\rm dust}=\sum_{k=1}^{N-1}{d_k(i)\over \alpha_k+1}\left[\left({a_{k+1}\over
a_k}\right)^{\alpha_k}a_{k+1}-a_k\right]
\end{equation}
where $\alpha_k=\log(d_{k+1}/d_k)/\log(a_{k+1}/a_k)$. For $\alpha_k-1 <
\epsilon \approx 10^{-4}$ a term in the sum is replaced by
$d_k(i)a_k\log(a_{k+1}/a_k)$, which is good to $O(\epsilon^2)$.

  To add dust to a particle, we simply add to the $d_k(i)$ values of the
assumed grain size distribution at $a_i$, such that the total mass
corresponding to the added distribution is equal to the desired
additional dust mass.  This in general will {\em not} give a set of
$d_k$ that yield the correct total mass (though it will be
close), so we `renormalize' the entire distribution, since the total
mass scales with a constant multiplying all the $d_k(i)$.  This
effectively yields a `best fit' of the two summed piece-wise power-law
distributions using a third with the same mass. A sample addition is
shown in Fig~\ref{fig-dusttest} (lower panel).

For dust destruction, we find the sputtered radius $a_s$ using the gas
temperature and density from the simulation, the time interval since
the last simulation output, and the thermal sputtering yields of Jones
et al. (1994). The grain distribution is then transformed by $dN(a)/da
\rightarrow dN(a+a_s)/da$, or in practice $d_k \rightarrow d_k'$,
where the latter is given by a power-law interpolation: $$
d_k'=d_j\left({a_k+a_s\over a_j}\right)^{\alpha_j}\left({a_k\over
a_k+a_s}\right)^3,$$ where $j$ is selected so that $a_j \le a_k+a_s <
a_{j+1}.$ For $j \ge N$ we use $j=N-1$, i.e.\ extrapolation.  The
fraction of dust destroyed could then be calculated by integrating
this new set of $d_k'$ and comparing to the old. This gives a good
approximation to the effect of thermal sputtering on the dust mass and
grain-size distribution.  We have not calculated non-thermal sputtering
of the grains (because it requires grain velocities), but it could be
important: the grain velocity $v_{\rm dust}$ would, in the grain's
rest frame, correspond roughly to a temperature of $6\times
10^5(v_{\rm dust}/100\kms)^2\,K$.

\ifthenelse{\boolean{apj}}{
\vbox{ \centerline{ \epsfig{file=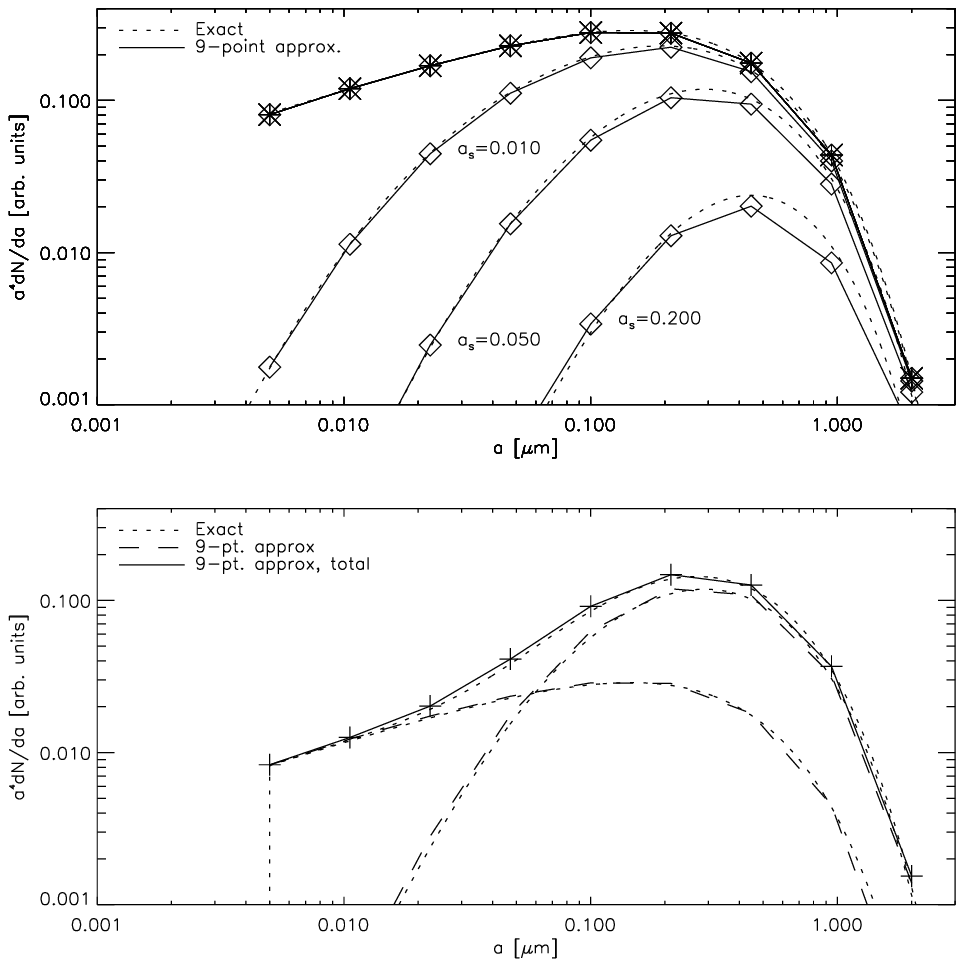,width=9.0truecm}}
\figcaption[]{ \footnotesize Demonstration of the dust treatment. {\bf
Top:} Unsputtered (stars) and sputtered (diamonds) grain-size
distributions, for several values of the sputtered radius, $a_s$, for
graphite grains. Dotted lines show the exact distribution; solid lines
show the simulations representation using 9 points.  {\bf Bottom:}
sputtered distribution with $a_s=0.05\mic$ added to an unsputtered
distribution with half the mass, yielding a new grain-size distribution
(solid).  Exact solutions are dotted, 9-point approximations are
dashed and solid.
\label{fig-dusttest}}}
\vspace*{0.5cm}
}{
\begin{figure}[tbp]
\centerline{\epsfxsize=5.0in
\epsffile{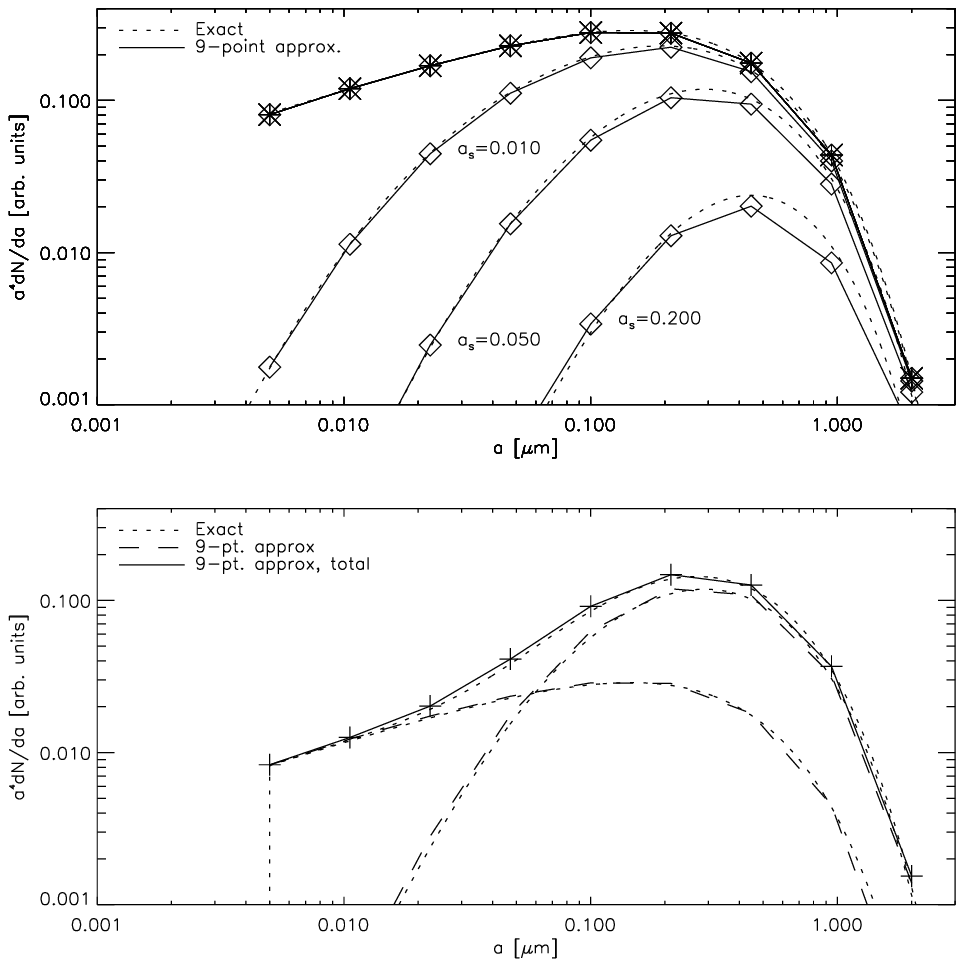}}
\caption{\baselineskip=12pt
Demonstration of the dust treatment. {\bf
Top:} Unsputtered (stars) and sputtered (diamonds) grain-size
distributions, for several values of sputtered radius $a_s$, for
graphite grains. Dotted lines show the exact distribution; solid lines
show the simulations representation using 9 points.  {\bf Bottom:}
sputtered distribution with $a_s=0.05\mic$, added to an unsputtered
distribution with half the mass yielding a new grain-size distribution
(solid).  Exact solutions are dotted, 9-point approximations are
dashed and solid.}
\label{fig-dusttest}
\end{figure}}

	In this study, for the grain-size distribution we use the
`power-law exponential decay' (PED) fits of Kim, Martin \& Hendry
(1994) of $N(a) \propto a^\alpha\exp(-a/a_c)$.  They give
$\alpha=-3.48, a_c=0.28\mic$ for graphite, and $\alpha=-3.06,
a_c=0.14\mic$ for silicates.

While the dust treatment employed here is fairly accurate, it currently has a few limitations:
\begin{enumerate}
\item{The dust destruction is underestimated in the simulations
because we have neglected non-thermal sputtering, and dust destruction
during ejection. Therefore, we only compute dust destruction correctly 
{\em after} the grains have come to rest in the IGM.}
\item{Dust destruction may also be
underestimated if dust encounters very dense, hot regions for shorter
time intervals than the interval between the simulation outputs.
for example, this could happen for dust cycling through cluster
cores.}
\item{We currently only treat one grain species at a time, and add
dust to gas with a fixed grain-size distribution, even though in a
realistic ejection scenario dust should be somewhat segregated by size
due to differences in grain absorption efficiencies and gas drag.}
\end{enumerate}

\begin{deluxetable}{cccc}
\singlespace
\footnotesize
\tablecaption{Parameters for Dust Treatment}
\tablehead{\colhead{Parameter} & \colhead{Description} & \colhead{Default value} & \colhead{Range}}
\startdata
$a_{\rm min}$ & minimal grain size & $0.005\mic$ & - \nl
$a_{\rm max}$ & maximal grain size & $2.0\mic$ & - \nl
dust type & dust chemical composition & graphite & graphite/silicate \nl
GSD & grain-size distribution & PED & - \nl \hline
\enddata
\label{tab-fidpargrain}
\doublespace
\end{deluxetable}

\section{Observations Concerning Cosmic Metals}
\label{sec-obssumm}

	Sections~\ref{sec-resdyn}-\ref{sec-resrad} will discuss the
models we have run, and the results obtained.  First, however, it will be
useful to briefly review the existing observations concerning cosmic
metallicity.

\subsection{Cluster and Group Metallicity}

Measurements of elemental abundances using X-ray line emission have
revealed that the intracluster medium of rich galaxy clusters is
highly enriched, to between 1/3 and 1/2 solar metallicity (e.g.,
Mushotzky et al. 1996).  Because the intracluster gas mass in a
typical rich cluster exceeds the stellar mass in cluster galaxies by a
factor of $\sim 5-10$ (e.g., Renzini 1997), this implies that cluster
galaxies have probably ejected a large fraction (possibly up to three
quarters) of their metals.\footnote{If a significant fraction of
cluster stars are intergalactic, the factor might be reduced, but
galaxies must still lose $\sim 1/2$ of their metals.}  It also seems
that a super-solar yield is necessary to account for all of the
metals, since dividing the total cluster metal mass by the total
stellar mass in the cluster gives a yield of $y_* \sim 2-4$ (Renzini
1997; see also Aguirre 1999).  It should be noted that measured
cluster metallicities are emission-weighted and favor the cluster
cores; strong radial metallicity gradients would imply a different
mean metallicity for the cluster.  Cluster abundance gradients are
currently somewhat inconclusive (see Renzini 2000), but seem to be
weak except in clusters with strong cooling flows (White 2000;
Finoguenov, David \& Ponman 1999).  

While there is general agreement that in the cores of
 clusters at $z \lsim 0.3$ the gas typically has roughly constant $Z/Z_\odot
 \approx 0.3-0.5$ for cluster temperatures down to $\sim 3.0\,$KeV
 (e.g., Renzini 1999), the observed enrichment of cooler clusters and
 groups is less secure.  Davis, Mulchaey \& Mushotzky (1999) find that
 $Z/Z\sim 0.3$ for intra-group gas persists down to temperatures $T
 \sim 1.5\,$KeV, but for cooler groups the metallicity -- and perhaps
 also the baryon fraction -- declines, perhaps implying that winds can
 remove metals even from groups.  Buote (2000), however, finds that
 two temperature models fit the X-ray data better than the
 one-temperature models used by Davis et al. and others---these 
 fits yield significantly higher abundances in groups
 ($Z/Z_\odot \sim 0.5-1$).  Well-resolved observations using {\em
 Chandra} should provide much firmer data on the enrichment properties
 of groups.

\subsection{Metals in Ly-$\alpha$ Absorbers}
    
    A very useful window into the chemical enrichment of the IGM is
provided by studies of quasar absorption lines.  It is now widely
accepted that the `forest' of Ly$\alpha$ lines found in the spectra of
$z\gsim 2$ quasars is due to absorption by a smoothly fluctuating
neutral hydrogen component of the IGM (e.g. Cen et al. 1994; Zhang,
Anninos \& Norman 1995; Hernquist et al. 1996).  Although these
absorbers were initially expected to be pristine, high resolution
spectroscopy has unambiguously identified metal lines (chiefly CIV and
SiIV) associated with $N(H\,I) \gsim 10^{14.5}\,{\rm cm\,^{-2}}$
absorbers (e.g., Cowie et al. 1995; Songaila \& Cowie 1996; Cowie \&
Songaila 1998; Ellison et al. 2000).  Applying an ionization
correction to the abundances derived from the line column densities
(relative to the absorbers' H I column densities), and using a
correlation between the Ly$\alpha$ column density and gas overdensity
from numerical simulations, these lines can give useful information
about the metallicity of the low-density component of the IGM.  All of
the absorption line studies essentially agree on an inferred
metallicity of $Z/Z_\odot \sim 10^{-2.5}$ for $2.5 \lsim z \lsim 3.5$
absorbers with $N(H\,I) \gsim 10^{14.5}\,{\rm cm\,^{-2}}$, with about
an order of magnitude scatter in the metallicity for different
absorbers (e.g., Songaila \& Cowie 1996; Rauch, Haehnelt \& Steinmetz
1997; Songaila 1997; Hellsten et al. 1997; Dav\'e et
al. 1998). However, the metallicity of lower column density regions,
corresponding to physical overdensities $\rho/\bar\rho \sim 1$, which
would represent the bulk of baryonic matter at $z\sim 3$, is more
uncertain. The metal lines corresponding to the low-column density
Ly-$\alpha$ absorbers are generally too weak to detect directly and
their presence can only be derived in a statistical sense, e.g., by
analyzing the median absorption per pixel (Cowie \& Songaila
1998). Recently, Ellison et al. (2000) used a very high quality quasar
spectrum to show that the CIV enrichment must extend to column
densities significantly lower than $N(H\,I) = 10^{14.5}\,{\rm
cm\,^{-2}}$. Schaye et al. (2000) demonstrated that OVI is a more
sensitive probe of the metallicity in low-density gas than CIV. Using
a pixel analysis, they detected OVI in gas with $\tau(HI) < 1$, which
corresponds to gas densities around the cosmic mean.  (See, also,
Hellsten et al. 1998; Dav\'e et al. 1998.)
 
        In summary, the typical metallicity of the high column density
IGM is $10^{-3}$ -- $10^{-2}$ solar. Very little is known about the
variation and the scatter of the metallicity as a function of
density. Although the presence of metals in the low-density IGM has
been established using statistical techniques, both the overall mean
metallicity of the IGM and the fraction of the IGM that is enriched
have yet to be determined.

\subsection{Metals in Galaxies}

	To develop a consistent theory of cosmic metallicity, one must
take into account not only metals in the IGM, but also metals
distributed in and near the metal-forming galaxies.  The
chemical properties of galaxies constitutes a vast subject (see,
e.g., Pagel 1997), and here we will only summarize some observations
that will be of use in evaluating the results of our simulations.

	The abundances of heavy elements in galactic gas and stars are
known to vary with galaxy mass or luminosity and galactic radius,
and perhaps galaxy type.  All three effects are evidenced in the
very useful compilation by Zaritsky, Kennicutt \& Huchra (1994).  They
note that nearly all observed galaxies demonstrate radial abundance
gradients, making it difficult to assign a particular metallicity to a
galaxy.  However, by choosing a `characteristic' radius (either an
isophotal radius or the disk exponential scale length), they can
compare abundance properties of various galaxies at that radius.  This
reveals a strong metallicity-luminosity relation (see
Eq.~\ref{eq-massmet} above and Zaritsky et al., Figure 13) for $[O/H]$
in spiral H II regions.  A similar but shallower M-Z relation
exists in $[Fe/H]$ in stars in ellipticals (Zaritsky et al. 1994;
Kobayashi \& Arimoto 1999).  In both cases the characteristic
metallicity can range from $\ll 1/10\,Z_\odot$ in the smallest
galaxies to several times solar in the largest.  These relations can
be meaningfully compared to the gaseous and stellar metallicities in
the simulation galaxies at $z=0$, although we do not have information
on the Hubble type of the simulation galaxies.

	At high redshifts, data concerning the chemical properties of
galaxies can be gleaned from observations of Lyman-break galaxies or from
studies of damped Ly$\alpha$ absorbers. Unfortunately, metallicity
information about Lyman-break galaxies is extremely limited (Pettini et
al. 2000) and we cannot make a meaningful comparison with our
simulation metallicities at $z=3$.  The damped Ly$\alpha$ absorbers
constitute the highest column density ($N(H I) \gsim 10^{20}\,{\rm
cm^{-2}}$) features of QSO absorption spectra.  Abundances have been
measured for many of these systems, giving metallicities of $\sim
1/30-1/10$ solar (e.g., Pettini et al. 1999; Prochaska \& Wolfe 2000).
It is unclear, however, exactly what sort of systems the damped Lyman
absorbers represent; they may arise from a diverse population of
objects (e.g., dwarf galaxies, outer disks of spirals, etc.; see
Pettini et al. 1999 and references therein). Since we cannot draw a
one-to-one relation between these objects and an overdensity or
average density of a galaxy, we will not compare our simulations to
these observations in the present study. 

\section{Galaxy Masses and Star Formation rates in the Simulations}
\label{sec-simsvsobs}

	In this study we apply the method described in
\S~\ref{sec-method} to three SPH simulations, performed using the
method described by Katz, Weinberg \& Hernquist (1996).  The
simulations themselves are described more specifically in Murali et
al. (2001). Briefly, all three use parameters
$\Omega_m=0.4=1-\Omega_\Lambda$, $\Omega_b=0.02h^{-2}$, $\sigma_8=0.8$
and $h=0.65$.  The first simulation has $144^3$ gas and $144^3$ dark
particles in a $(50\,h^{-1}\,{\rm Mpc})^3$ box, giving a dark particle
mass of $6.3\times10^9\msol$ and a gas particle mass of
$8.5\times10^8\msol$.  The other simulations are both
$(11\,h^{-1}\,{\rm Mpc})^3$ in volume, and use $2\times 128^3$ and
$2\times 64^3$ particles, respectively.  The $128^3$ simulation has
dark and gas particle masses of $9.8\times10^7\msol$ and
$1.3\times10^7\msol$, respectively.  The smaller simulation boxes stop
at $z=3$, and most conclusions regarding the universe at $z=3$ will be
drawn from these.  The $144^3$ simulation runs down to $z=0$.  The
$128^3$ and $64$ simulations include an ionizing background of from
Haardt \& Madau (1996) (attenuated by a factor of two to adjust for
our assumed $\Omega_b$; see Weinberg, Hernquist \& Katz 2000), whereas
the $144^3$ simulation does not.

	For our method to yield useful results, it must be applied to
simulations that can reasonably reproduce the mass, luminosity, and
SFR distributions of observed galaxies.  Here we briefly discuss the
comparison of the simulation mass function and SFR to observations and
their role in our predictions (for more details see Weinberg et al.\
1999, 2000; Murali et al. 2001; Katz et al., in preparation, Bullock
et al., in preparation).  The luminosity function, which is generated
in our calculations of radiation-pressure ejection, is discussed in
\S~\ref{sec-resrad}.

	The simulations yield a mass function of galaxies directly.
An `observational' mass function can be constructed using an observed
luminosity function and a prescription for converting light to mass.
Using either a constant $M/L_B=7.5$ (in solar units) or a $M/L_B$
function derived by Salucci \& Persic (1999; Persic \& Salucci 1997),
the mass function from the 2dF survey (Folkes et al.\
1999)\footnote{The ESO slice project results of Zucca et al. (1997) are
very similar.} survey shows quite good agreement -- in both shape and
normalization -- with the $144^3$ simulation for galaxy (gas+star)
masses between $10^{10.5}\,M_\odot$ and $10^{11.75}\,M_\odot$ (Katz et
al., in preparation). At lower masses, galaxies are unresolved by this
simulation: as shown by Weinberg et al. (1999), only galaxies with
$\gsim 60$ SPH particles are well resolved; this corresponds to
baryonic masses of $5.1\times10^{10}\msol$ in the $144^3$ simulations
and $7.9\times 10^8\msol$ in the $128^3$ runs.  At higher masses the
simulation exhibits a significant excess of galaxies.  This may result
from inaccuracies in the mass-to-light conversion for very massive
galaxies, or from incompleteness of the surveys (e.g., due to surface
brightness effects; see Impey \& Bothun 1997), or due to differences
in the way masses of real and simulated galaxies are estimated, or due
to overproduction of massive galaxies in the simulations.  The last
two uncertainties are exacerbated by the special environments of the
largest galaxies, most of which are found in cluster cores.

When integrated, the simulation mass function for the $144^3$ runs
yields a cosmic density in stars of $\Omega_*(z=0)=0.011$.  This is
somewhat higher than the value of $\Omega_* = 0.004 \pm 0.002$ derived
by Fukugita et al. (1998) from observed luminosity functions, or from
the integrated SFR of Steidel et al. (1999), which yields
$\Omega_*\approx 0.006$.  Likewise, the simulation SFRs both at low
and high redshift tend to exceed the observed values (see Weinberg et
al. 1999).  A similar discrepancy was noted based on earlier
simulations by Katz et al. (1996) and by Pearce et al. (1999).  These
discrepancies may be caused by an observational underestimate of
$\Omega_*(z=0)$ and of the SFR (due, for example, to surface
brightness effects or high production of low-mass, dim stars), and/or
by simulation galaxies being too massive by some roughly constant
factor, and/or by the overproduction of certain galaxies in the
simulations.\footnote{ The high-resolution runs seem to over-produce
small galaxies at $z=3$ as compared to the luminosity function
computed by Steidel et al. (1999).}

In summary, all of the simulations fail to resolve the fraction of the
mass function below the simulation resolution limit.  At the same
time, the simulations may nevertheless {\em over}predict the number
and/or mass of galaxies that are resolved.  We will address these
issues in detail in future papers (Katz et al., in preparation;
Bullock et al., in preparation; see Weinberg et al. 1999 for
preliminary discussion), but note for the present that the large SFR
and possible overabundance of galaxies we find could result from
the single-phase description of the gas in the simulations, or from
the lack the of strong feedback.  Feedback could suppress star
formation both in the wind-driving galaxy and in nearby galaxies (see
Scannepieco \& Broadhurst 2000 for some discussion of the latter
effect), but there is no good way to model such feedback using our
current methodology.  If enrichment were caused mainly by dust
ejection, dynamics or metal-rich, quiescent winds, our neglect of the
winds' effect on the galaxies would be more self-consistent, though
effects other than strong feedback would then be required to suppress
any excess of small galaxies.

	Any real disagreements between the simulations and reality
will clearly affect the accuracy of the IGM enrichment calculations of
this study.  We test a possible way of compensating for such
disagreements by including a parameter $\epsilon_*$ that can multiply
the SFR (which effectively sets the supernova rate in our method), the
yield $y_*$, and the luminosity of the groups.  With $\epsilon_* =
\Omega_*^{\rm obs}/\Omega_*^{\rm sim}$, this would be an accurate
adjustment if the discrepancy were due to hidden low mass stars, since
most of the radiation, metal and energy comes from the most massive
stars.  If simulation galaxies are simply more massive by a constant
factor $\Omega_*^{\rm sim}/\Omega_*^{\rm obs}$, the same adjustment
would properly account for the different galaxy luminosities, galaxy
SFRs, and overall enrichment of the universe.  However, since the
galaxies would still have their original unadjusted masses, we would
effectively {\em over}estimate the effect of gravitation as well as
underestimate the SFR per unit mass in the galaxies.  Finally, if the
`observed' values are underestimated (and the simulations correct), no
adjustment is necessary; the galaxies {\em that the simulations
resolve} would be like their counterparts in reality.  However, as
noted above it seems likely that the low-mass galaxies (at least)
are significantly overproduced in the high resolution simulations at
$z \ge 3$.

\section{Results for Dynamical Removal}
\label{sec-dresults}

	The removal of metal-enriched gas from galaxies by dynamical
processes has been studied in the context of clusters (e.g., Fukumoto
\& Ikeuchi 1996; Abadi, Moore \& Bower 1999; Quilis, Moore \& Bower 2000),
and in the cosmological
context (Gnedin \& Ostriker 1997; Gnedin 1998; Cen \& Ostriker 1999).
Gas may be removed during a close interaction between two or more
galaxies in which a fraction of the gas attains escape velocity (see,
e.g., Barnes \& Hernquist 1992). Alternatively, loosely bound gas may
be stripped from a galaxy by the ram pressure of the IGM through which
the galaxy is moving.  Removal of metals via mergers or tidal
disruption may occur in clusters or in the general IGM, whereas
ram-pressure stripping is probably only important in massive clusters.

	We have applied the method of \S~\ref{sec-locmeth} to our
simulations to examine the dynamical removal of metals.  Metals are
deposited only in bound groups, thus by calculating the IGM enrichment
at later times we can directly assess the amount of metals that can be
removed by dynamics alone.  The results of these trials are shown in
Table~\ref{tab-dynmodrun} and Figs.~\ref{fig-denrich} and
\ref{fig-zgal}.  The table gives the
overall fraction of metals that are outside of bound groups, the
metallicity of the IGM at its mean density 
and at an overdensity $\delta\equiv \rho_{\rm gas}/\langle\rho_{\rm
gas}\rangle=100$, and the mean stellar metallicity in galaxies.
Unless otherwise noted, we include both gaseous metals and dust in
computing the metallicities (for dynamical and wind enrichment, this
is at very worst a factor of two larger than if dust is not included;
for dust ejection we show dust and gaseous metal fractions
independently).

Figure~\ref{fig-denrich} shows the mean metallicity $\langle Z_\delta
\rangle$ of the IGM as a function of $\delta$, and the mean
metallicity of hot ($> 5\times 10^6\,$K) bound cluster/group gas as a
function of gas temperature.  Since the clusters tend to have radially
declining temperature gradients, the plotted cluster temperatures are
lower than the core temperatures; we have also plotted the mean
temperature of the ICM within 100\,kpc for the hottest five clusters.
Note also that the cluster gas metallicity is slightly overestimated
because some hot gas associated with galaxies is included, but an
examination of radial metallicity profiles indicates that the effect
is small.  

The mean metallicity vs. overdensity should be interpreted with care;
for a highly inhomogeneous metal distribution, the mean can be
dominated by a few highly enriched particles and should not
necessarily be compared to the `typical' metallicity found in
Ly$\alpha$ absorbers.  The median particle metallicity gives a better
estimate of typical {\em particle} metallicities, but since each observed
line would correspond to gas represented by a number of particles, the
median metallicity of absorption systems should probably be higher
than the median for particles (for a highly inhomogeneous
distribution).  Thus we expect that the mean and median particle
metallicities should lie above and below the metallicity that should
be compared to `typical' observed metallicities at a given column
density (see Aguirre et al. 2001b for more discussion). In this paper
we display mainly mean metallicities, which should be considered upper
limits.

	The results show that dynamical removal is rather inefficient,
removing only a few percent of a typical galaxy's metals over its
lifetime.  The enrichment, as clearly shown in Fig~\ref{fig-denrich}
(dot-dashed line), is insufficient to account for the metallicity of
Ly-$\alpha$ absorbers at $z\sim 3$, which have $Z\approx
10^{-2.5}\,Z_\odot$ down to at least $N(H I)\approx 10^{14.5}{\,\rm cm^{-2}}$.

These results are in substantial agreement with simulations of
individual galaxy interactions, which tend to show that only a few
percent of an interacting galaxy's mass can attain escape energy
during an encounter (Barnes 1988; Barnes \& Hernquist 1992; Hernquist
1992, 1993).  Our 

\ifthenelse{\boolean{apj}}{ \vbox{ \centerline{
\epsfig{file=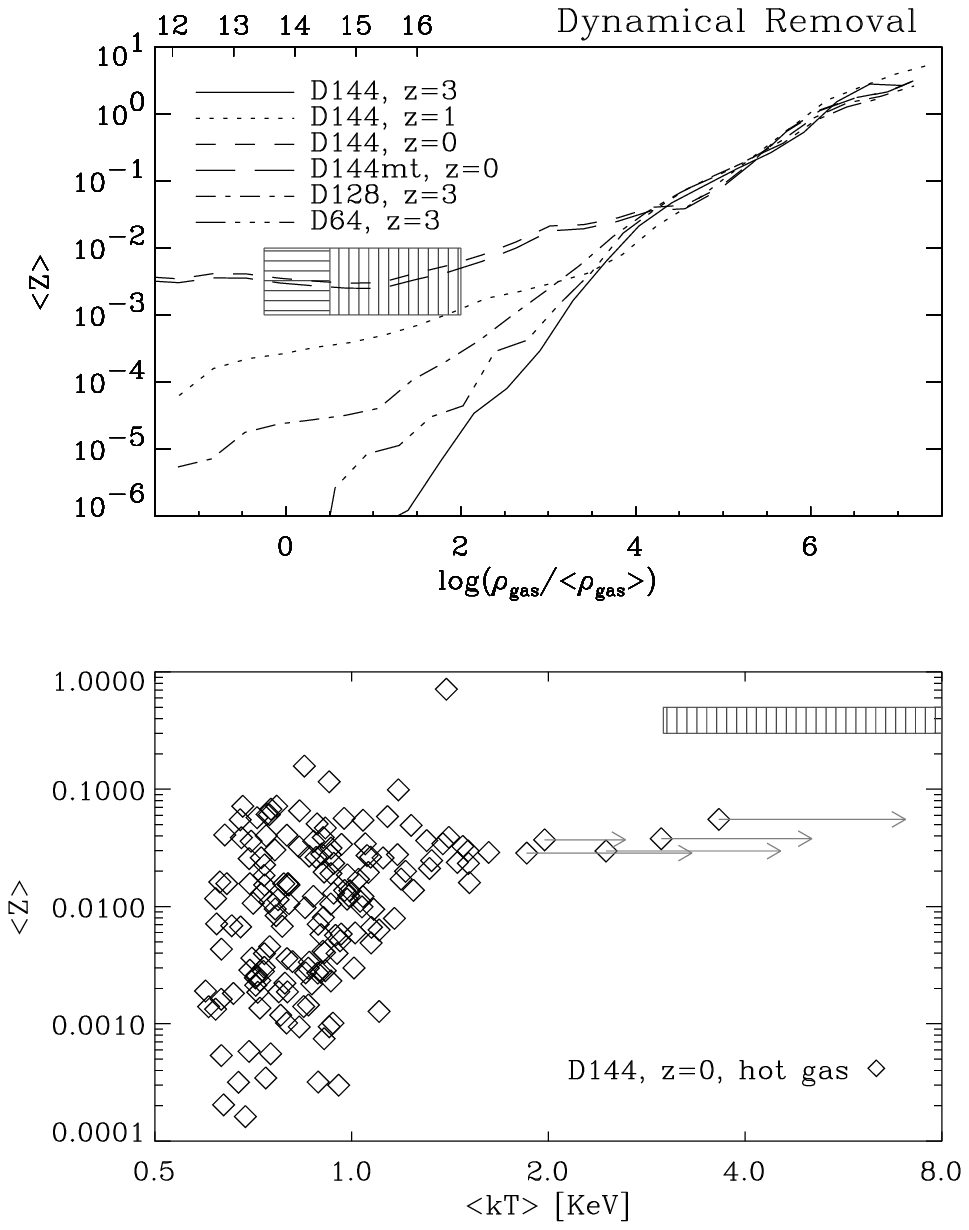,width=9.0truecm}} \figcaption[]{ \footnotesize
Enrichment of the IGM for dynamical removal of metals. {\bf Top:} Mean
metallicity $\langle Z\rangle$ vs. overdensity for the three
simulations at $z=3$, and for the $144^3$ simulation at $z=0$.  Also
plotted is the $z=0$ `D144mt' model as a time-resolution test. The top
axis is $\log N(H\,I)$ (cm$^{-2}$) of an absorber corresponding to
the bottom-axis overdensity {\em at $z=3$}, using the Dav\'e et
al. (1999a) relation.  The vertically striped box outlines the
approximate current constraints from the Ly$\alpha$ forest at
$z\approx 3$ while the horizontally striped box shows an extension of
these constraints to lower densities (see discussion in
\S~\ref{sec-simsvsobs}).  {\bf Bottom:} Group/cluster mean metallicity
(in hot gas) vs. mean temperature of hot bound gas at $z=0$.  Arrows
show the mean temperature of hot gas in the central 100\,kpc for the
hottest clusters.  The vertically striped box outlines the approximate
constraints from cluster X-ray studies.
\label{fig-denrich}}}
\vspace*{0.5cm}
}{
\suppressfloats\begin{figure}[tbp]
\centerline{\epsfxsize=6.0in%
\epsffile{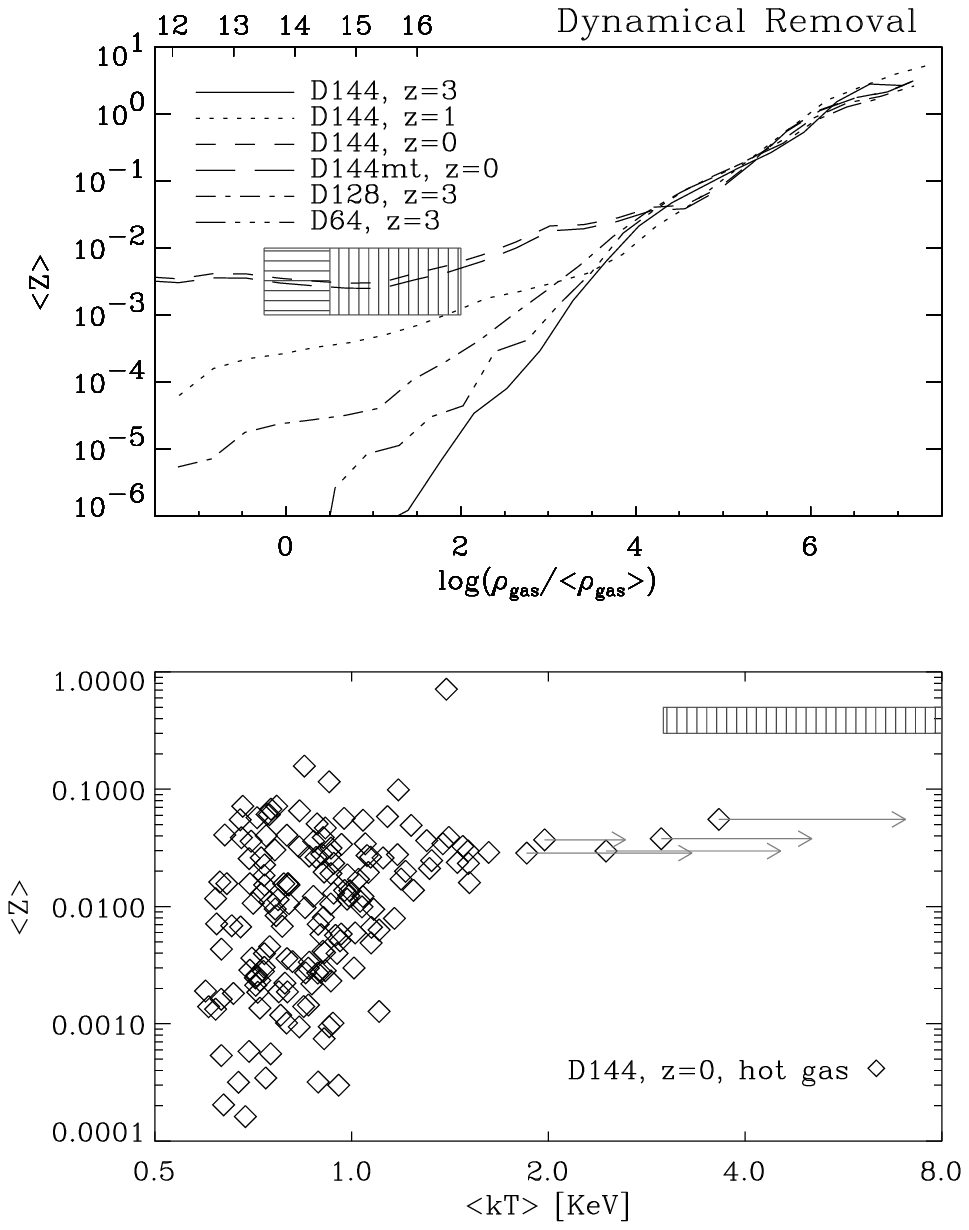}}
\caption{\baselineskip=12pt
Enrichment of the IGM for dynamical removal of metals. {\bf Top:} Mean
metallicity $\langle Z\rangle$ vs. overdensity for the three
simulations at $z=3$, and for the $144^3$ simulation at $z=0$.  Also
plotted is the $z=0$ `D144mt' model as a time-resolution test. The top
axis is $\log N(H I)$ (in cm$^{-2}$) of an absorber corresponding to
the bottom-axis overdensity {\em at $z=3$}, using the Dav\'e et
al. (1999) relation.  The vertically striped box outlines the approximate
current constraints from the Ly$\alpha$ forest at $z\approx 3$ while
the horizontally striped box shows an extension of these constraints
to lower densities (see discussion in \S~\ref{sec-simsvsobs}).
{\bf Bottom:} Group/cluster mean metallicity (in
hot gas) vs. mean temperature of hot bound gas at $z=0$.  Arrows show the mean
temperature of hot gas in the central 100\,kpc for the hottest
clusters.  The vertically striped box outlines the approximate
constraints from cluster X-ray studies.
}
\label{fig-denrich}
\end{figure} 
}

\noindent results are somewhat in disagreement with those of Gnedin
(1998), who found -- using high resolution SLH-P$^3$M simulations at
$z \gsim 4$ -- that dynamical enrichment was more effective than
supernova-driven ejection, and also sufficient to enrich the
low-density IGM.  While our results support the general idea that {\em
some} metals escape into the $z=3, \delta < 100$ IGM through dynamics
alone, we find that the amount is negligible compared to the other
mechanisms we consider.  We cannot rule out the possibility that small
(unresolved by our simulations), quickly merging galaxies at $z \gg 5$
could enrich the IGM more than our calculations suggest; this would,
however, require more {\em efficient} ejection (or a very super-solar
yield at high $z$), since 
if only a few percent of metals are ejected
and mixed thoroughly, at least $10\%$ of gas must form stars with
solar yield to give a uniform enrichment of $\sim 10^{-2.5}\zsol$.

The metallicity of the gas in rich clusters in the $144^3$ run is an
order of magnitude below the observed value of $\approx 0.3-0.5\zsol$
(see \S~\ref{sec-obssumm}), from which we can conclude that dynamical
removal of metals from massive galaxies ($\gsim 10^{10.5}\,M_\odot$)
cannot account for the metallicity observed in cluster gas.  Since
ram-pressure stripping should be most efficient in small galaxies, it
is important to note that we cannot directly address the importance of
pressure-stripping of dwarf galaxies in ICM enrichment.  However, if
this process is to account for the observed ICM enrichment, it must
happen at rather high redshift, since the mass contribution of $M \lsim
10^{10.5}\,\msol$ galaxies at present is quite small, i.e.\ the
enrichment would have to happen at a high enough redshift that the
galaxy mass function had a significant fraction of mass in the small
galaxies.  Note also that we have neglected enrichment from Type Ia
supernovae in intergalactic cluster stars. We do not expect this to be
significant in the field because only a few percent of the simulation
stars are intergalactic (having been moved into the IGM dynamically),
but it may be more important in clusters, which may have more IG stars.
This effect would be best treated by introducing `delayed enrichment'
into our simulations.
Although inefficient (for the galaxies resolved), dynamical removal is
more effective in clusters than in the field.  Assuming that the ratio of cluster
stars to cluster gas is comparable to the field value of
$\Omega_*/\Omega_b \simeq 0.23$, the ejection fraction $f_{\rm
ej,cl}$ in rich clusters is
\begin{equation}
f_{\rm ej,cl} \simeq {\langle Z_{\rm cl}\rangle \over
(\Omega_*/\Omega_b) y_*} \simeq 0.16
\label{eq-fejcl}
\end{equation}
in the `D144' model, 
as compared to 0.04 for all galaxies.
(Though if stars form more efficiently in clusters, 
this
fraction would be smaller.)

\ifthenelse{\boolean{apj}}{
\begin{figure*}
\vbox{ \centerline{ \epsfig{file=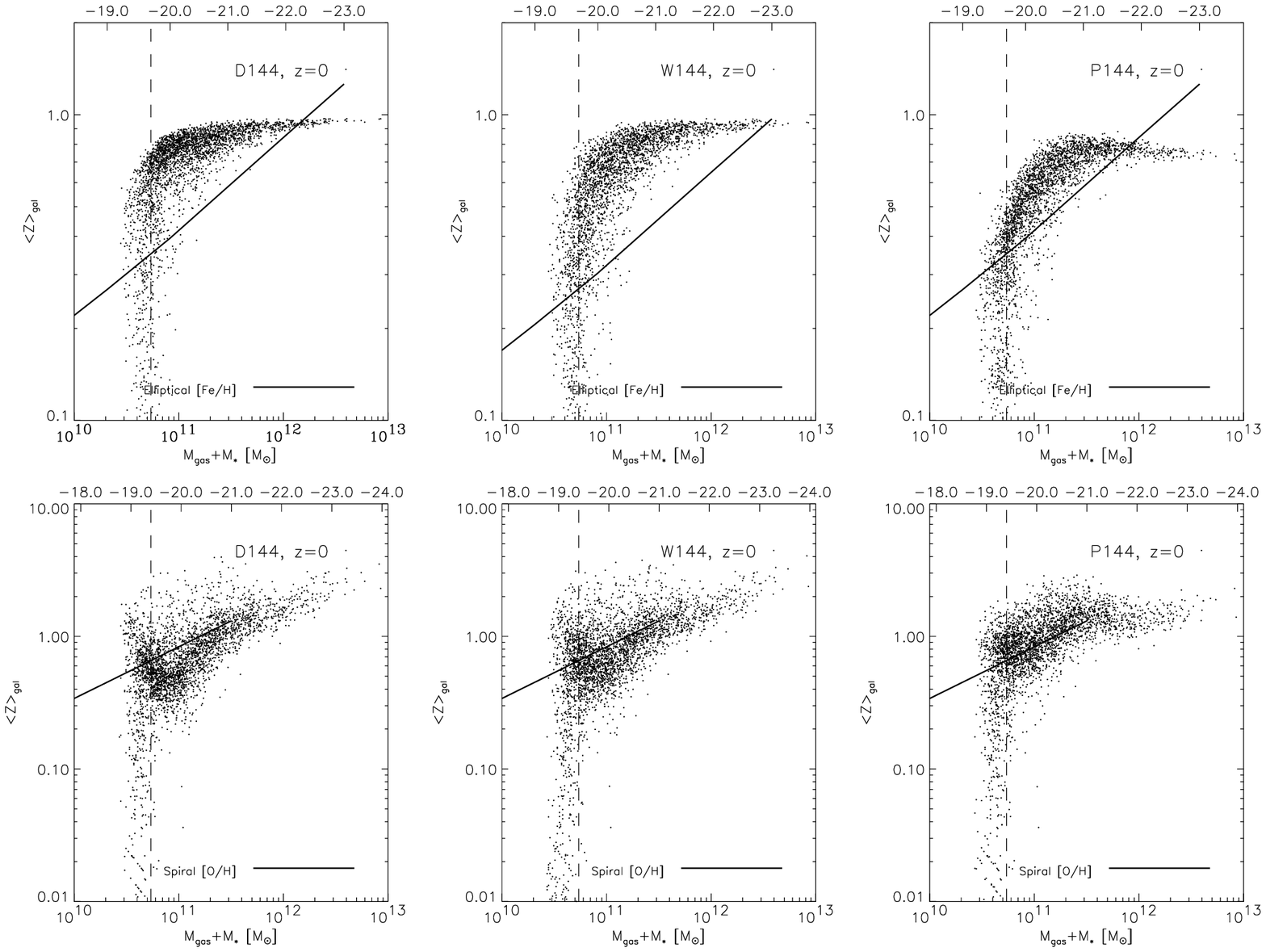,width=18.0truecm}}
\figcaption[]{ \footnotesize {\bf Top three panels:} Mean stellar (top
three panels) metallicity of galaxies in the fiducial dynamical, wind,
and dust (resp. `D144', `W144' and `P144') models at $z=0$.  Bottom
axis shows stellar+gas mass, top axis shows $B$-band magnitude,
converted from the mass using the elliptical galaxy relation from
Persic \& Salucci (1997). The solid line shows the elliptical [Fe/H]
M-Z relation taken from Zaritsky et al. (1994). {\bf Bottom three:}
Average gas metallicity of galaxies at $z=0$ for the same three
runs. Bottom axis shows stellar+gas mass, top axis shows $B$-band
magnitude, converted from the mass using the spiral galaxy relation
from Salucci \& Persic (1999); solid line is the spiral M-Z relation
from Zaritsky et al. (1994). The mass of 64 gas particles is indicated
by the vertical dashed line.
\label{fig-zgal}}}
\vspace*{0.5cm}
\end{figure*}%
}{
\suppressfloats\begin{figure}[tbp]
\centerline{\epsfxsize=6.0in%
\epsffile{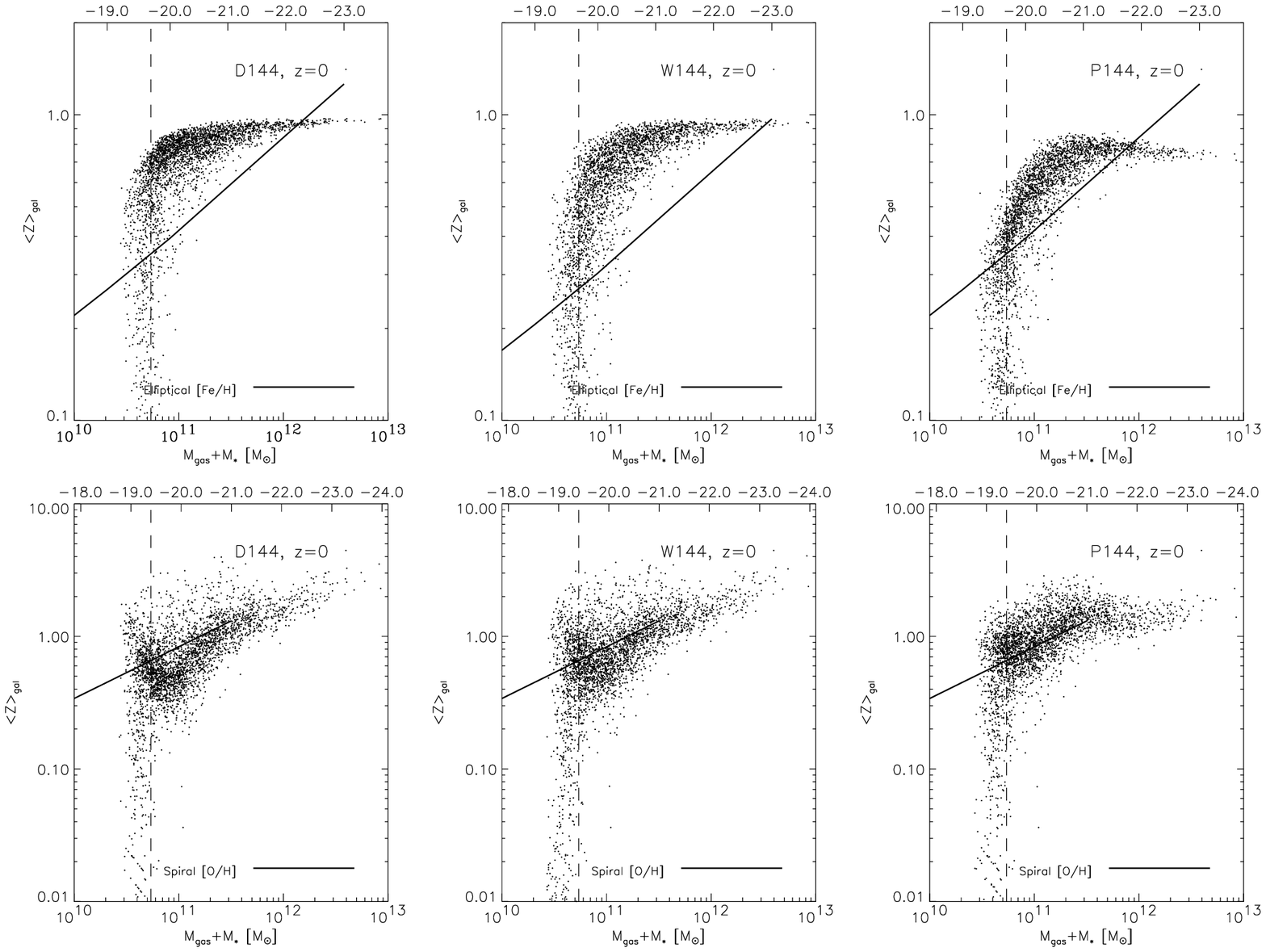}}
\caption{\baselineskip=12pt
 {\bf Top three panels:} Mean stellar (top
three panels) metallicity of galaxies in the fiducial dynamical, wind,
and dust (resp. `D144', `W144' and `P144') models at $z=0$.  Bottom
axis shows stellar+gas mass, top axis shows $B$-band magnitude,
converted from the mass using the elliptical galaxy relation from
Persic \& Salucci (1997). The solid line shows the elliptical [Fe/H]
M-Z relation taken from Zaritsky et al. (1994). {\bf Bottom three:}
Average gas metallicity of galaxies at $z=0$ for the same three
runs. Bottom axis shows stellar+gas mass, top axis shows $B$-band
magnitude, converted from the mass using the spiral galaxy relation
from Salucci \& Persic (1999); solid line is the spiral M-Z relation
from Zaritsky et al. (1994). The mass of 64 gas particles is indicated
by the vertical dashed line.
}
\label{fig-zgal}
\end{figure} 
}

Panels 1 and 4 of Figure~\ref{fig-zgal} give the average metallicity
of stars and gas, respectively, as a function of galaxy mass.
Interestingly, stellar and gaseous M-Z relations exist even though
galaxies retain nearly all of their metals.  The relations occur both
because the smallest galaxies have higher gas fractions, and because
they tend to be younger.  These can be compared to the plotted lines
that are rough fits to the mass-metallicity relations found by
Zaritsky et al. (1994) for ellipticals and spirals spanning $\sim~15$
B magnitudes.\footnote{ Since our simulations do not provide types for
the galaxies, we treat all galaxies as spirals when comparing to the
gas metallicities, and all as ellipticals when comparing to the
stellar metallicities.}  Unfortunately the simulations only have
enough dynamic range to probe the brightest four magnitudes.  One
should also be cautious about the properties of the smallest galaxies,
near the resolution limit (vertical line).  Nevertheless, the `D144'
model does seem to exhibit an M-Z relation that is too weak in stars,
and the stars have significantly higher metallicity than observations
indicate.  More effective feedback would prevent stars from forming in
high-metallicity regions but cannot cure the problem, since the metals
would then be present in the gas -- but the gaseous M-Z relation is
quite close to the observations.  Thus it seems that the observed M-Z
relation very likely indicates metal ejection beyond pure dynamics.  We
can also see from these figures that although all the results are
shown for a solar yield, the results cannot be scaled by changing
$y_*$ by a significant amount without clearly violating the observed
abundances.  Thus while it might be argued that dynamical removal of
metals could pollute the IGM more than we predict given a much higher
yield (presumably due to an IMF biased toward massive stars), this
argument would require the higher yield to apply only at high-$z$,
before the bulk of cosmic metals are formed.\footnote{Note also that
Pettini et al. (2000) find in their study of the $z=2.73$ galaxy MS
1512-cB58 no evidence for a non-standard IMF.}

Our results concerning dynamical removal of metals are not weakened by
the large uncertainties in the assumed parameters, since in this
prescription only the yield $y_*$ is important, and can be constrained
within a factor of two.
However, our predictions also are subject to some
numerical uncertainties.

First, if the timescales for some dynamical processes that remove
metals from galaxies were shorter than the time interval between the
simulation outputs used, our prescription might not accurately treat
their importance.  We have checked this by repeating our calculation
using 27 timesteps (model `D144mt') rather than 19 (model `D144'); the
additional steps were chosen to roughly halve the interval between
snapshots during the epoch when most star formation takes place ($1\lsim
z\lsim 3$).  The differences between these two models are quite small
(see Figure~\ref{fig-denrich}), indicating that even with 19 timesteps our
calculations have numerically converged.

A numerical uncertainty that is harder to address is that caused by
the limited resolution of the simulations themselves.  An accurate
assessment of the dynamical removal of metallic gas -- whether during
mergers, through tidal disruption, or via ram pressure stripping --
depends on the ability of the simulations to accurately treat both the
IGM and the structure of the galaxies.  The mean physical
inter-particle spacing in our simulations is
$133\delta^{-1/3}(1+z)^{-1}\,$kpc for the $128^3$ runs and
$534\delta^{-1/3}(1+z)^{-1}\,$kpc for the $144^3$ runs (where $\delta$
is the gas overdensity), large compared to the scale of a typical
galaxy.  Moreover, the central 200 kpc of a cluster with $100 \lsim
\delta \lsim 1000$ has only 200-2000 SPH particles; it is therefore
doubtful that ram pressure effects on cluster galaxies are treated
accurately (see Abadi, Moore \& Bower 1999 for
discussion). Unfortunately we cannot yet perform resolution tests as
we have only one simulation complete to $z=0$.  Our simulations treat
galaxy-galaxy interactions more accurately but still with limitations.
Large, thick galaxies are probably represented well, whereas low mass
galaxies and thin disks will not be captured.  Therefore, while {\em
in principle} our type of investigation can assess the efficiency of
dynamical removal quite well, in practice we expect limitations due to
limited resolution.

We have attempted to test this effect by comparing the efficiency of
dynamical metal removal in the $64^3$ and $128^3$ runs with the
restriction that metals are only added to gas particles in galaxies of
a fixed mass range.  For example, we may add metals only to galaxies
with $60-120$ (gas+star) particles in the $64^3$ run and only to
galaxies with $ 480-960$ particles in the $128^3$ run, to compare the
relative efficiency of dynamical removal in galaxies with baryonic
mass $ 6\times10^9-1.2\times 10^{10}\msol$ with different resolution.
In this case, we find that $\approx 0.15\%$ of metals are lost by
well-resolved galaxies in the $64^3$ run by $z=3$, whereas $\approx
1.7\%$ are lost in the $128^3$ run.  Curiously, we find that as we
increase the mass cut for galaxies that receive metals, the ejection
fraction {\em increases} in the $128^3$ run, but {\em decreases} in
the $64^3$ run.  The difference between the two runs can be accounted
for either by a resolution effect (i.e.\ galaxies of the same mass lose
different amounts of metals depending on the number of particles
constituting them), or by the difference in the mass function of
galaxies (i.e.\ the presence or absence of galaxies small compared to
those with the metals).  Were we to {\em assume} that in both runs
most metals are lost from interactions between galaxies of comparable
mass, then it would necessarily be a resolution effect.  But if it
were purely a resolution effect, it is very difficult to see why the
ejection fraction would decrease with the number of particles in the
$64^3$ run while increasing in the $128^3$ run.  Thus we suspect that
metal-loss from the massive galaxies is dominated by interactions with
lower-mass galaxies in the $128^3$ run, but dominated by interactions
with other well-resolved galaxies in the $64^3$ run (the small
galaxies being absent). This dependence on the presence or absence of
small galaxies makes our resolution test inconclusive.

\begin{deluxetable}{cccccccc}
\singlespace
\footnotesize
\tablecaption{Dynamical Models Run}
\tablehead{\colhead{Model} & \colhead{Variation} &
 \colhead{$f_{\rm IGM}$} &
 \colhead{$\langle Z_{\delta=1}\rangle$} &
 \colhead{$\langle Z_{\delta=100}\rangle$} & 
\colhead{$\langle Z_{\rm cl}\rangle$} & 
\colhead{$\langle Z_{\rm gal}\rangle$}}
\startdata
D144 & - & 0.038 & 0.0035 & 0.0064 & 0.036 & 0.87   \nl
D144mt & more outputs & 0.037 & 0.0030 & 0.0051 & 0.032 & 0.87  \nl
D64 & $64^3$ & 0.0025 & $1.9\times10^{-7\dagger}$ & $4.6 \times 10^{-5}$ & - & 0.63 \nl
D64mt & $64^3$, more outputs  & 0.0023 &  $3.5\times10^{-9\dagger}$ & $4.3 \times 10^{-5}$ & - & 0.69 \nl  
D128 & $128^3$ run & 0.017 & $2.4 \times 10^{-5}$ & 0.00035 & - & 0.75 \nl
D128lt & $128^3$, less outputs & 0.017 & $2.5\times10^{-5}$ & 0.00040 & - & 0.61 \nl

\enddata 
\tablenotetext{}{Note: all results are given at $z=0$ for the
$144^3$ run, and at $z=3$ for the other two runs.\\
$\dagger$In these runs average metallicities at $\delta \lsim 10$
should not be trusted, since only a few particles have
nonzero metallicity in this density range.}
\label{tab-dynmodrun}
\doublespace
\end{deluxetable}

\label{sec-resdyn}

\section{Models and Results for Winds}
\label{sec-reswind}

\subsection{Fiducial Wind Model}

	As we discussed in \S~\ref{sec-windmeth} there have been
numerous investigations of galactic winds and their possible role in
the enrichment of the IGM, and the chief empirical data concerning
this process comes from observations of starburst-driven superwinds.
Our fiducial wind model, labeled `W' in Table~\ref{tab-windmodrun} and
in the figures, assumes an outflow velocity of $600\pm200\,{\rm
km\,s^{-1}}$, with wind efficiency (defined by Eq.~\ref{eq-mdotchi})
$\chi=1$, an entrainment fraction $\epsilon_{\rm ent} = 0.1$, and a
critical SFR/(area) for driving a strong wind of SFR$_{\rm crit} =
0.1\msol{\rm\,yr^{-1}}\kpc^{-2}$.  As described in
\S~\ref{sec-windmeth}, these are chosen to match `typical' values
derived from the observations, where possible.  Quantities that are
not directly observable, such as $\epsilon_{\rm ent}$, the ejection
fraction $Y_{\rm ej}$, and $\alpha$ (which controls the distribution
that governs the placement of metals in the IGM) are given reasonable
values that we vary in subsequent trials.  The fiducial parameter
values are listed in Table~\ref{tab-fidparwind}.

	We show the results of our fiducial wind model in
Tables~\ref{tab-windmodrun} and~\ref{tab-windmodrun128}, and in
Figs.~\ref{fig-zgal}, \ref{fig-wqz1.5}, \ref{fig-wenrich} and
\ref{fig-intsampz1.5}.  We first focus on results at $z=0$ from the
$144^3$ simulation. As listed in the first row of
Table~\ref{tab-windmodrun}, the galaxies resolved by the $144^3$
simulation lose about 6\% of their metal mass to the IGM over their
lifetimes.  At $z=0$, the mean metallicity of the IGM at its mean
density, $\langle Z_{\delta=1} \rangle = 0.008\,Z_\odot$, $\langle
Z_{\delta=100} \rangle = 0.02\,Z_\odot$, and the mean ICM metallicity
for hot clusters is $0.04\,Z_\odot$. Figure~\ref{fig-wenrich} allows
the most direct comparison to the principal results from the dynamical
model, shown earlier in Figure~\ref{fig-denrich}, and reveals that
winds can enrich the low-density IGM must more effectively than can
the dynamical removal of metals.

\ifthenelse{\boolean{apj}}{
\begin{figure*}
\vbox{ \centerline{ \epsfig{file=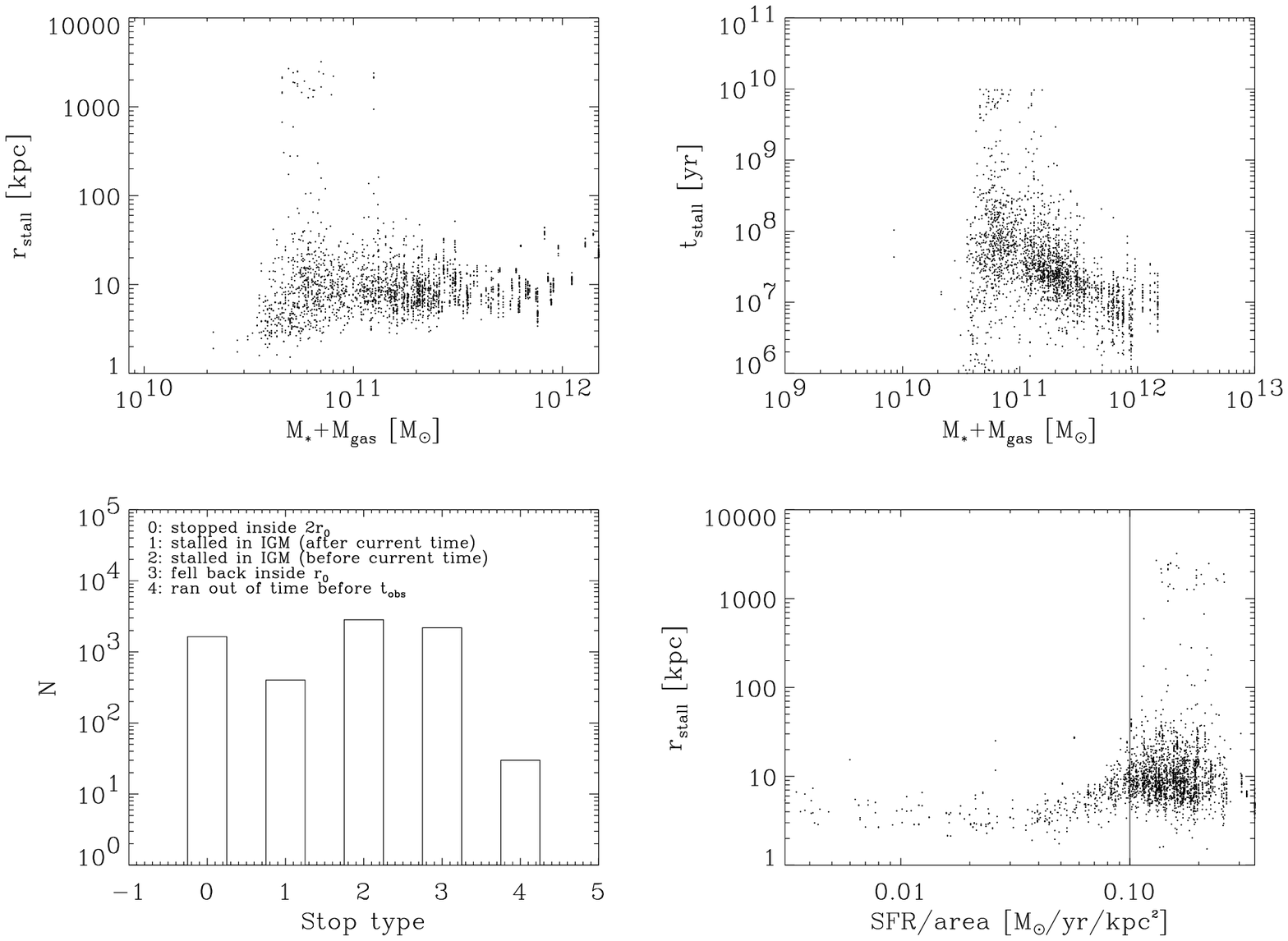,width=18.0truecm}}
\figcaption[]{ \footnotesize
Quantities in wind ejection for fiducial
`W' model at $z=1.5$. {\bf Panel 1:} Physical shell stalling radius
vs. galaxy mass.  {\bf Panel 2:} Time
between the launching of a shell from $r_0$ and its stalling in the
IGM.  The cutoff at $\sim 10^{10}\,$yr corresponds to $z=0$.
  {\bf Panel 3:} Histogram of the final fate of propagating shells.
{\bf Panel 4:} Shell radius vs. SFR/(area).  The
solid line indicates the critical SFR/(area). 
\label{fig-wqz1.5}}}
\vspace*{0.5cm}
\end{figure*} 
}{
\suppressfloats\begin{figure}[tbp]
\centerline{\epsfxsize=6.0in%
\epsffile{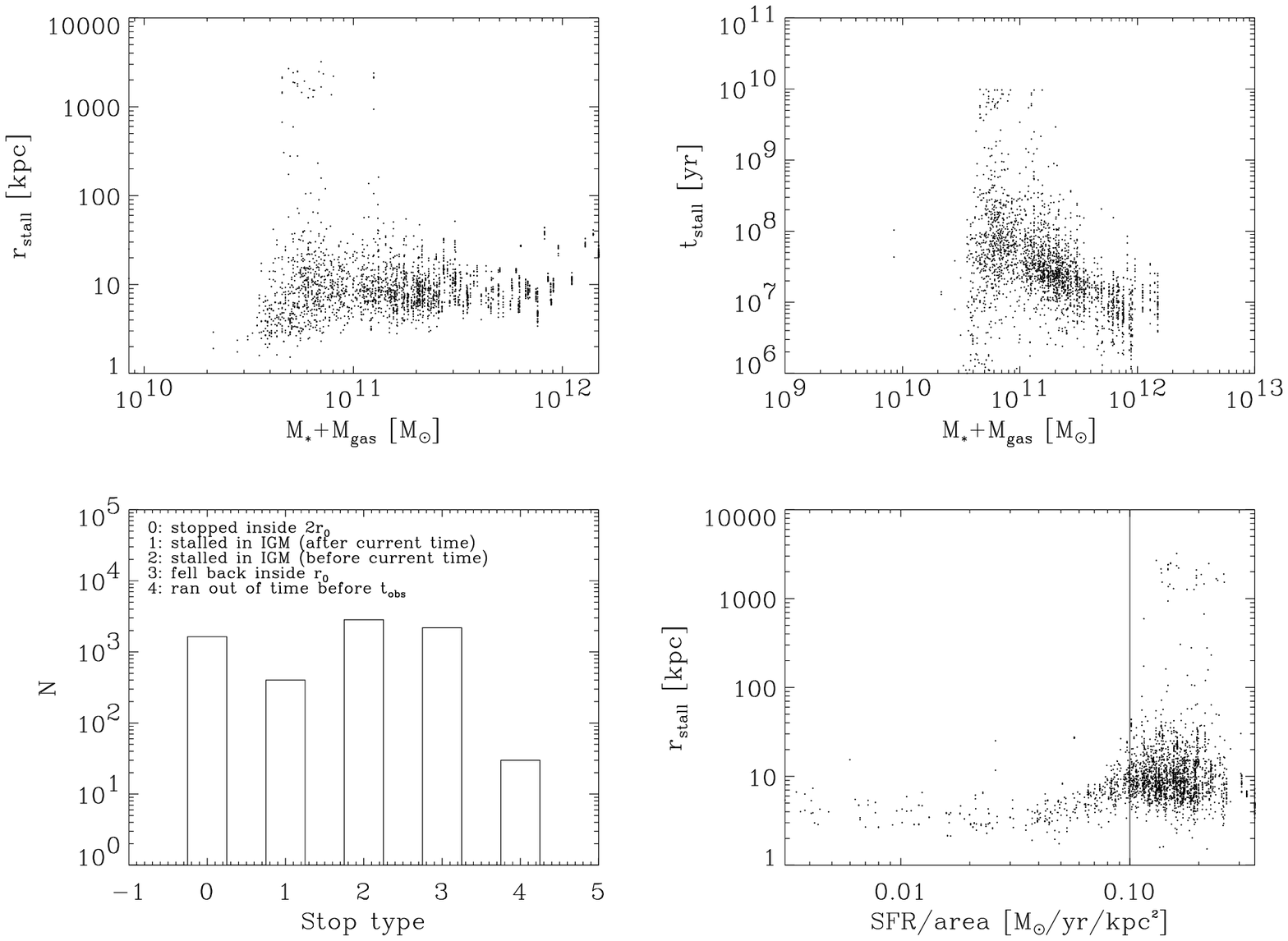}}
\caption{\baselineskip=12pt 
Quantities in wind ejection for fiducial
`W' model at $z=1.5$. {\bf Panel 1:} Physical shell stalling radius
vs. galaxy mass.  {\bf Panel 2:} Time
between the launching of a shell from $r_0$ and its stalling in the
IGM.  The cutoff at $\sim 10^{10}\,$yr corresponds to $z=0$.
  {\bf Panel 3:} Histogram of the final fate of propagating shells.
{\bf Panel 4:} Shell radius vs. SFR/(area).  The
solid line indicates the critical SFR/(area). 
}
\label{fig-wqz1.5}
\end{figure} 
}

Figure~\ref{fig-wqz1.5} shows some details as to how this enrichment
occurs.  The first panel plots the wind stalling radius $r_{\rm
stall}$ as a function of galaxy mass, and shows that large galaxies
with deep potential wells can retain their metals. Most galaxies
resolved by the $144^3$ simulation cannot drive winds past a few tens
of kpc (smaller than the galaxies themselves), though a small fraction
can drive metals hundreds or thousands of kpc into the IGM, where they
eventually stall after flowing for up to a few Gyr (see panel 2).  We
have plotted all angles for each galaxy, so the vertical `stripes'
demonstrate the range of radii to which the shells propagate in
different directions 

\ifthenelse{\boolean{apj}}{
\vbox{ \centerline{ \epsfig{file=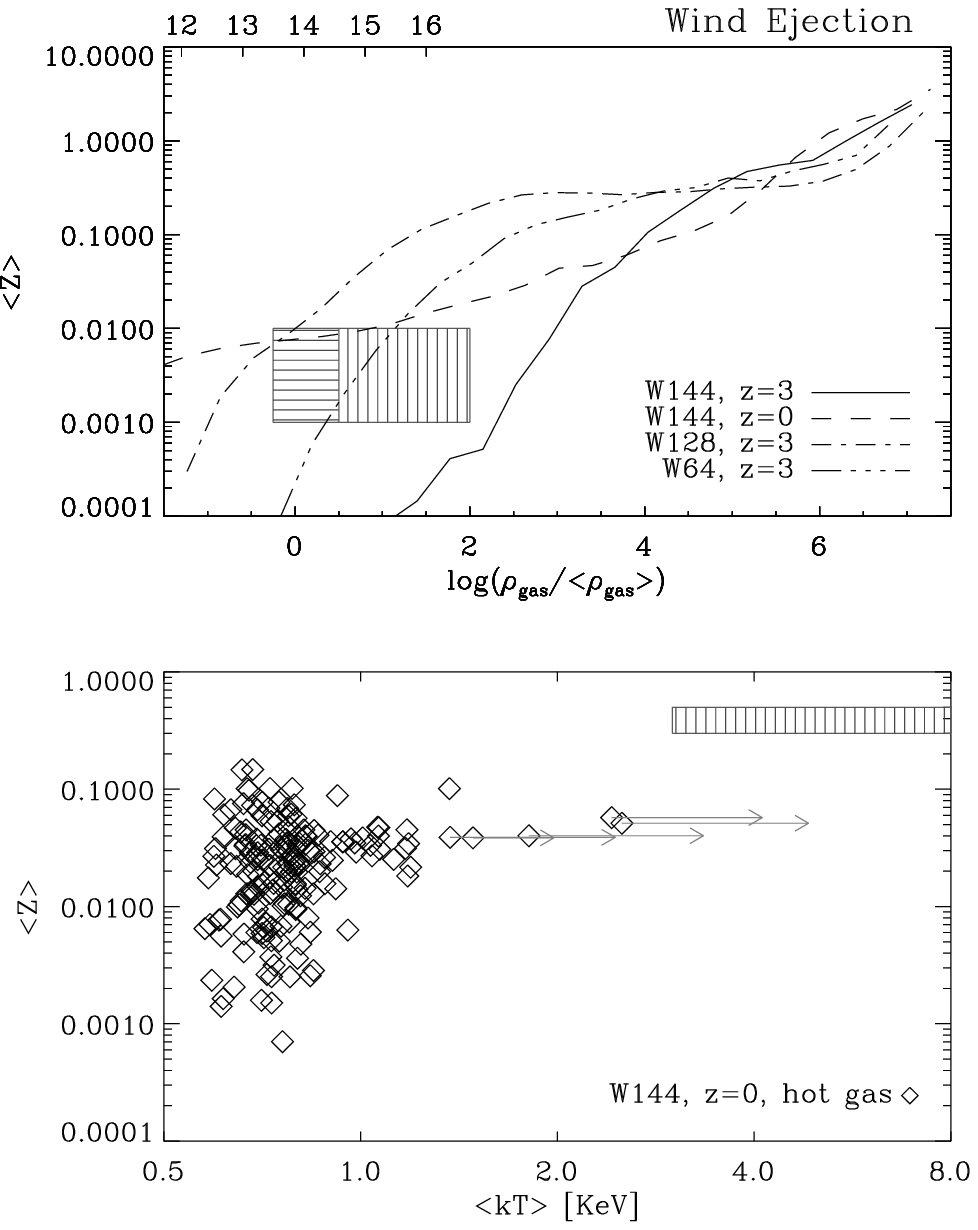,width=9.0truecm}}
\figcaption[]{ \footnotesize
Enrichment of the IGM for the wind ejection of metals in the fiducial
`W' model (see Tables~\ref{tab-fidparwind} \& \ref{tab-windmodrun}).
Plotted quantities are as in Fig~\ref{fig-denrich}. 
\label{fig-wenrich}}}
\vspace*{0.5cm}
}{
\suppressfloats\begin{figure}[tbp]
\centerline{\epsfxsize=6.0in%
\epsffile{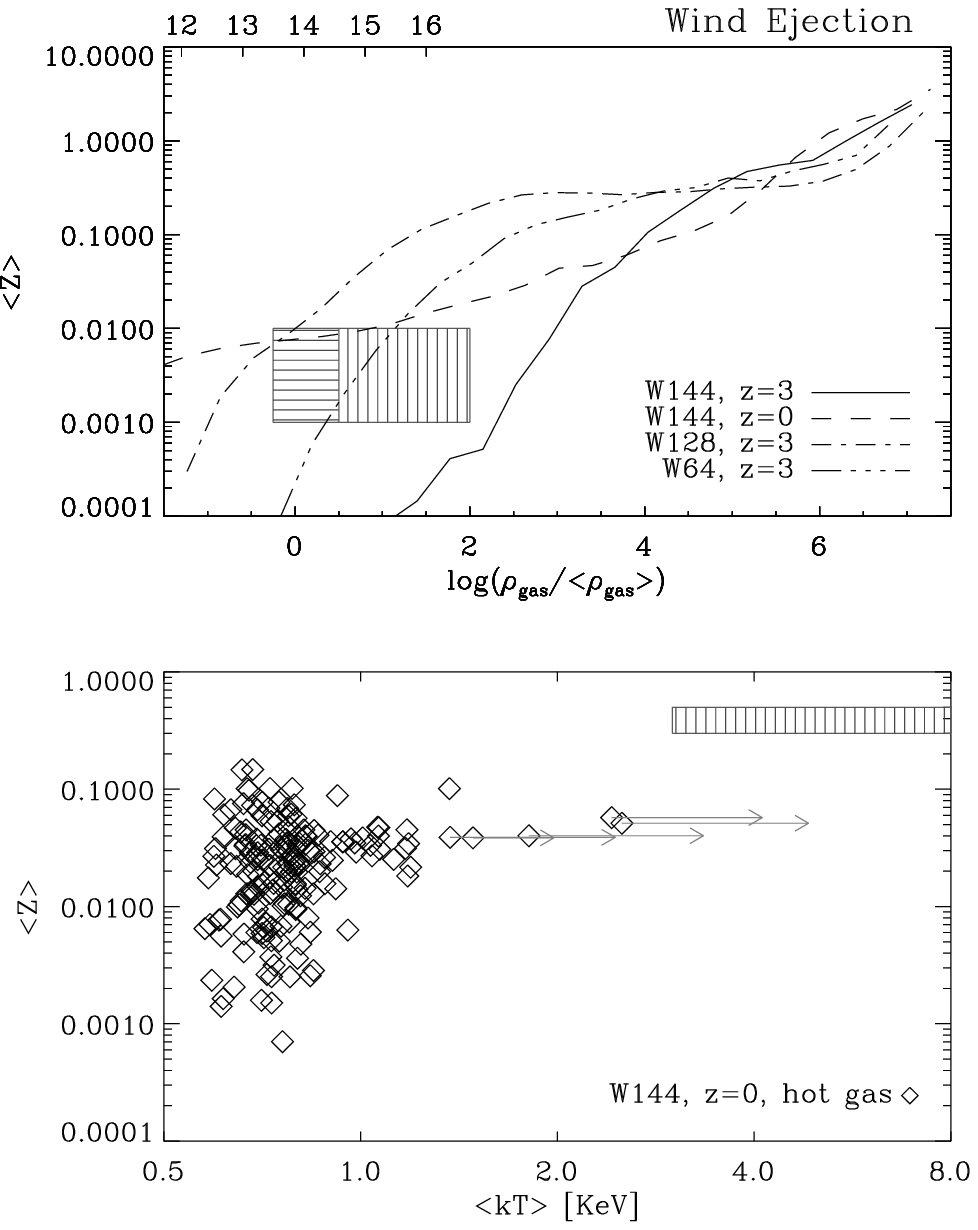}}
\caption{\baselineskip=12pt
Enrichment of the IGM for wind ejection of metals in fiducial
`W' model (see Tables~\ref{tab-fidparwind}, \ref{tab-windmodrun}).
Plotted quantities are as in Fig~\ref{fig-denrich}.
}
\label{fig-wenrich}
\end{figure} 
}

\noindent (typically 2-32 angles are used for each galaxy).
Panel 3 histograms the final state of the shells ejected between
$z=1.75$ and $z=1.5$.  About 2/3 of the shells either stall within
$2r_0$ or turn around and fall back to within $r_0$.  Another third
stall after $z=1.5$, and a smaller fraction stall before $z=1.5$ but
after reaching $2r_0$.  Finally, a small fraction are still
propagating at $z=0$.
As shown in the last panel, most galaxies at $z=1.5$ in the fiducial
model are assumed to be driving winds (i.e.\ their SFR exceeds the
assumed critical SFR), but even galaxies with very high areal SFRs may
not drive an {\em effective} outflow, if they are very massive.  

Figure~\ref{fig-intsampz1.5} shows the details of the radial
integration for one angle of one galaxy at $z=1.5$.  The first panel
gives the shell, wind, infall and Hubble velocities and the IGM sound speed versus
the shell radius.  The shell, starting at $\sim 800\kms$ at
$r_0=1.4$\,kpc, is quickly decelerated by gravity and by the sweeping
up of matter, leaves the $\sim 15\,$kpc galaxy at $\sim 200\kms$ with
about 6 times its initial mass (see the thin, solid line).  Outside of
the galaxy, the same two factors continue to decelerate the wind (see
panels 2 and 3), though now the wind ram pressure is also important,
imparting enough force to keep the shell at roughly constant velocity
out to several hundred kpc.  In this example, the wind coasts for a
long time, eventually stopping after running into a nearby mass
concentration at $\sim 600\,$kpc after $\sim 3\,$Gyr.

The winds in the fiducial model enrich the ICM little, bringing the
cluster metallicity to $\sim 0.04\,\zsol$, not much more than dynamics
alone (see Table~\ref{tab-windmodrun} and Figure \ref{fig-wenrich}).
Because the overall metal ejection fraction for winds is
about 50\% higher than for dynamics alone, this indicates that the
cluster environment is suppressing wind escape.  More massive
groups/clusters are enriched to a rather uniform level, whereas cooler
groups show a large scatter in their metallicity.  This scatter, which
may or may not be supported by observations (see
\S~\ref{sec-obssumm}), contrasts with the more uniform enrichment by
radiation-pressure ejection (as discussed in \S~\ref{sec-resrad}
below) and persists even in models in which the metal ejection efficiency
is much higher (as in the `W144max' model described below).

	The first row of Table~\ref{tab-windmodrun128} shows $z=3$
results for the fiducial model using the $128^3$ run.  Here, galaxies
have lost nearly half of their metals, enriching the $\delta=1$ IGM to
a mean metallicity of 1\% solar, somewhat higher than the high end of
the observed metallicity of the Ly$\alpha$ absorbers of similar or
greater density (see Fig.~\ref{fig-wenrich}).\footnote{Recall that the
{\em mean} metallicity is not necessarily comparable to the observed
metallicities if the enrichment is non-uniform; see
\S~\ref{sec-resdyn} and Aguirre et al. (2001b) for discussion.}
Relative to $z=0$, the greater escape fraction of metals occurs
because the galaxy mass spectrum is shifted toward smaller mass
galaxies at high redshift; galactic escape velocities thus become
small compared to the fiducial (`W') model's typical outflow velocity.
As discussed in Aguirre et al. (2001b), we find that the enrichment of
low-density regions is limited primarily by the time available for the
shells to propagate into the IGM.

	The dependence of the metal escape efficiency on galaxy mass
leads to a slightly steeper stellar M-Z relation than does dynamical
removal alone, as is evident in Figure~\ref{fig-zgal}.  The second and
fourth panels show galaxy stellar and gaseous metallicity for the
fiducial wind model, and the enhanced M-Z relation that wind ejection
induces on the galaxies.  We see a similar but stronger effect in the
$z=3$ galaxy metallicities of the `W128' run (not plotted), but there
the M-Z relation is evident mainly in the enhanced metallicity of
$\gsim 10^{\rm 10}\,M_\odot$ galaxies where the escape speed
approaches the assumed wind velocity.

As in the case of dynamical removal, we have checked some numerical
details of the calculation using a number of additional models.
First, we have verified that using roughly half or twice as many
simulation outputs changes all results insignificantly (i.e.\ at a
similar level as for the same test in the dynamics-only prescription;
see Table~\ref{tab-dynmodrun} and Fig.~\ref{fig-denrich}).  We have
checked that all the results are similarly insensitive to the time
step used in the integration, the radius over which we average when
integrating (as long as it is not large), and the details of the
stalling criterion for the shell.  The accuracy of the radial
integration itself has been checked by propagating shells with all
forces except gravity turned off; comparing the shell velocity
(computed via integration) to the wind velocity (computed using energy
conservation as per Eq.~\ref{eq-vworf}) tests the integration
accuracy.

A few more numerical details make small but noticeable differences in
our results.  Halving or doubling the galaxy mass contained within the
initial shell radius changes the results of our fiducial model only
slightly (at worst by a factor of two in mean metallicity at the
lowest densities).  The difference between $r_{\rm stall}$ and $h_{\rm
wind}$, which adjusts for 

\ifthenelse{\boolean{apj}}{
\vbox{ \centerline{ \epsfig{file=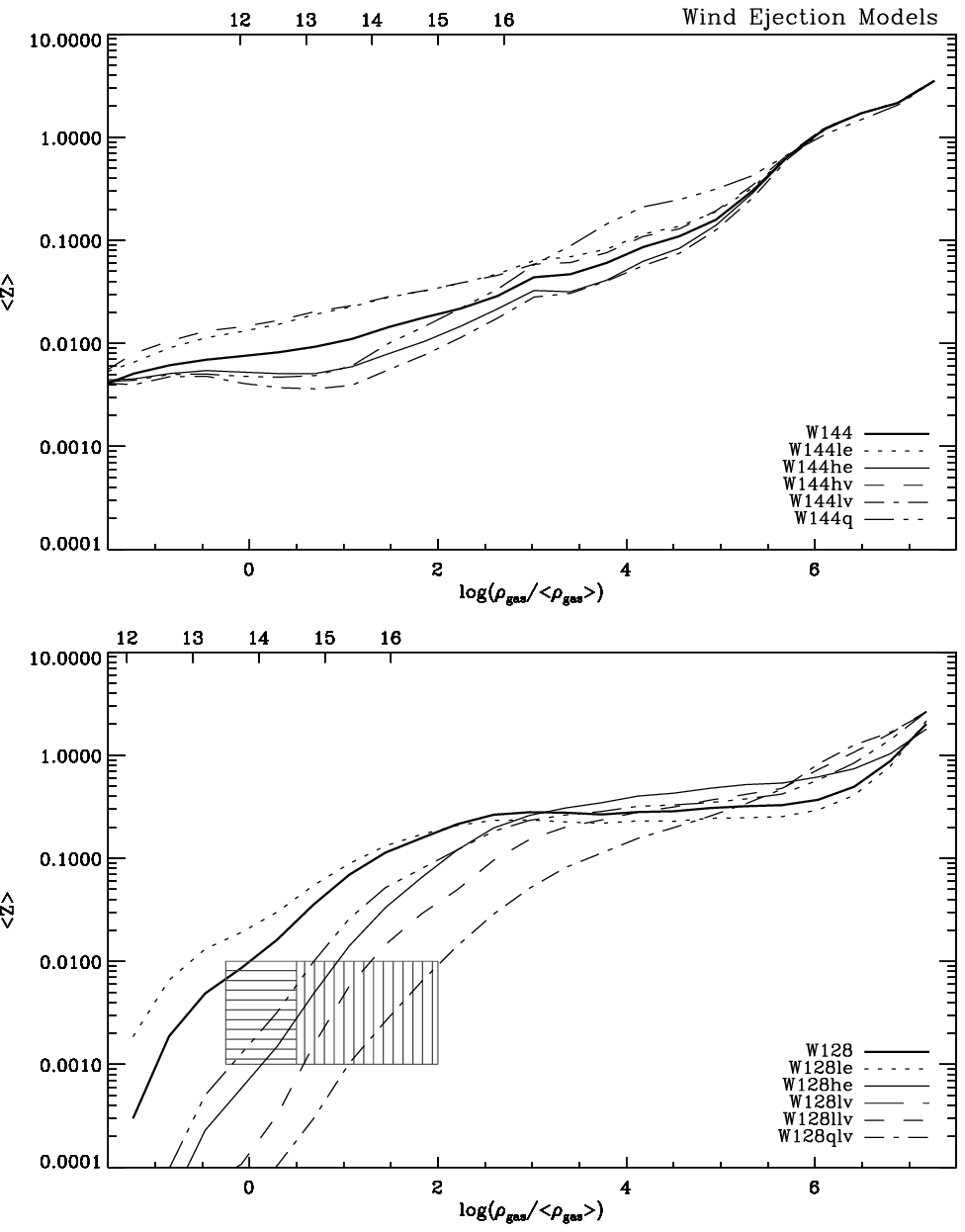,width=9.0truecm}}
\figcaption[]{ \footnotesize
Enrichment of the IGM for wind ejection of metals for several
wind models (see Table~\ref{tab-windmodrun}) at $z=0$ (top panel) and
at $z=3$ (bottom panel).  Plotted quantities are as in the top panel of
Fig~\ref{fig-denrich}.
\label{fig-wmods}}}
\vspace*{0.5cm}
}{
\suppressfloats\begin{figure}[tbp]
\centerline{\epsfxsize=6.0in%
\epsffile{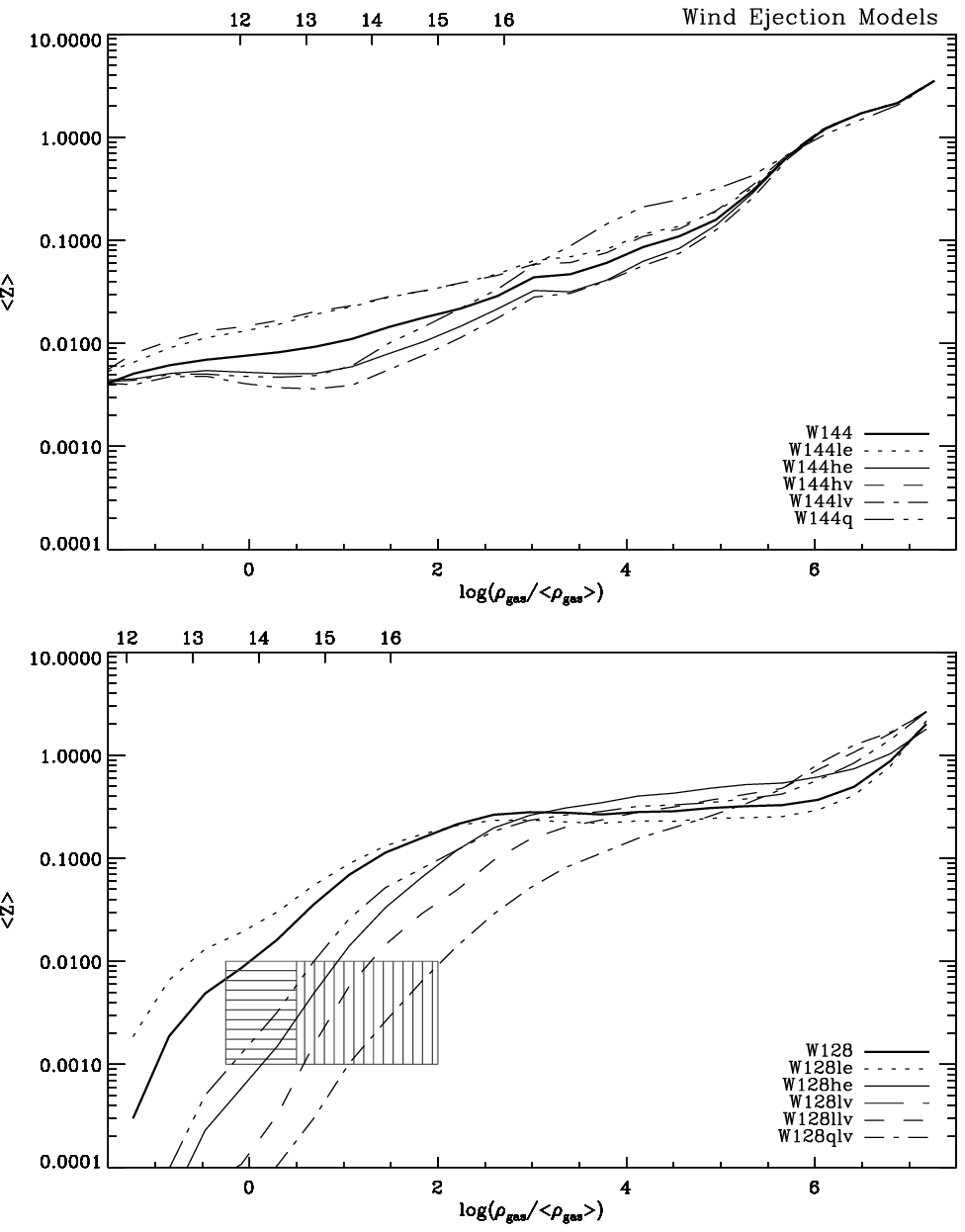}}
\caption{\baselineskip=12pt
Enrichment of the IGM for wind ejection of metals for several
wind models (see Table~\ref{tab-windmodrun}) at $z=0$ (top panel) and
at $z=3$ (bottom panel).  Plotted quantities are as in top panel of
Fig~\ref{fig-denrich}.
}
\label{fig-wmods}
\end{figure} 
}

\noindent unphysical movement of metals, turns out to have little
effect; all results of both the `W128' and the `W144' models are
changed by $< 10\%$ if this adjustment is removed.  A third numerical
issue, raised in \S~\ref{sec-windmeth}, is that while we calculate the
SFR by dividing the stellar mass formed in a galaxy between time $\tp$
and $\tc$ by $\tc-\tp$, we assume that this SFR applies for the entire
shell propagation time, which may exceed $\tc-\tp$.  Since the SFRs we
compute are smooth (i.e.\ don't depend on short episodes of
star-formation) they should vary on roughly the Hubble time.  We have
run trials in which the wind is turned off entirely a Hubble time
after its launch, and find unimportant changes in the results.  This
insensitivity would also extend to episodic star formation, as long as
the episodes are spaced more frequently than the timescale for the
wind to decelerate (so that it does not stall between episodes).

A final numerical detail is the number of angles we use, $N_a$.  As
described in \S~\ref{sec-windmeth}, we choose $N_a$ to ensure $\sim
16$ gas particles per angle in the galaxy so that the spacing between
successive particles in radius is smaller than the scale over which
the physical properties of the shell change. This typically results in
2-32 angles per galaxy.  Using the $144^3$ and $128^3$ simulations, we
have run trials using 16, 32, 64 and 128 particles per angle.  The
results are changed very little, indicating that enrichment is quite
similar whether we use $\sim 2$ or $\sim 16-32$ angles per galaxy.
This insensitivity is probably due to the fairly spheroidal shape of
galaxies and their halos in the simulations, and to the fact that
winds tend to be either confined, or escape to large radii where the
distribution radius is limited primarily by the time constraint.
Thus it seems that our results are not significantly affected by lack
of angular resolution.  

\ifthenelse{\boolean{apj}}{
\begin{figure*}
\vbox{ \centerline{ \epsfig{file=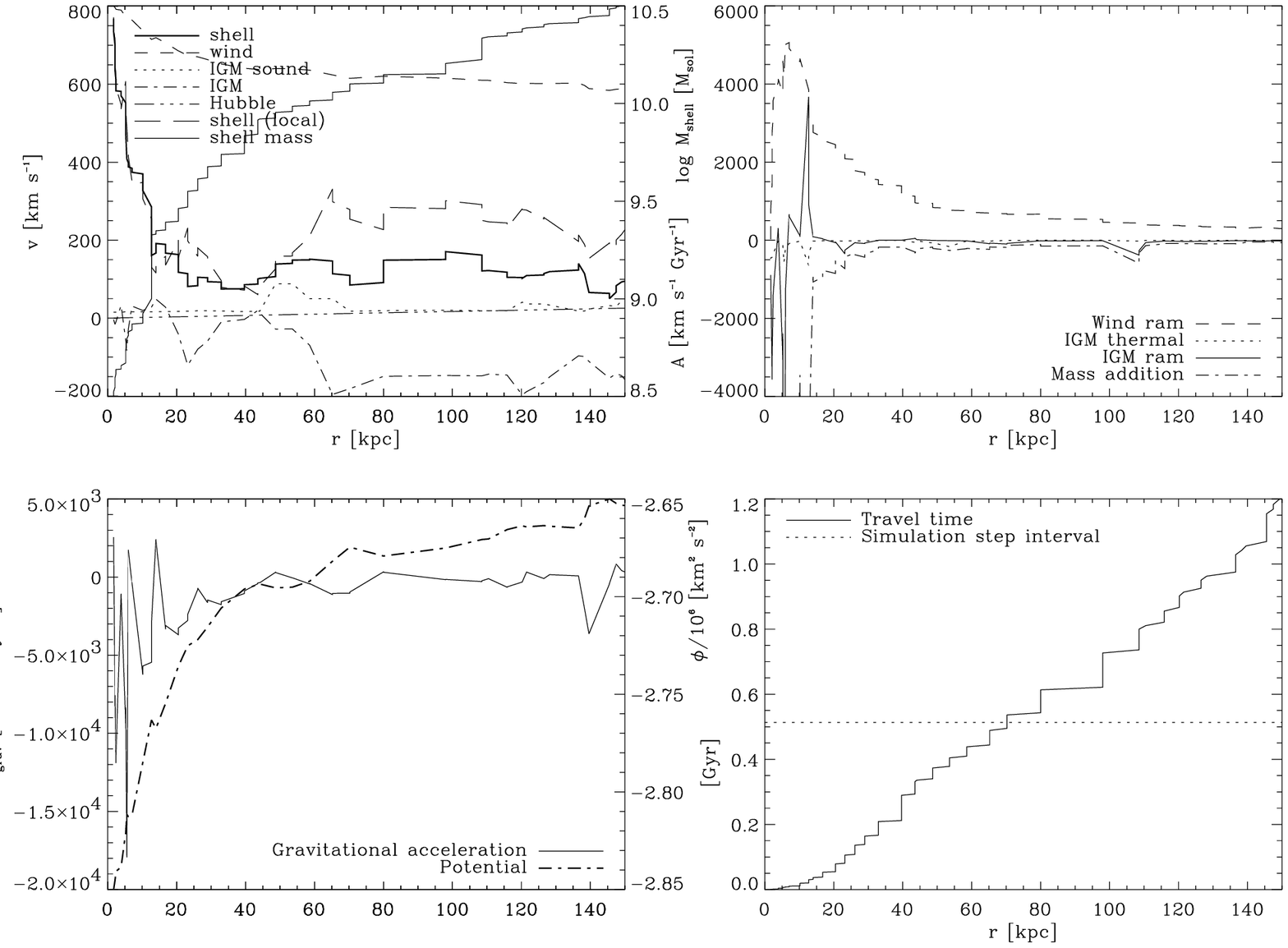,width=18.0truecm}}
\figcaption[]{ \footnotesize Sample for shell propagation at $z=1.5$,
for a shell with initial velocity of $\sim 800\kms$ at initial radius $5.6\,$kpc in a
galaxy of baryonic mass $1.3\times 10^{11}\msol$.  {\bf Panel 1:}
physical radial velocities (with respect to the galaxy center where
appropriate) of the shell, the outflowing wind, the Hubble flow, and
the IGM.  We give also the local sound speed of the IGM, and
the shell velocity in the frame of the ambient gas, as well as the
mass of the shell (right axis).  {\bf Panel 2:} acceleration of the
shell due to the ram pressure of the IGM, the ram pressure of the
wind, the thermal pressure of the IGM, and the acceleration due to the
addition of mass to the shell (i.e.\ the term $(v/m)(dm/dt)$ where $v$ and
$m$ are the velocity and mass of the shell).  {\bf Panel 3:}
Acceleration due to gravity (left axis) and gravitational potential
(right axis). {\bf Panel 4:} Elapsed time since launch at initial
radius.  The dotted line indicates the time corresponding to the next
simulation output used.
\label{fig-intsampz1.5}}}
\vspace*{0.5cm}
\end{figure*}
}{
\suppressfloats\begin{figure}[tbp]
\centerline{\epsfxsize=6.0in%
\epsffile{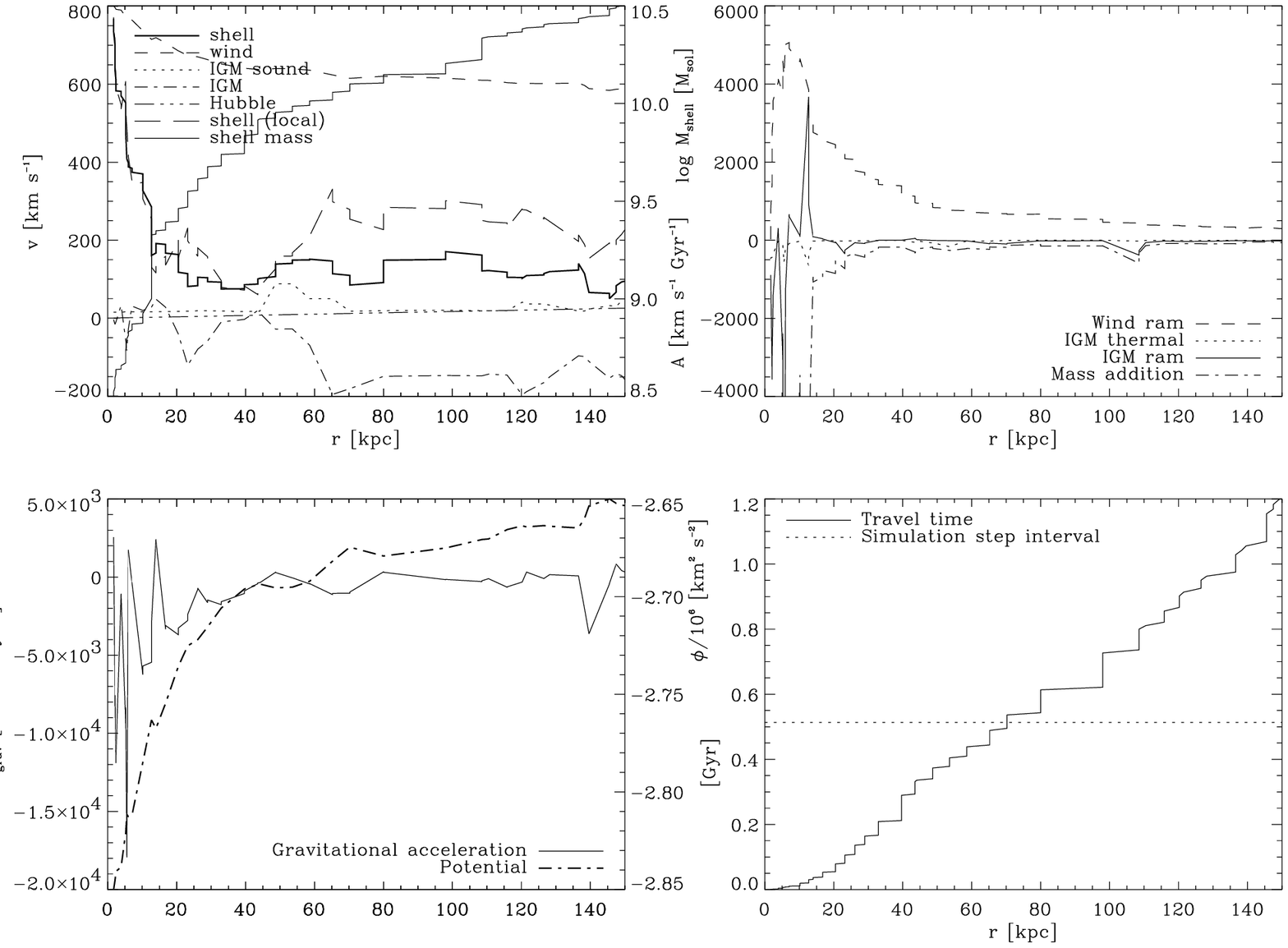}}
\caption{\baselineskip=12pt
 Sample for shell propagation at $z=1.5$,
for a shell with initial velocity of $\sim 800\kms$ at initial radius $5.6\,$kpc in a
galaxy of baryonic mass $1.3\times 10^{11}\msol$.  {\bf Panel 1:}
physical radial velocities (with respect to the galaxy center where
appropriate) of the shell, the outflowing wind, the Hubble flow, and
the IGM.  We give also the local sound speed of the IGM, and
the shell velocity in the frame of the ambient gas, as well as the
mass of the shell (right axis).  {\bf Panel 2:} acceleration of the
shell due to the ram pressure of the IGM, the ram pressure of the
wind, the thermal pressure of the IGM, and the acceleration due to the
addition of mass to the shell (i.e.\ the term $(v/m)(dm/dt)$ where $v$ and
$m$ are the velocity and mass of the shell).  {\bf Panel 3:}
Acceleration due to gravity (left axis) and gravitational potential
(right axis). {\bf Panel 4:} Elapsed time since launch at initial
radius.  The dotted line indicates the time corresponding to the next
simulation output used.
}
\label{fig-intsampz1.5}
\end{figure} 
}

\subsection{Other Wind Models}

	Having examined the fiducial model, we now turn to a set of
models in which the simulation parameters have been varied.  In
analyzing these variations we may divide their effects on metal
distribution into three aspects.  First, the yield $y_*$, when
combined with the simulation's star-formation efficiency, determines
the {\em total} metal content in the simulation volume and the rough
normalization of the $M-Z$ relation.  Second, $v_{\rm out}^{\rm fid}$,
SFR$_{\rm crit}$, $\epsilon_{\rm ent}$, $\chi$ and $Y_{\rm ej}$
determine the fraction of metals that escape into the IGM; $Y_{\rm
ej}$ controls this directly, whereas the other three control whether
outflows occur, whether they are strong enough to escape the
galaxies and whether the metals get far enough away from the galaxy so that they
do not fall back.
These parameters, therefore, directly affect the ratios $f_{\rm IGM}$
and $\langle Z_{\rm cl} \rangle / \langle Z_{\rm gal} \rangle$, as
well as affecting the M-Z relation and the level of enrichment of the
IGM.  Third, $v_{\rm out}^{\rm fid}$, $\dot m_{\rm out}/\Omega_{\rm
out}$ and $\epsilon_{\rm ent}$ affect how {\em far} metals travel from their
progenitor galaxies; this is most strongly reflected in $\langle
Z_{\delta=1}\rangle/\langle Z_{\delta=100}\rangle$ or $\langle
Z_{\delta=1}\rangle /\langle Z_{\rm cl}\rangle$, and in the slope of
the curves in Figs.~\ref{fig-wenrich} and \ref{fig-wmods}.

The effects of changing $y_*$ are clear; the output metallicities of
the IGM and galaxies are all proportional to this parameter.
Comparison of the M-Z relation of well-resolved simulation galaxies to
the observed relation (see Figure~\ref{fig-zgal}) indicates $y_* \sim 1$
is appropriate, and values differing from this by more than a factor
of $\sim 2$ could not account for the metallicity of observed
galaxies unless $f_{\rm ej} \ga 1/2$.

	The effects of changing the parameters $v_{\rm out}^{\rm
fid}$, SFR$_{\rm crit}$, $\epsilon_{\rm ent}$ and $\chi$ are a
bit more complicated and we describe such variations each in turn.

	As discussed in \S~\ref{sec-windmeth}, superwinds in nearby
starburst galaxies appear to be characterized by outflow velocities of
$\sim600\kms$, but it may be that faster or slower wind velocities
better capture the real effect of winds.  Models `lv' and `hv' (see
Figure~\ref{fig-wmods} and Tables~\ref{tab-windmodrun}
and~\ref{tab-windmodrun128}) assume wind velocities of $300\pm 100\kms$
and $1000\pm200\kms$, respectively.  The high velocity model exhibits
the features one would expect: more metals escape because the outflows
are less easily confined to small radii, and metals travel farther
from their progenitor galaxies.  Also, the M-Z relation becomes
somewhat stronger (the mass threshold above which galaxies tend to
retain their metals increases).  Both trends are reversed in the low
velocity model.  Interestingly, the `W144lv' model shows that some
metals reach very low-density regions quite well even if the outflow
velocities only drive metals to $\la 100\,$kpc.  At $z=3$ in the high
resolution simulations the results are similar.  IGM enrichment is
quite high for the `W128hv' model, and the enrichment would be far
higher if not for the time constraint.  This is not surprising since
the model assumes $1000\kms$ winds flowing from dwarf galaxies; it is
not clear that such winds could be sustained for long without a
catastrophic effect on their hosts. Note, however, that the value of
SFR$_{\rm crit}$ is such that the galaxies only drive strong winds
early on; thus galaxies cannot eject all of their metals. The `W128lv'
model is interesting for it shows that even low velocity winds can
enrich the $\delta=1$ IGM to $\> 1/1000\,Z_\odot$ (assuming that the
overestimation of the SFR and $\Omega_*$ by the simulation is not too
severe; see \S~\ref{sec-simsvsobs}).  We have also tested even lower
outflow velocities at high $z$; model `W128llv' assumes outflow
velocities of $100\pm50\kms$.  Even this model enriches the IGM at
$\delta > 10$ to a {\em mean} level of $\ga 1/1000\zsol$.

	As discussed in \S~\ref{sec-windmeth}, galaxies with
SFR/(area) $\lsim 0.1\msol{\rm\,yr^{-1}}\kpc^{-2}$ do not seem to
drive observable superwinds, whereas starburst galaxies that {\em do}
drive such winds can have much higher areal SFRs.  But since we cannot
formulate a rigorous criterion for which galaxies in our simulations
are driving winds, it is useful to check how strongly the enrichment
predictions depend on SFR$_{\rm crit}$.  Models `W144lcrit' and
`W144hcrit' (not plotted) reveal that halving or doubling SFR$_{\rm
crit}$ changes the results in a predictable way. For SFR$_{\rm
crit}=0.05$, all galaxies at $z=2$ drive outflows, as do a small
fraction of $z=0$ galaxies.  The ejection efficiency is significantly
increased, as is $\langle Z_{\delta=1}\rangle$; there is a somewhat
smaller effect on the IGM metallicity.  Further decreasing the
critical SFR would have only a small effect. These trends are roughly
reversed when SFR$_{\rm crit}$ is doubled, although in this case yet
higher values would lead to progressively more suppression of winds.

	An important uncertainty in our assumptions is the entrainment
fraction $\epsilon_{\rm ent}$, introduced to account for the fact that
an expanding wind-driven shell almost certainly fragments and may
either leave a large fraction of itself behind or significantly reduce
its covering factor, or, even if fragmentation is not severe, may not
sweep up all of the ambient medium if the gas is clumped.  In
principal this parameter could be deduced from numerical simulations,
but current simulations do not follow the shell far enough into the
IGM or resolve small-scale structure in the gas well enough to do
this.  We have tried values of $\epsilon_{\rm ent}$ of 1\%, 10\% and
100\% in model `le', the fiducial model, and `he' respectively.  These
reveal that the entrainment fraction is quite important (see
Figure~\ref{fig-wmods} and
Tables~\ref{tab-windmodrun},~\ref{tab-windmodrun128}), especially in
the lower-mass galaxies (the high mass galaxies tend to retain their
winds gravitationally).  The differences are particularly large in the
enrichment of the low-density IGM.  As in the case of the wind
velocity, this sensitivity is actually useful, as it could be used to
constrain the properties of winds given the observed enrichment of the
IGM (if winds are assumed to be responsible).

	Another parameter that is somewhat uncertain is the wind
efficiency $\chi$ that fixes the constant of proportionality between
the wind energy and the SFR.  We have calibrated $\chi$ to reproduce
the approximate observed mass outflow rates. Models `l$\chi$' and
`h$\chi$' show the effect of changing this efficiency by a factor of
two.  The effect is significant, showing that continual driving by
the wind is important in the shell's propagation.  The importance of
$\chi$ decreases with increasing $\epsilon_{\rm ent}$ (as the shell
propagation becomes dominated by conservation of the initial shell
momentum).

	Two final parameters that might be varied (and for which {\em
a priori} values are hard to justify) are $\alpha$ (controlling the
steepness of the radial profile of metal distribution) and $\beta$
(the sharpness of the energy attenuation below SFR$_{\rm crit}$).
Trials with $\alpha = 1, 2, 3$ and 4 have at most $\sim 20\%$
differences in the output quantities listed in
Table~\ref{tab-windmodrun}.  Different values of $\beta$ give
important differences in the results, but entirely predictable
ones. For $\beta=0$ the results are similar to the `Q' model described
below (i.e.\ they simulate SFR$_{\rm crit}=0$).  Choosing larger
values of $\beta$ has no effect for small SFR$_{\rm crit}$, a drastic
effect if all of the galaxies have SFR $<$ SFR$_{\rm crit}$, and an
effect similar to raising SFR$_{\rm crit}$ itself for intermediate
values.

In summary, we find that the fraction of metals that escape galaxies,
expressed as $f_{\rm IGM}$ or $\langle Z_{\rm cl} \rangle / \langle
Z_{\rm gal} \rangle$, is fairly sensitive to the assumed outflow
velocity and to the entrainment fraction (varies by a factor of 2-3
within our assumed range), is slightly less sensitive to the critical
SFR/(area) and wind efficiency (varies by less than a factor of 2),
roughly scales with $Y_{\rm ej}$, and is insensitive to other
parameters such as $\alpha$ and $\beta$.  Very similarly, the
enrichment of low-density regions, which requires metals to travel farther
from their progenitor galaxies, is also sensitive to $v_{\rm out}^{\rm
fid}$ and $\epsilon_{\rm ent}$ ($\langle Z_{\delta=1}\rangle$ varies
by a factor of up to $\sim 10$).  Low-density enrichment is somewhat
less sensitive to the assumed critical SFR and wind efficiency,
again roughly scaling with $Y_{\rm ej}$, and is not sensitive to
$\alpha$ or $\beta$.  Cosmologically-averaged metal mass is determined
by $y_*$ but can be constrained strongly by comparing the calculations
to the observed M-Z relation and cluster metallicities.

	After varying many of the parameters individually, we have
 also varied combinations of parameters to model various physically
 distinct possibilities.  First, we have performed trials with
 $\epsilon_* = 0.004/\Omega_*^{\rm sim}(z=0) = 0.36$; this is
 equivalent to dividing SFR$_{\rm crit}$, and multiplying $\chi$ and
 $y_*$, by the same value.  As described in \S~\ref{sec-otherpar},
 $\epsilon_*$ is introduced to account for possible differences between the
 simulated and observed SFR, mass function and $\Omega_*$ of galaxies.
 Setting a low $\epsilon_*$ accurately mimics an IMF in which most
 mass goes into low-mass stars and brown dwarfs, and somewhat less
 reliably adjusts for over-efficient star formation in the
 simulations.  The `lsfr' models show significantly less ejection and
 a smaller enrichment of the IGM.  The smaller ejection fraction is
 largely due to the effectively larger SFR$_{\rm crit}$.  Winds reach
 smaller radii, due to the lower effective value of $\chi$. 
The very low
 metallicity of the low density IGM is a product of the lowered $y_*$
 and the lower ejection efficiency.  The former effect is realistic if
 $\Omega_* / \Omega_b$ is significantly overestimated by the
 simulations.  However, if we do assume a steep IMF so that many
 low-mass objects are present, we see that $y_* > 1$ would be required
 if galaxies are to have reasonable metallicities, and this would tend
 to cancel the effect.

	As pointed out in \S\ref{sec-windmeth}, we have formulated our
model to simulate powerful winds from galaxies with the highest SFRs.
But even galaxies that are {\em not} undergoing violent star formation
may drive winds (c.f. \S~\ref{sec-windmeth}), though these must have
$\chi\la 1$ most of the time, or they would disassemble the entire
galaxy over time (the mass outflow rate given by Eq.~\ref{eq-mdotchi}
would always be larger than the SFR).  The `q' and `qlv' models were
chosen to represent such winds; SFR$_{\rm crit}$ is set low enough so
that almost all galaxies drive winds.  The wind efficiency $\chi$ is
1/10th its fiducial value, and $Y_{\rm ej}=0.5$.  At low-z and from
massive galaxies, IGM enrichment in model `W144q' is almost as
effective as in model `W144', both because the wind stopping radius
depends fairly weakly on $\chi$, and the larger fraction of galaxies
driving winds compensates for the less effective ejection.  At
high-$z$ the `W128q' model enriched significantly less than the
fiducial model; this occurs because most galaxies at $z > 3$ are
driving winds even for the fiducial value of SFR$_{\rm crit}$.  The
`qlv' model assumes also that the winds are relatively slow-moving.
The enrichment in this model is very weak at both high and low $z$,
thus if `quiescent' winds are to enrich the IGM significantly, they
must have fairly high velocity.

	Given the assumption that wind speed does not depend on galaxy
mass, low mass galaxies (mostly unresolved by the $144^3$ simulation)
should eject metals most efficiently.  Thus we expect our $144^3$
simulations to under-predict the IGM enrichment due to winds.
Nevertheless it is interesting that neither the fiducial model nor any
of its minor variants can account for the metal enrichment of
groups/clusters.  Because we resolve most of the observed $z=0$ mass
function, this indicates that something other than winds provides the
bulk of the enrichment, or that the enrichment happens at fairly high
redshift (where the mass function shifts to lower mass galaxies not
resolved by the $144^3$ simulation), or that winds are described by
parameters somewhat different than those we have assumed. To make this
point more robust, we have generated a `maximal' model, `W144max', which
combines a high wind velocity, high wind efficiency, low entrainment
fraction and low critical SFR.  While galaxies in this model eject
$\sim 20\%$ of their metals, the enrichment of clusters is still only
about 1/10th solar, several times smaller than observed.  Moreover,
this is achieved at the cost of a mass-metallicity relation
significantly steeper than that observed.  In fact, it does not appear
possible to fit both the M-Z relation and the cluster metallicity for
any set of parameters (i.e.\ without modifying the method).  Because of
the low resolution of our $z=0$ simulation it is premature to draw
strong implications for the enrichment of clusters, but we hope to
return to this topic in a future study.

	An interesting physical effect we can examine with our
calculations is the effect of winds on {\em dynamical} enrichment: if
metals are moved into galaxy halos, it seems likely that they will be
more easily removed by dynamical processes. This effect has been seen
in detailed simulations by Murakami \& Babul (1999) of galactic winds
in clusters, and discussed in the context of `general' IGM enrichment
by Ferrara et al. (2000).  To investigate this effect we have run wind
models in the $128^3$ and $144^3$ simulations with wind velocities of
$300\pm100\kms$ and $1000\pm200\kms$, respectively.  In the first, we
generate metals only at $z=5$, and examine their distribution at $3
\le z\le 5$.  In the second, we generate metals only at $z=2$, and
examine their distribution at $0 \le z\le 2$.  The results are shown
in Fig.~\ref{fig-diff}, with the $128^3$ runs on the left and the
$144^3$ on the right.  The top panels shown median metallicities; mean
metallicities are given in the bottom panels.  The right panels show
that at low $z$ the effect exists; some metals deposited at $\delta
\gg 100$ find themselves at $\delta \la 100$ at $z=0$.  But the effect
is slight, and is strongest (particularly as evinced by the median
metallicities) at $\delta \sim 100$, suggesting that it may be happening
primarily in groups and clusters.  Indeed the left panels show that
between $z=5$ and $z=3$ metals tend to migrate from low- to
high-density regions (the metals which go `missing' from the plots are
those absorbed by stars).  Thus it seems that in the `general' IGM,
dynamics tend to move metals from moderate density regions into
galaxies and other higher-density regions, rather than distributing
them more widely into low-density regions.

\ifthenelse{\boolean{apj}}{
\begin{figure*}
\vbox{ \centerline{ \epsfig{file=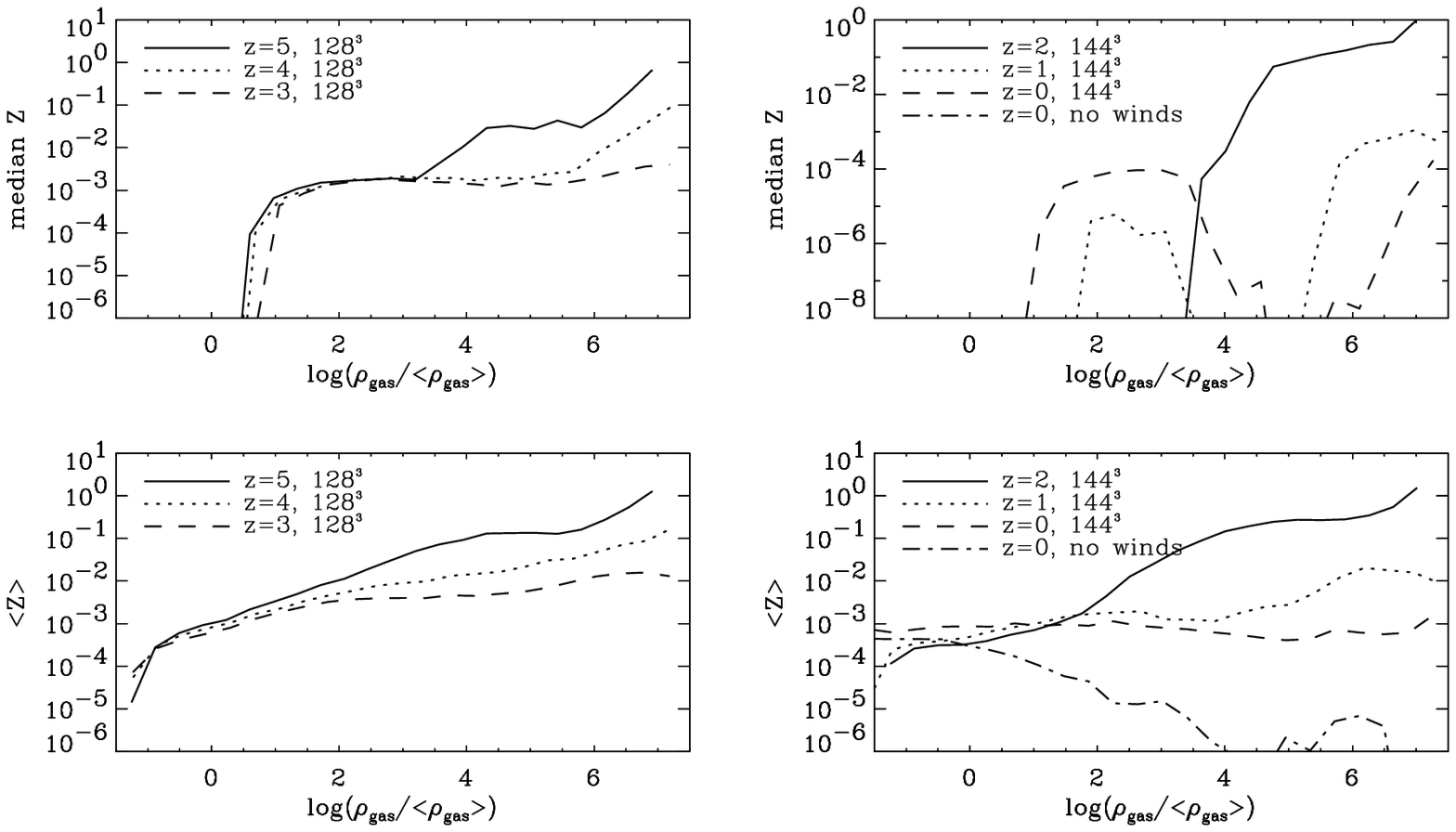,width=18.0truecm}}
\figcaption[]{ \footnotesize Movement of metals by gas processes after
their distribution near galaxies by winds.  Only the timestep at $z=5$
(left panels) or at $z=2$ (right panels) was enriched.  The median
(top panels) and mean (bottom panels) metallicity (vs. density) at
lower redshifts is plotted, showing the flows of enriched gas between
regions of different density.  Metals which seem to `disappear' with
decreasing redshift are those incorporated into stars.  We also plot
the $z=0$ results for local enrichment (no winds) at $z=2$ (right
panels).  This line is missing from the top right panel because the
median gas metallicity is everywhere zero.
\label{fig-diff}}}
\vspace*{0.5cm}
\end{figure*}
}{
\begin{figure*}[tbp]
\centerline{\epsfxsize=5.0in
\epsffile{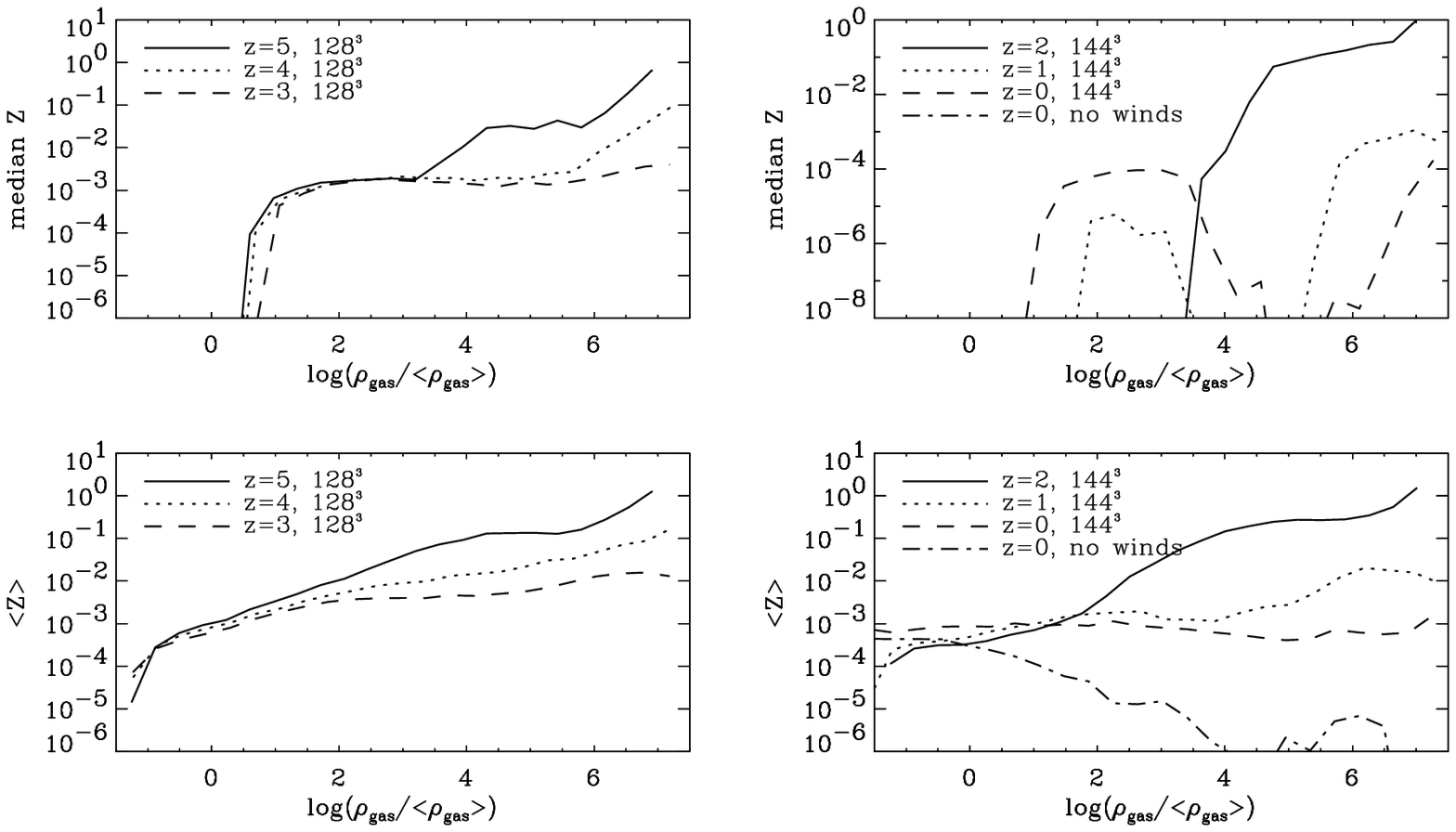}}
\caption{\baselineskip=12pt
Movement of metals by gas processes after
their distribution near galaxies by winds.  Only the timestep at $z=5$
(left panels) or at $z=2$ (right panels) was enriched.  The median
(top panels) and mean (bottom panels) metallicity (vs. density) at
lower redshifts is plotted, showing the flows of enriched gas between
regions of different density.  Metals which seem to `disappear' with
decreasing redshift are those incorporated into stars.  We also plot
the $z=0$ results for local enrichment (no winds) at $z=2$ (right
panels).  This line is missing from the top right panel because the
median gas metallicity is everywhere zero.
}
\label{fig-diff}
\end{figure*}}

\subsection{Summary and Discussion}

In summary, we find that many physically distinct `types' of winds can
enrich the IGM to comparable levels.  For example, quiescent winds
(weak winds from all galaxies) can enrich the IGM at all densities to
a similar level (within a factor of two) as winds with very strong
outflows from only the galaxies with the very highest SFRs.  The
predictions are, however, different enough (e.g., in the M-Z relation)
that with refined simulations and better observational data our
calculations could be used to constrain the physical models of winds
themselves.

While the emphasis of this paper is methodology, we can draw some
general conclusions regarding wind enrichment of the IGM using our
calculation.  We repeat that we have {\em not} shown that winds can
develop in any particular type of galaxy (as in
e.g., Mac Low \& Ferrara [1999] or Strickland \& Stevens [2000]). Instead,
we assume the wind properties based on observations, amalgamating the
physical criteria for a wind into the methodological criterion of a
high enough areal SFR.  Under this assumption, galaxies can lose a
significant fraction of their
metals to the IGM.  Averaged over all galaxies, we find that typically
5-50\% of metals are expelled into the IGM, and that the fraction is
highly degenerate in the assumed parameters.

Our calculations indicate that galactic winds can enrich even the
low-density regions of the IGM quite effectively.  At $z=3$, winds
escape to large radii, and the ensuing enrichment is sufficient to
roughly account for the metallicity seen in low-density Ly$\alpha$
absorbers (see Aguirre et al. 2001b for more discussion).  Enrichment
of the IGM becomes progressively higher, and spreads to progressively
lower density regions of the IGM, as redshift decreases; this is in
agreement with the findings of Cen \& Ostriker (1999).  At $z=0$, we
predict that even quite underdense regions of the IGM are enriched to
a mean metallicity of $Z \gsim 0.005\zsol$ unless the parameters are
pushed to unreasonable values.

	The main implication we can draw for the enrichment of cluster
gas is that large galaxies can enrich the gas to $\sim 1/10$ of its
observed metallicity given our fiducial model assumptions about the
wind properties. (Some of this enrichment is direct, and some probably
occurs because metals are moved into galactic halos where they are
more easily removed by dynamical processes.)  This is probably an
underestimate of the overall importance of winds to cluster enrichment
because we cannot address the importance of low-mass galaxies at high
redshift, which would eject metals more efficiently.  If winds are to
account for the metals in clusters, we find that the enrichment must
happen at relatively high redshift, or that one of our assumptions
regarding winds must be modified.

Our simulations also make predictions about the properties of
galaxies.  While our range in galaxy masses is too small for a
conclusive comparison, we find that the M-Z relation is nevertheless a
good way to break the degeneracy between our parameters -- especially
that between outflow speed and wind efficiency.  While there seems to
be an M-Z relation in observed bright ($M_B \lsim -21$) galaxies, we
have difficulty reproducing this relation.  Several effects may
contribute to this difficulty.  First, our assumptions that all galaxy
types have the same outflow velocity may be flawed; but
experimentation reveals that we would have to change this assumption
drastically to alter the high-mass M-Z relation, partially because the
highest mass galaxies tend to be in clusters where the ICM suppresses
their winds.  Second, the brightest galaxies may be too extended for
our method to treat well (i.e.\ the winds are stopped inside and by the
galaxy itself).  Third, the `average' metallicity determined by
observations may include fewer of the outlying, low-metallicity stars
than in the simulations.  Finally, it may be that something besides
wind ejection causes the observed M-Z relation in massive galaxies.

The results reported in this section point to a few general
conclusions regarding our methodology.  First, like semi-analytic
theories of galaxy formation, the method employs a number of
parameters that strongly affect the calculation predictions and that
cannot be ascribed definite values using observational data.  This
limits the number of specific general conclusions we can draw.
Nevertheless, the method is excellent for generating predictions {\em
given} a specific model (e.g., the model based on locally observed
starburst-driven superwinds).
Further studies, either observational or theoretical, producing more
robust physical models of winds will generate correspondingly more
robust predictions of the cosmic metal distribution using our method.

\begin{deluxetable}{ccccccc}
\singlespace
\footnotesize
\tablecaption{Wind Models, $144^3$ simulation} 
\tablehead{\colhead{Model} &
\colhead{Variation} & 
\colhead{$f_{\rm IGM}$} &
\colhead{$\langle Z_{\delta=1}\rangle$} &
\colhead{$\langle Z_{\delta=100}\rangle$} &
\colhead{$\langle Z_{\rm cl}\rangle$} &
\colhead{$\langle Z_{\rm gal}\rangle$}}
\startdata 
W144 & none & 0.057 & 0.008 & 0.019 & 0.037 & 0.83 \nl
W144he & $\epsilon_{\rm ent}=1$ & 0.044 & 0.0052 & 0.012 & 0.031 & 0.84 \nl
W144le & $\epsilon_{\rm ent}=0.01$ & 0.086 & 0.013 & 0.035 & 0.054 & 0.79 \nl
W144lsfr & $\epsilon_* = 0.355$ & 0.035 &
0.0015 & 0.0025 & 0.0070 & 0.31 \nl 
W144$_{np32}$ & 32 galactic gas particles/angle & 0.056 & 0.0073 & 0.019 & 0.036 & 0.83\nl
W144$_{np128}$ & 128 galactic gas particles/angle & 0.056 & 0.0073 & 0.018 & 0.036 & 0.83 \nl
W144hv & $v_{\rm out}^{\rm fid} =
1000\kms$ & 0.088 & 0.015 & 0.035 & 0.057 & 0.79 \nl 
W144lv & $v_{\rm out}^{\rm fid} = 300\kms$
& 0.038 & 0.0040 & 0.0091 & 0.025 & 0.85 \nl
W144hcrit & SFR$_{\rm crit}=0.05$ & 0.039 & 0.0048 & 0.0088 &
0.023 & 0.86 \nl 
W144lcrit & SFR$_{\rm crit}=0.2$ & 0.091 &
0.012 & 0.039 & 0.051 & 0.79 \nl 

W144h$\chi$ & $\chi=2.0$ & 0.073 & 0.011 & 0.028 & 0.046 & 0.81 \nl
W144l$\chi$ & $\chi=0.5$ & 0.047 & 0.0057 & 0.014 & 0.032 & 0.84 \nl
W144max & hv, lcrit, h$\chi$, le & 0.22 & 0.042 & 0.10 & 0.12 & 0.66 \nl
W144q & SFR$_{\rm crit}=0.001$, $\chi=0.1$ & 0.052 & 0.0048 & 0.017 & 0.032 & 0.83 \nl 
W144qlv & q+lv & 0.036 & 0.0039 & 0.0082 & 0.022 & 0.85 \nl 
\enddata
\tablenotetext{}{Note: all results are given at $z=0$.}
\label{tab-windmodrun}
\doublespace
\end{deluxetable}

\begin{deluxetable}{cccccc}
\singlespace
\footnotesize
\tablecaption{Wind Models, $128^3$ simulation} 
\tablehead{\colhead{Model} &
\colhead{Variation} & 
\colhead{$f_{\rm IGM}$} &
\colhead{$\langle Z_{\delta=1}\rangle$} &
\colhead{$\langle Z_{\delta=100}\rangle$} &
\colhead{$\langle Z_{\rm gal}\rangle$}}
\startdata 
W128 & none & 0.47 & 0.01 & 0.18 & 0.34 \nl
W128he & $\epsilon_{\rm ent}=1$ & 0.19 & 0.00079 & 0.09 & 0.52 \nl
W128le & $\epsilon_{\rm ent}=0.01$ & 0.58 & 0.022 & 0.19 & 0.26 \nl
W128lsfr & $\epsilon_* = 0.355$ & 0.18 &
0.0017 & 0.021 & 0.18 \nl 
W128$_{np32}$ & 32 galactic gas particles/angle & 0.47 & 0.0093 & 0.19 & 0.33 \nl
W128$_{np128}$ & 128 galactic gas particles/angle & 0.54 & 0.0099 & 0.23 & 0.28 \nl
W128hv & $v_{\rm out}^{\rm fid} =
1000\kms$ & 0.57 & 0.017 & 0.20 & 0.27 \nl 
W128lv & $v_{\rm out}^{\rm fid} = 300\kms$
& 0.22 & 0.0017 & 0.098 & 0.51 \nl 
W128llv & $v_{\rm out}^{\rm fid} = 100\kms$
& 0.086 & 0.00016 & 0.038 & 0.61 \nl 

W128hcrit & SFR$_{\rm crit}=0.05$ & 0.32 & 0.0080 & 0.12 & 0.40 \nl 
W128lcrit & SFR$_{\rm crit}=0.2$ & 0.48 &
0.010 & 0.19 & 0.33 \nl 
W128h$\chi$ & $\chi=2.0$ & 0.56 & 0.016 & 0.21 & 0.28 \nl
W128l$\chi$ & $\chi=0.5$ & 0.36 & 0.0056 & 0.15 & 0.41 \nl
W128q & SFR$_{\rm crit}=0.001$, $\chi = 0.1$ & 0.083 & 0.00031 & 0.036 & 0.63 \nl 
W128qlv & q+lv & 0.033 & $5.8\times 10^{-5}$  & 0.010 & 0.68 \nl 
\enddata
\tablenotetext{}{Note: all results are given at $z=3$.}
\label{tab-windmodrun128}
\doublespace
\end{deluxetable}

\section{Models and Results for Dust Ejection}
\label{sec-resrad}

\subsection{Fiducial Dust Model}

	Radiation-pressure driven efflux of dust from galaxies
provides an interesting alternative to galactic winds to pollute the
IGM with metals.  A number of studies have pointed out that typical
spiral galaxies could eject a significant amount of their dust {\em
if} the dust can decouple from bound gas and magnetic fields (e.g.,
Chiao \& Wickramasinghe 1972; Ferrara et al. 1990; Shustov \& Vibe
1995; Davies et al. 1998; Simonsen \& Hannestad 1999).  Our method
does not directly address the question of whether dust can decouple
but {\em can} yield a reasonable estimate of the equilibrium radius of
dust for galaxies with properties given by the simulations.  If dust
can escape the inner galaxy there should be no obstacle to its
reaching the equilibrium radius, provided it can do so in a short
enough time. We can, therefore, give plausible estimates of the radii
to which dust grains might be ejected by galaxies of various masses,
ages and types, at various redshifts, and track the distribution of
metals after they are deposited, even if they only reach the halos of
galaxies.

Our fiducial dust-ejection model, labeled `P144' and `P128' in
Table~\ref{tab-dustmodrun} and in the figures, assumes graphite grains
with $Q_{\rm pr}/a\rho = 19$ (see eq.~\ref{eq-qpr}); this is the
maximum absorption efficiency from graphite grains subject to a 12,000
K blackbody spectrum.  We assume a Scalo IMF in calculating the
stellar luminosities, and a dust correction depending on metallicity,
adjusted so that the integrated background light has equal parts in
FIR and UV-optical-NIR.  We assume that half of each galaxy's
metals are distributed non-locally ($Y_{\rm ej}=0.5$), which is an
upper limit since only about half of a typical galaxy's metal mass
can be in dust.  (Also, the connection between the metal outflow rate
and the SFR is much less clear for dust ejection than for supernova
winds.)

	The results of the fiducial model are given in
Table~\ref{tab-dustmodrun} and Figs.~\ref{fig-zgal},
\ref{fig-penrich}, \ref{fig-pqz2}, and \ref{fig-pqz0}.  The table and
the first two figures give the same quantities that were presented for
winds, while the last two figures give details about the ejection from
individual galaxies. 

Under our assumptions, dust ejection is fairly efficient.  In the
$144^3$ simulation, $z=0$ galaxies have lost 16\% of their metals,
meaning that $32\%$ of the dust that we assume {\em can} escape
actually does escape, with the rest being retained either because it
falls back into galaxies or because the galaxies have insufficient
light/mass ratios to eject it.  The ejected dust enriches the
intracluster gas to $\sim 1/5\,Z_\odot$, and the $z=0$ IGM at mean
density is enriched to $0.02\,Z_\odot$.  These values can be roughly
linearly scaled to lower values of $Y_{\rm ej}$.  Dust ejection at
high $z$ pollutes the $\delta=1$ IGM at $z=3$ to $\sim
10^{-4}\,Z_\odot$. The values given are for {\em all} metals (solid
and gaseous).  Figure~\ref{fig-pmods} also shows the amount of dust
converted to gaseous metals by $z=3$ due to thermal sputtering, in a
number of models including the fiducial model. Nonthermal sputtering
and dust destruction during ejection would also destroy some dust, so
the actual level of observable gaseous metals should lie between the
`gaseous metals+dust' curves and the `gaseous metals only' curves.

	A few aspects of the results for the fiducial model merit
attention.  First, the enrichment of low-mass groups is much more
uniform than in the wind or dynamics-only model (compare figures~\ref{fig-denrich},~\ref{fig-wenrich}
and~\ref{fig-penrich}).  For dust ejection, the metallicity of hot gas
declines steadily with gas temperature, leveling off for the hottest
clusters.  For wind ejection -- even for runs with $f_{\rm ej} \sim
1/5$ -- there is a large scatter in the metallicity of groups.  It is
currently unclear which of these cases has more observational support.
Second, as is evident from Figure~\ref{fig-zgal}, the M-Z relation of
galaxies is enhanced by dust ejection over the dynamical removal case
and is more similar to the observed relation (the gas metallicities
match the observations particularly well).
The M-Z relation is, however, somewhat different than for winds;
although the general trend is similar, there is more scatter of
low-mass galaxies into high metallicities, and a somewhat less abrupt
drop at masses $M \lsim 10^{11}\msol$ (especially compared to wind
models with a high escape fraction).  As in the wind case, the stellar M-Z
relation for the most massive galaxies is flatter than observed.

	We can investigate the enrichment process in more detail by
examining the properties of the galaxies driving the dust outflows.
Figures~\ref{fig-pqz2} and \ref{fig-pqz0} show the relations between
galaxy mass $M$, metallicity $Z$, mass-to-light ratios $M/L_B$ and
$M/L_{\rm bol}$, and maximum radius of dust ejection $h^{\rm dust}$.
Interestingly, at $z \gsim 1$, fairly large galaxies drive dust to the
largest radii, whereas at $z \lsim 1$, dust ejection is most efficient
in the smallest and the largest galaxies (see panel 4 of both
figures).  This occurs because large galaxies have a relatively larger
ratio of stars to gas and dark matter.  At lower redshifts, this
effect is overwhelmed in the smallest galaxies by the dust correction
(which increases $M/L$ more in large galaxies; see panels 2 and 5),
and by the tendency of smaller galaxies to be younger.

 At high $z$, graphite dust can be driven to $\sim 1\,$Mpc by at least
some galaxies, whereas at low $z$ galaxies can drive dust only to a
few hundred kpc.  The $z=0$ galaxies have typical bolometric (dust
corrected) $M/L$ values 

\ifthenelse{\boolean{apj}}{ \vbox{ \centerline{
\epsfig{file=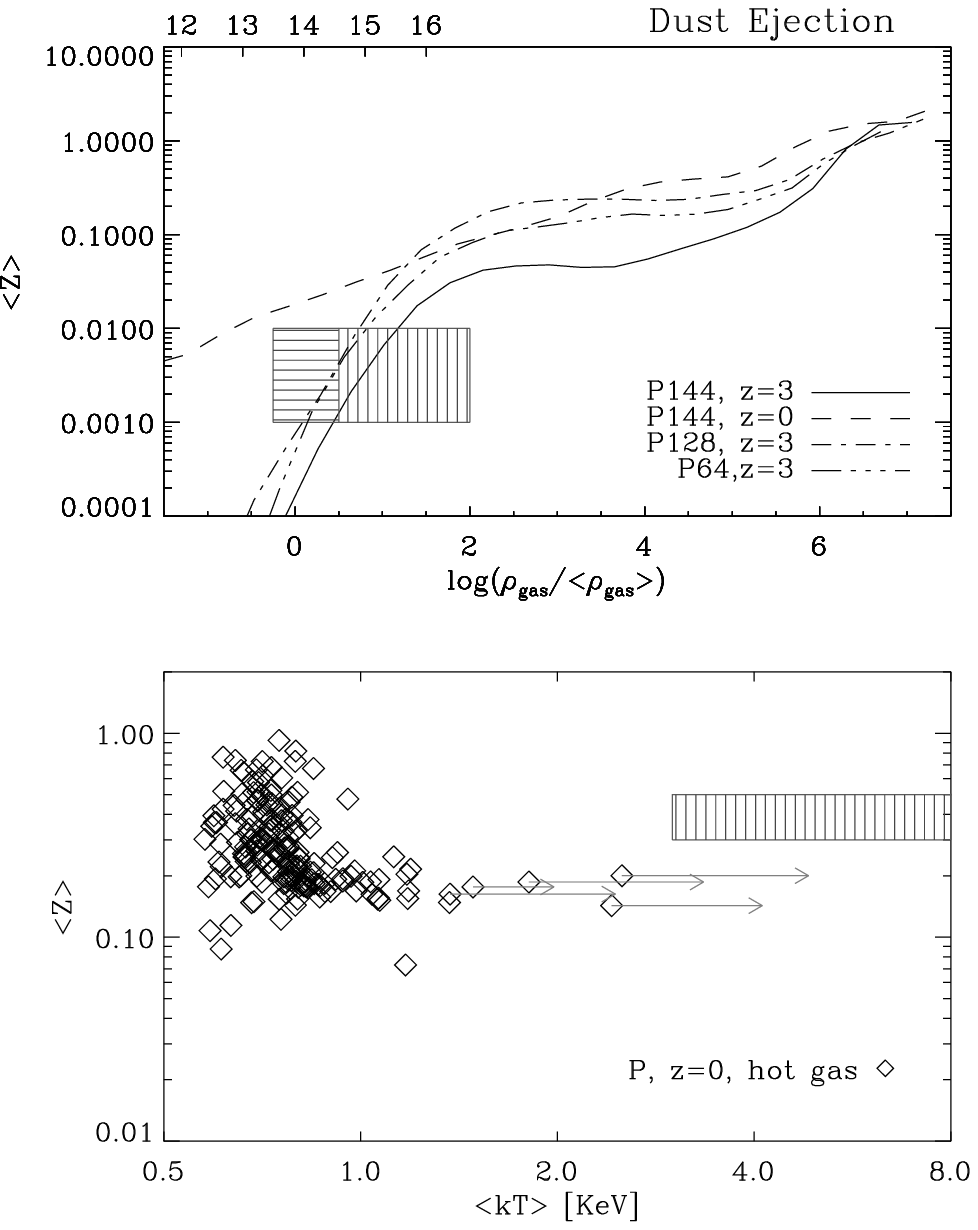,width=9.0truecm}} \figcaption[]{ \footnotesize
Enrichment of the IGM for radiation pressure ejection of dust in the
fiducial `P144' model (see Tables~\ref{tab-fidpardust} and
\ref{tab-dustmodrun}).  Plotted quantities are as in
Fig~\ref{fig-denrich}.
\label{fig-penrich}}}
\vspace*{0.5cm}
}{
\suppressfloats\begin{figure}[tbp]
\centerline{\epsfxsize=6.0in%
\epsffile{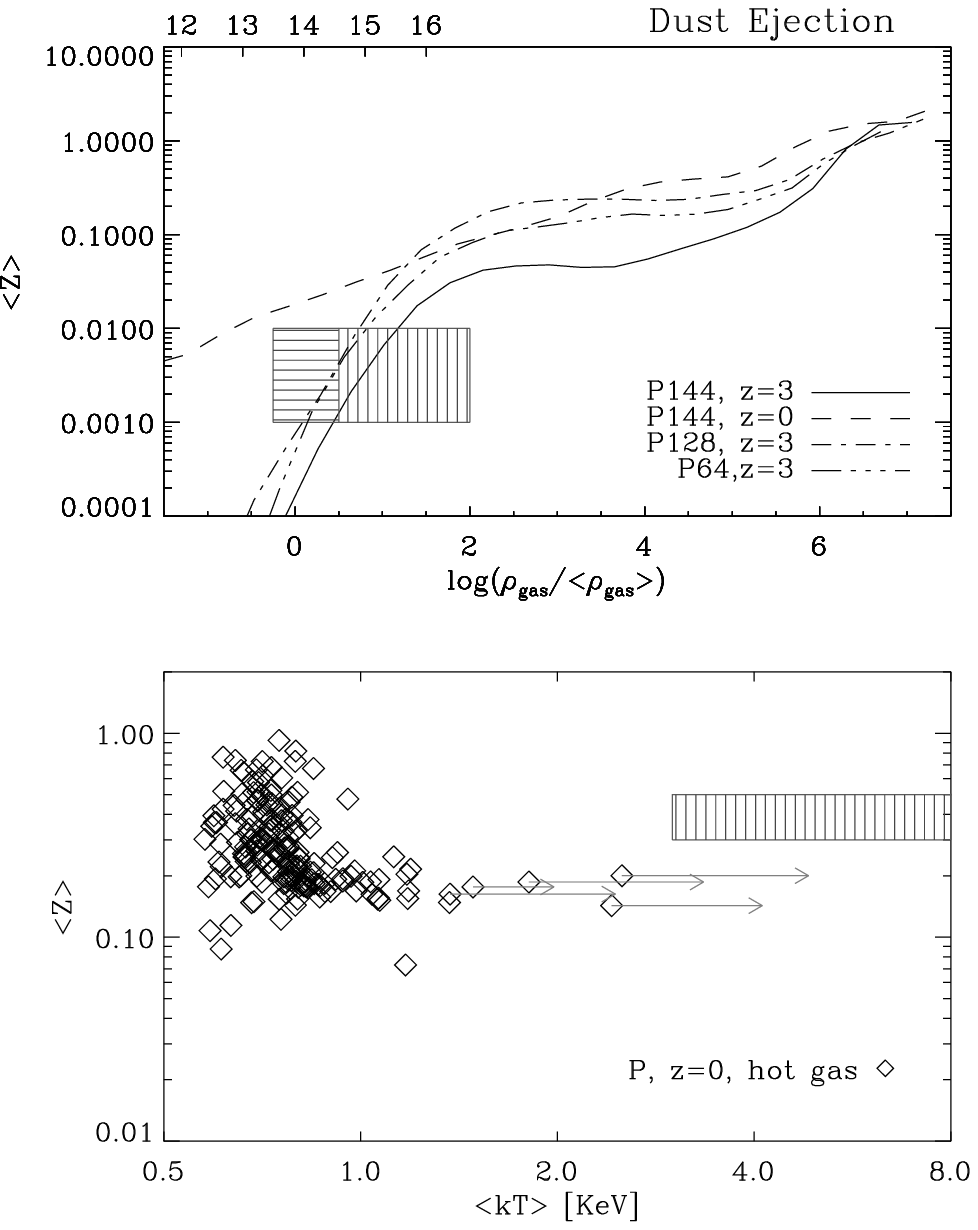}}
\caption{\baselineskip=12pt
Enrichment of the IGM for radiation pressure ejection of dust in the
fiducial `P144' model (see Tables~\ref{tab-fidpardust} and
\ref{tab-dustmodrun}).  Plotted quantities are as in
Fig~\ref{fig-denrich}.
}
\label{fig-penrich}
\end{figure} 
}

\noindent of $\sim 1-5$ and {\em uncorrected} $M/L_B$
ratios of $\sim 1-4$.  Panel 6 of Figure~\ref{fig-pqz0} shows the
B-band luminosity function of the $z=0$ simulation galaxies computed
using the simple spectral synthesis described in~\ref{sec-radmeth},
and compares it to the 2dF observed luminosity function (dashed line).
The simulated luminosity function with no dust correction (light solid
line) is a poor fit to the observations.  The dark solid line shows
the same data with the metallicity-dependent dust correction (given in
Eq.~\ref{eq-zcorr2}) used in calculating dust ejection.  Applying this
correction requires a choice of $C_f$, where $10^{C_f}$ (in this
context) is the ratio of B-band extinction to UV (1900$\,\AA$)
extinction.  The curve shown assumes $C_f=-0.5$; with such a choice
the simulations can roughly fit the observed luminosity function.  The
actual relation between UV and blue extinction depends not only on the
dust absorption curve but also on the amount and distribution of dust
in the galaxy (see, e.g., Charlot \& Fall 2000) and is uncertain; but
$C_f=-0.5$ is not an unreasonable value, indicating that the dust
correction and spectral synthesis methods can probably produce
satisfactory (good to within a factor of a few) luminosities for our
simulation galaxies.

\ifthenelse{\boolean{apj}}{
\begin{figure*}
\vbox{ \centerline{ \epsfig{file=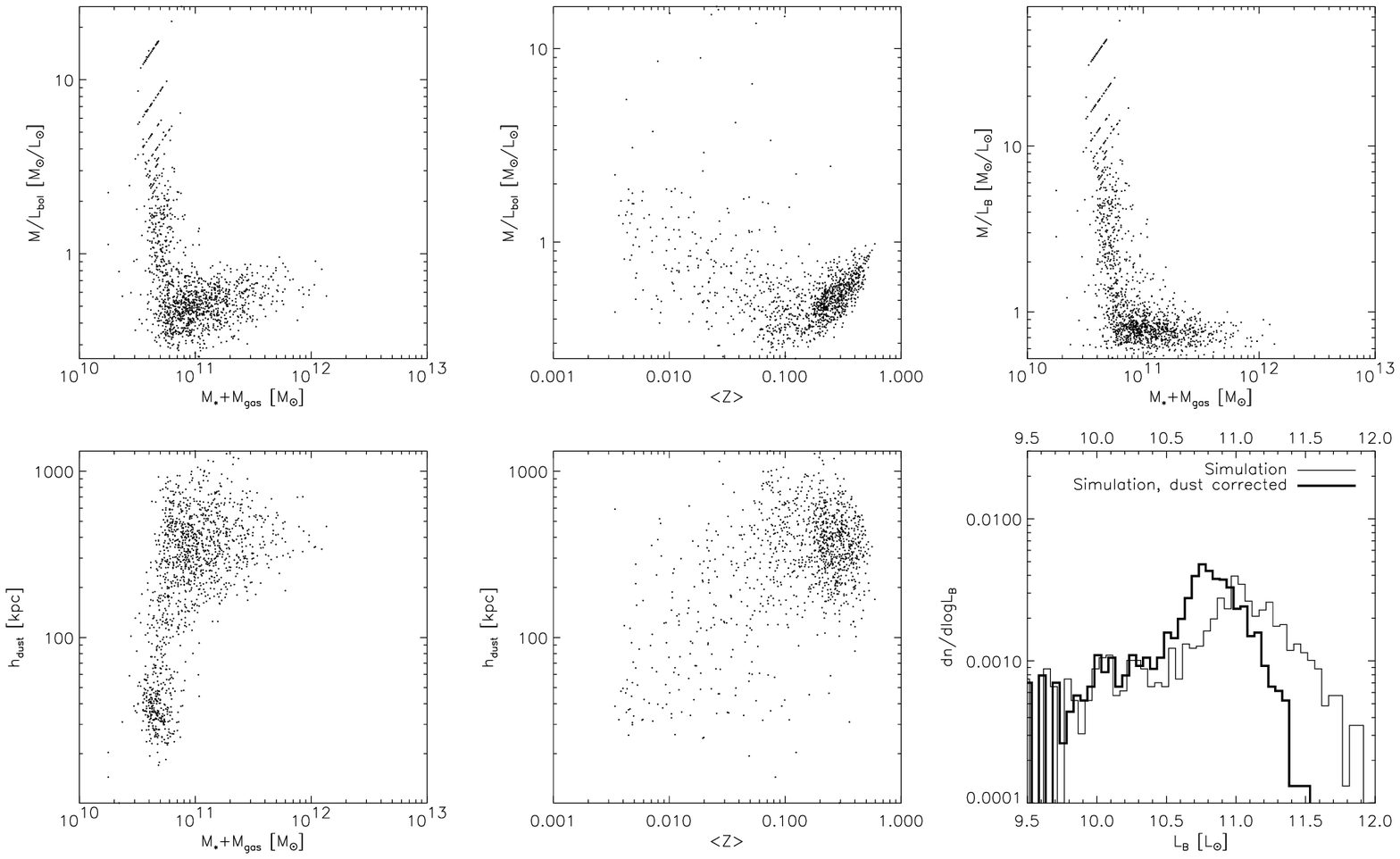,width=18.0truecm}}
\figcaption[]{ \footnotesize
Quantities used in dust ejection for the fiducial `P144' model at
$z=2$. {\bf Panel one:} Bolometric mass/light ratio $M/L_{\rm bol}$
vs. galaxy mass. {\bf Panel two:} $M/L_{\rm bol}$ vs. galaxy average
metallicity.  {\bf Panel three:} B-band $M/L_B$ vs. galaxy mass. {\bf
Panel four:} Maximal dust ejection radius $h^{\rm dust}$ vs. galaxy
mass. {\bf Panel five:} $h^{\rm dust}$ vs. mean metallicity. {\bf Panel
six:} B-band luminosity function of simulated galaxies, uncorrected
for dust (solid, thin) and corrected for dust using metallicity (solid, thick).
Top axis gives B-band magnitude.
\label{fig-pqz2}}}
\vspace*{0.5cm}
\end{figure*}
}
{
\suppressfloats\begin{figure}[tbp]
\centerline{\epsfxsize=6.0in%
\epsffile{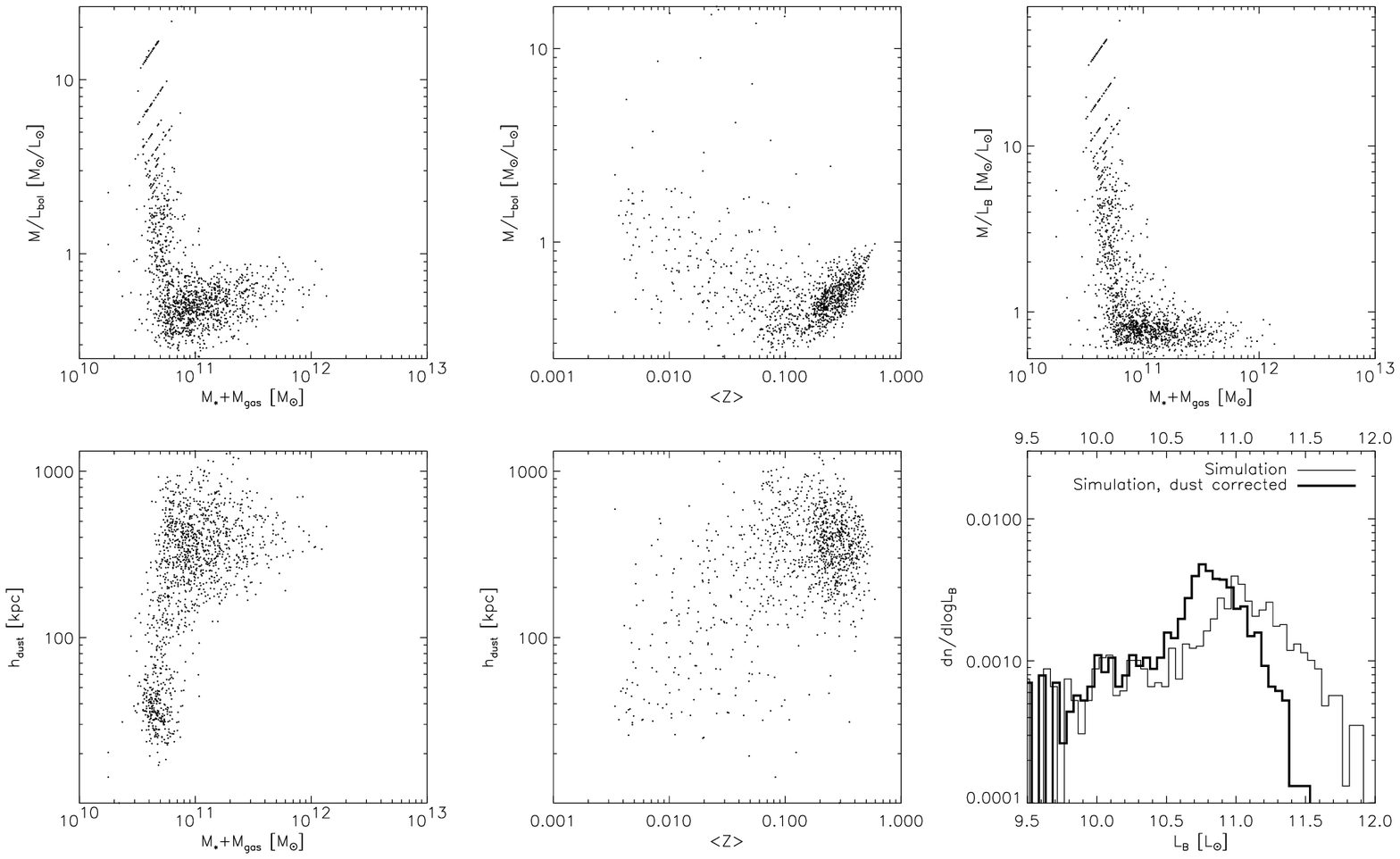}}
\caption{\baselineskip=12pt
Quantities used in dust ejection for the fiducial `P144' model at
$z=2$. {\bf Panel one:} Bolometric mass/light ratio $M/L_{\rm bol}$
vs. galaxy mass. {\bf Panel two:} $M/L_{\rm bol}$ vs. galaxy average
metallicity.  {\bf Panel three:} B-band $M/L_B$ vs. galaxy mass. {\bf
Panel four:} Maximal dust ejection radius $h^{\rm dust}$ vs. galaxy
mass. {\bf Panel five:} $h^{\rm dust}$ vs. mean metallicity. {\bf Panel
six:} B-band luminosity function of simulated galaxies, uncorrected
for dust (solid, thin) and corrected for dust using metallicity (solid, thick).
Top axis gives B-band magnitude.
}
\label{fig-pqz2}
\end{figure} 
}

\ifthenelse{\boolean{apj}}{
\begin{figure*}
\vbox{ \centerline{ \epsfig{file=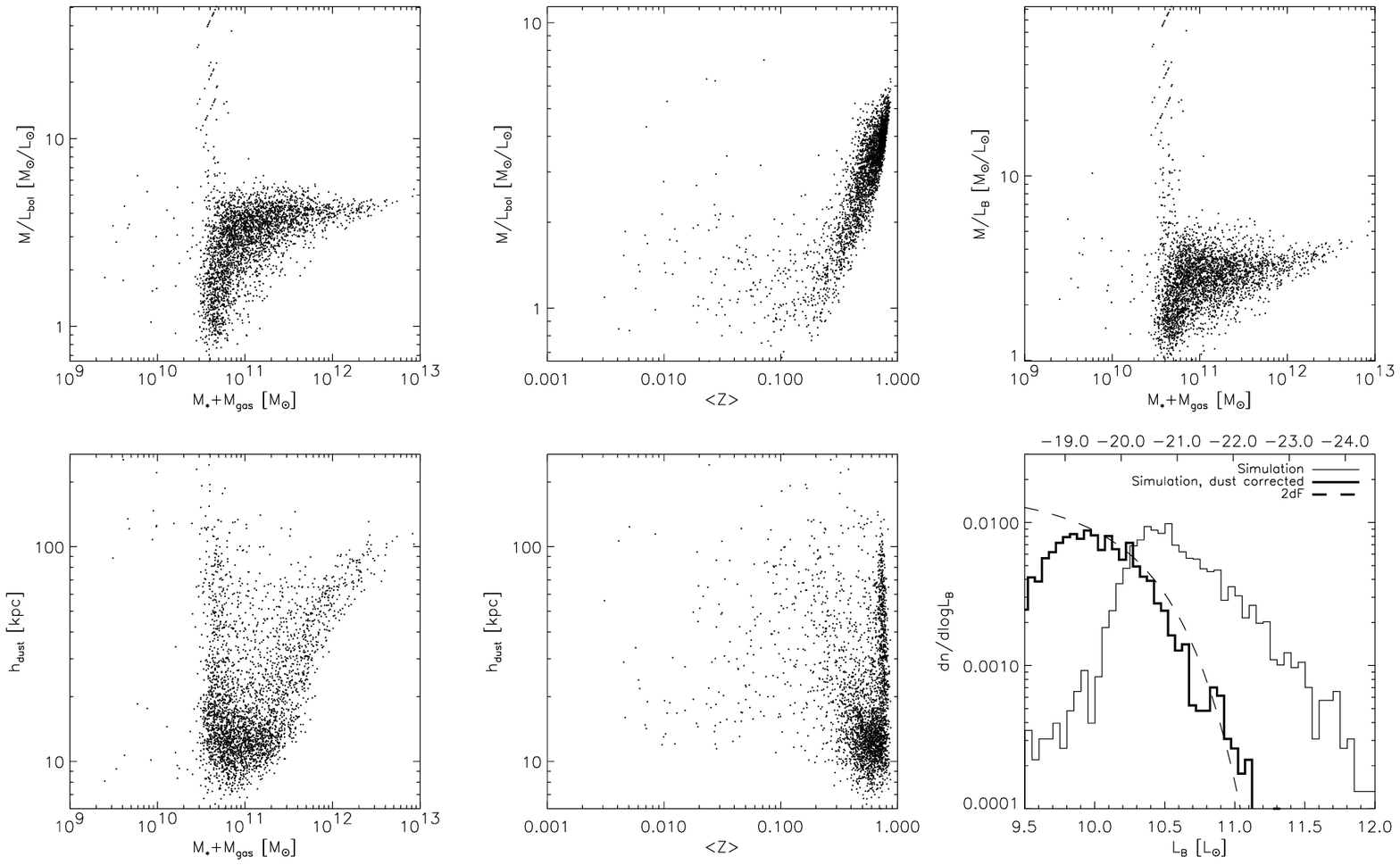,width=18.0truecm}}
\figcaption[]{ \footnotesize Quantities used in dust ejection for the fiducial
`P144' model at $z=0$. Plotted quantities are as in Figure~\ref{fig-pqz0},
except that we include the observed 2dF B-band luminosity function in panel
6 for comparison with the simulation.
\label{fig-pqz0}}}
\vspace*{0.5cm}
\end{figure*} 
}{
\suppressfloats\begin{figure}[tbp]
\centerline{\epsfxsize=6.0in%
\epsffile{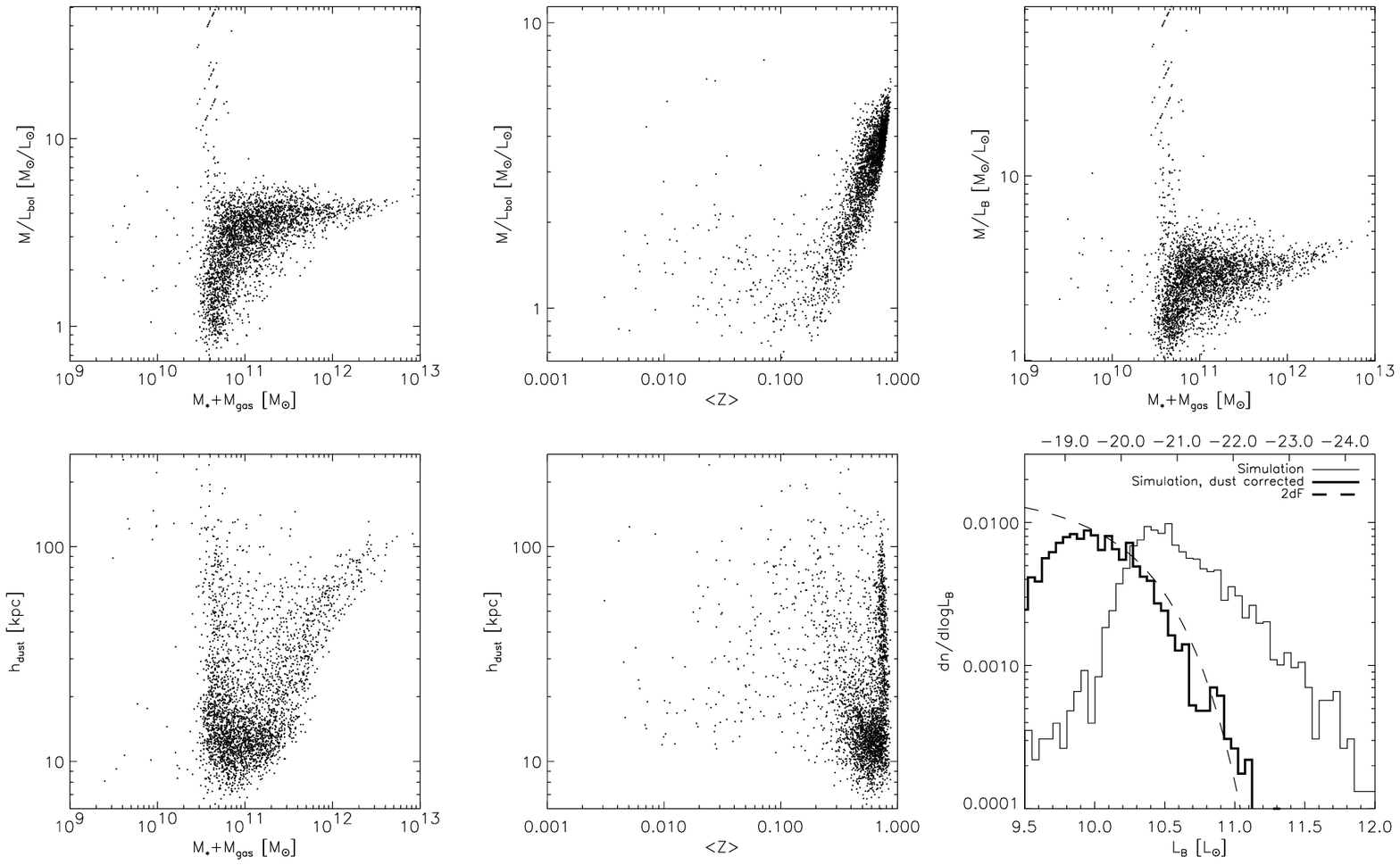}}
\caption{\baselineskip=12pt
 Quantities used in dust ejection for the fiducial
`P144' model at $z=0$. Plotted quantities are as in Figure~\ref{fig-pqz0},
except that we include the observed 2dF B-band luminosity function in panel
6 for comparison with the simulation.
}
\label{fig-pqz0}
\end{figure} 
}

\subsection{Other Dust Models}

	We now investigate the effects of variations in the model
parameters. As in the wind models, $y_*$ determines
the total simulation metal mass.  The amount of metals driven from
galaxies depends directly on $Y_{\rm ej}$, less directly on $Q_{\rm
pr}/a\rho$, the $M/L$ ratio from stellar synthesis, and the dust
correction, and quite weakly on the form of $W_{\rm grp}$;
the last four quantities also determine how far the dust travels.  In
turn, $Q_{\rm pr}/a\rho$ depends primarily on the dust type, $M/L$
values depend on the IMF, and the dust correction depends on
the FIR/optical ratio of the extragalactic background
light.

	The absorption efficiency of dust grains depends mostly on their
composition; in this study we have used opacities for both graphite
grains and less efficiently absorbing silicate grains.  Model `sil'
assumes the latter.  Since the maximal $Q_{\rm pr}/a\rho$ is 1/5 that
of graphite, ejection is significantly less efficient (see
Table~\ref{tab-dustmodrun} and Figure~\ref{fig-pmods}).  This is
important because the majority of dust mass is locked in silicates in
most carbon/silicate grain models (e.g., Weingartner \& Draine 1999;
Duley, Jones \& Williams 1989; Mathis \& Whiffen 1989).  Thus a model
in between the fiducial model and the `sil' model is a more accurate
representation of a realistic dust distribution for a two-component
model.

\ifthenelse{\boolean{apj}}{
\begin{figure*}
 \vbox{ \centerline{
\epsfig{file=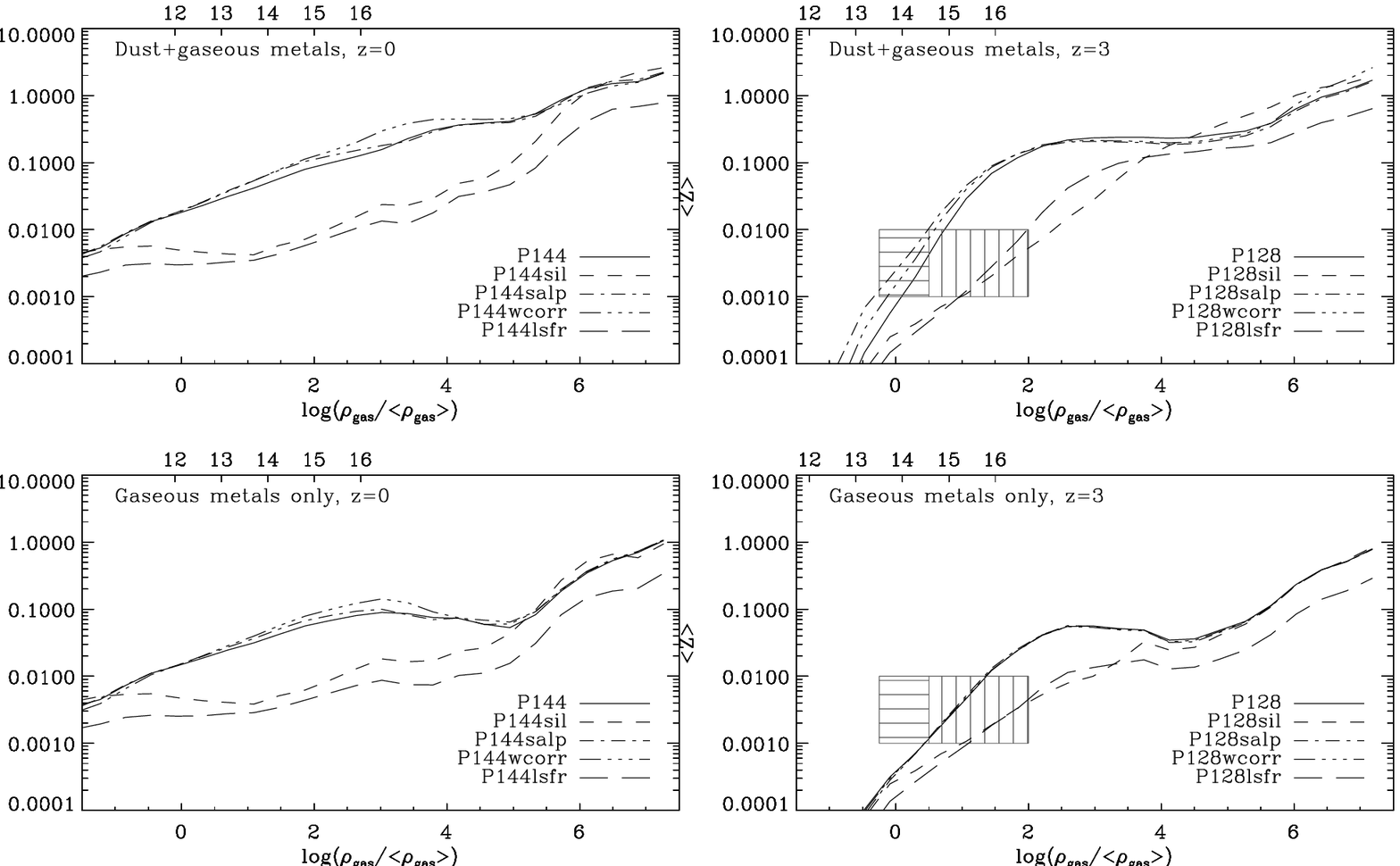,width=18.0truecm}} \figcaption[]{ \footnotesize
Enrichment of the IGM for radiation pressure driven dust ejection, for
several models (see Table~\ref{tab-dustmodrun}) at $z=0$ (left panels)
and at $z=3$ (right panels).  Plotted quantities are as in the top
panel of Fig~\ref{fig-denrich}. Top panels show total (gas+dust)
abundances and bottom panels show only gas phase abundances (after
thermal sputtering).
\label{fig-pmods}}}
\vspace*{0.5cm}
\end{figure*}
}{
\suppressfloats\begin{figure}[tbp]
\centerline{\epsfxsize=6.0in%
\epsffile{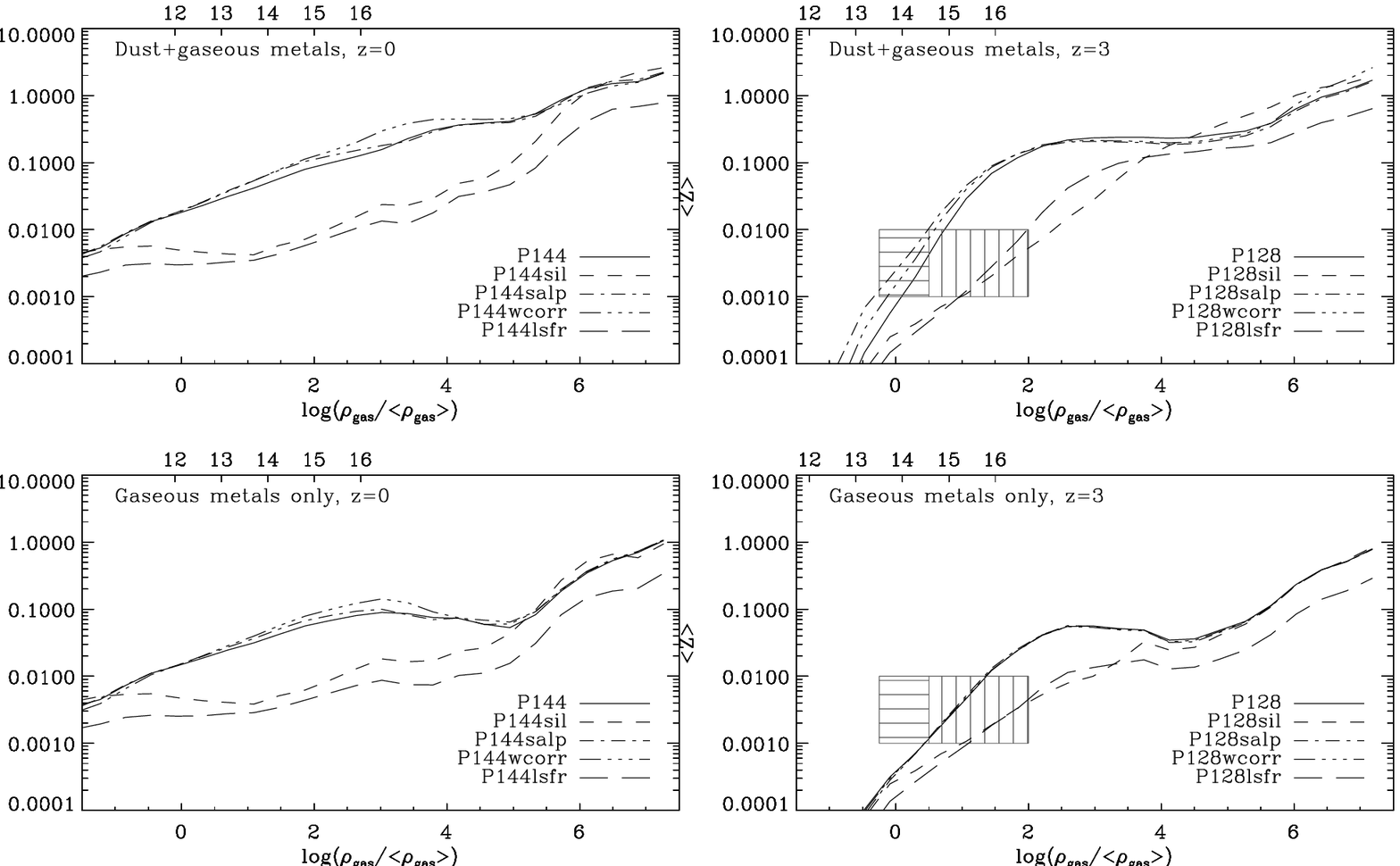}}
\caption{\baselineskip=12pt
Enrichment of the IGM for radiation pressure driven dust ejection, for
several models (see Table~\ref{tab-dustmodrun}) at $z=0$ (left panels)
and at $z=3$ (right panels).  Plotted quantities are as in the top
panel of Fig~\ref{fig-denrich}. Top panels show total (gas+dust)
abundances and bottom panels show only gas phase abundances (after
thermal sputtering).
}
\label{fig-pmods}
\end{figure} 
}
	We next tried a model with an assumed Salpeter IMF, with
$0.1\msol \le M \le 125\msol$.  This resulted in somewhat brighter
galaxies at $z \gsim 0.5 $, and somewhat dimmer galaxies at $z \lsim 0.5$;
the overall effect is to slightly enhance dust ejection.
The effect on the M-Z relation is also weak.  The simulation results
are also somewhat sensitive to the choice of the low-mass cutoff in
the IMF, as this changes the stellar mass with little effect on the
luminosity. For example, starting the Salpeter IMF at $0.2\msol$
rather than $0.1\msol$ would lower $M/L$ by $\approx 24\%$ and the
low-mass IMF advocated by Gould, Flynn \& Bahcall (1996) would
decrease it even more. These uncertainties are, however, unlikely to
change $M/L$ by more than a factor of two and are hence contained
within the range of variations we try.

	The assumed dust correction is more important than the chosen
IMF, and comparable in importance to the choice of grain properties.
Maintaining the dust correction based on metallicity, we have varied
$C_f$ to reproduce a present-day ratio of $1/3 \le F_{\rm FIR}/F_{\rm
opt} \le 2$ in the $z=0$ extragalactic background (where $F_{\rm opt}$
includes UV and NIR light also).  The extremes of this range are shown
in models `wcorr' and `scorr' in Table~\ref{tab-dustmodrun}. 
Although $h^{\rm dust}$ is generally several times higher in the `wcorr' model
than in `scorr', the enrichment of the low-density IGM is quite similar.  
This indicates that very low-density IGM regions can be spatially
close to high-density metal-forming regions.  We also modeled a
constant dust correction with $F_{\rm FIR}/F_{\rm opt}=1$, in model
`ccorr', and a luminosity-dependent dust correction (model `lcorr').
The results are only slightly different from the fiducial case.  As in
the other prescriptions, the dust ejection calculations are
insensitive to the number of time steps employed.

	Ejection of dust by radiation pressure requires that the dust
decouple from the gas.  Even in the brightest galaxies, radiation
pressure cannot overcome the gravitational force on both dust and an
associated gas mass more than $100$ times as large.  We have verified
this in our simulations using a trial with $Q_{\rm pr}/a\rho = 0.19$,
and no dust correction.  The results are nearly identical to the
dynamical removal run, indicating that radiation pressure is in this
case ineffective even at moving dust into the halos of galaxies.

\subsection{Summary and Discussion}

	The fiducial dust ejection model can account for both the
low-density IGM metallicity at $z=3$ and a significant fraction of
cluster metals. It also fits the stellar M-Z relation
reasonably well at intermediate masses, but not at high masses (where
none of our models predict high enough metallicity).

Dust can be ejected from galaxies by radiation pressure if three basic
conditions hold. First, the luminosity/mass ratio toward the galaxy
must be large enough that radiation pressure exceeds the gravitational
force.  Second, the gas drag on dust must be small enough that dust
can pass through the dense gaseous part of the galaxy and into the
halo/IGM before being destroyed.  Third, the dust must not be confined
by magnetic fields that are bound to the gas.  In this study, we have
addressed only the first condition, distributing the dust at the
radius where gravitational and radiation pressure forces balance.  We
have {\em not} addressed whether dust can truly decouple from the gas
or, if so, what dust outflow rate ensues. 

The trials presented in this section indicate that with reasonable
assumptions about the grain properties, spectral synthesis and dust
correction and with neglect of extra light from quasars and extra force on grains from
the photoelectric effect
 -- each of which gives approximately a factor of two
uncertainty in the radiation pressure force -- a significant fraction
of the dust can be expelled from the simulated galaxies.  This conclusion
would change only if several of the uncertainties conspired to make
the force an order of magnitude or more smaller than we have
estimated.

The main weakness of our method is a poor understanding of dust
ejection itself, and in particular how well magnetic fields can
confine grains to galaxies.  Assuming that grains can escape, our
method can provide strong predictions of the spatial distribution,
grain-size distribution and destruction of dust, as well as the extinction
and reddening due to IG dust.  These predictions will be presented in
detail in a future study.

\begin{deluxetable}{cccccccc}
\singlespace
\footnotesize
\tablecaption{Radiation Pressure Models Run}
\tablehead{\colhead{Model} & \colhead{Variation} & \colhead{$f_{\rm IGM}$} &
 \colhead{$\langle Z_{\delta=1}\rangle$} &
 \colhead{$\langle Z_{\delta=100}\rangle$} & 
\colhead{$\langle Z_{\rm cl}\rangle$} &
 \colhead{$\langle Z_{\rm gal}\rangle$} & \colhead{$C_f$}}
\startdata
P144 & none & 0.16 & 0.018 & 0.087 & 0.19 & 0.71 & -0.96 \nl
P64 & $64^3$ run & 0.32 & 0.00072 & 0.08 & - & 0.39 & -0.96  \nl
P128 & $128^3$ run & 0.27 & 0.00089 & 0.14 & - & 0.48 & -0.96 \nl
P144salp & Salpeter IMF & 0.19 & 0.019 & 0.11 & 0.20 & 0.68 & -0.81\nl
P128salp & SALP, $128^3$ & 0.33 & 0.0028 & 0.15 & - & 0.45 & -0.81 \nl
P144sil & silicate grains & 0.04 & 0.005 & 0.0084 & 0.03 & 0.86 & -1.05 \nl
P128sil & SIL, $128^3$ & 0.033 & 0.00028 & 0.0055 & - & 0.65 & -1.05 \nl
P144lsfr & $\epsilon_* = 0.355$ & 0.061 & 0.003 & 0.0065 & 0.021 & 0.30 & -0.33 \nl
P128lsfr & $\epsilon_* = 0.355$, $128^3$ & 0.098 & 0.00018 & 0.012 & -& 0.20 & -0.33\nl
P144scorr & $F_{\rm FIR}/F_{\rm opt} = 2$ & 0.11 & 0.015 & 0.052 & 0.12 & 0.77 & -0.64 \nl
P128scorr & sCORR, $128^3$ & 0.24 & 0.00064 & 0.12 & - & 0.50 & -0.64\nl
P144wcorr & $F_{\rm FIR}/F_{\rm opt} = 1/3$ & 0.20 & 0.019 & 0.11 & 0.22 & 0.67 & -1.26 \nl
P128wcorr & $F_{\rm FIR}/F_{\rm opt} = 1/3$ & 0.29 & 0.0012 & 0.15 & - & 0.47 & -1.26 \nl
P144lcorr & Luminosity correction & 0.15 & 0.016 & 0.073 & 0.16 & 0.73 & -1.09 \nl
P128lcorr & Luminosity correction & 0.28 & 0.0011 & 0.14 & - & 0.48 & -1.09 \nl
P144ccorr & const. correction & 0.15 & 0.016 & 0.077 & 0.18 & 0.73 & 0 \nl
P128ccorr & const. correction & 0.22 & 0.00054 & 0.12 & - & 0.51 & 0 \nl\hline
\enddata
\tablenotetext{}{Note: all results are given at $z=0$ for the $144^3$ run, and at $z=3$ for the other two runs.}
\label{tab-dustmodrun}
\doublespace
\end{deluxetable}

\section{Conclusions}
\label{sec-conclusions}

	We have developed a new method of calculating the chemical
evolution of galaxies and of the IGM, with particular emphasis on the
physical mechanisms that remove metals from galaxies.  The method is
applied to already-completed cosmological simulations, so it can be run
quickly to test changes in the assumed prescriptions and parameters.
In the method, metals are instantaneously
placed in gas that is in and nearby galaxies, according to
parameterized physical prescriptions that estimate where the metals
would `land' in the gas after $\sim 10^8-10^9\,$yr.  We have discussed
in detail the prescriptions used in this study, pointing out which
physical effects are captured by them.  The method can be used to
predict the cosmological distribution of metals {\em given}
assumptions about their ejection, or conversely to study the ejection
assumptions by comparing our calculations to observations in detail.

In this paper we have applied the method to several cosmological
simulations.  Using these results we can draw some conclusions
both about the methodology we have developed, and about metal ejection
in the real universe.

In our simulations, removal of metals by purely dynamical processes
such as ram-pressure stripping or tidal disruption of galaxies, is
relatively ineffective at polluting the IGM.  Averaged over their mass
function, galaxies of mass $\gsim 10^{10.5}\msol$ (i.e.\ those
resolved by our $144^3$ particle simulation) lose $\sim 4\%$ of their
metals by $z=0$.  This accounts for only about $1/12$th of the metal
density observed in the gas of rich clusters, though we cannot address
the dynamical enrichment of clusters by smaller galaxies that could
have comprised a significant fraction of the mass function at high
$z$.  Dynamical removal alone also cannot account for the
mass-metallicity relation observed in present-day galaxies of all
masses.  Galaxies of mass $M \gsim 10^{8.5}\msol$ also lose $\sim 2\%$
of their metals by $z=3$, enriching Ly$\alpha$ absorbers with $N(H\,I)
\sim 10^{14.5}{\rm\,cm^2}$ to $\sim 10^{-4.5}\zsol$, about $100$ times
less than observed.  These results indicate that {\em if} dynamical
removal were to account for metals in the low-density Ly$\alpha$
absorbers (Gnedin \& Ostriker 1997; Gnedin 1998), it must have been by
smaller galaxies, presumably at very high redshifts ($z \gg 6$).  We
also find that the dynamical removal of metals is enhanced if the
metals are moved into galaxy halos (i.e., by winds or dust-ejection),
but only slightly and only at low $z$; at high $z$ the net effect of
dynamics is to move metals from low- to high-density regions.

	If the metals in the IGM came from fairly massive galaxies at
$z\la 6$, some mechanism other than dynamical removal must have played
an important role. Supernova-driven winds are a plausible candidate.
Our prescription assumes that winds develop at a critical SFR/area,
with a fixed velocity, and an energy in the wind proportional to the
SFR.  We find that the degree of IGM enrichment is not very sensitive
to the wind efficiency (unless it is very different than we have
assumed) or how the metals are distributed within the wind `stopping
radius'.  The results {\em are} sensitive to the assumed fraction of
the ambient material entrained by the wind, the wind outflow velocity,
and the critical SFR.  If the latter is chosen so that a significant
fraction of high-$z$ galaxies drive winds (as indicated by
observations of Lyman-break galaxies; see Steidel et al. 2000 and
Pettini et al. 2001), then winds with outflow velocities of $\sim
200\kms$ or more (as also indicated by the observations) can escape to
large distances, and enrich the low-density IGM to roughly the level
observed at $z\sim3$.  Whether the enrichment can match the observed
metallicities in detail, and whether the wind process itself would
disturb the low-density IGM more than allowed by observations is an
important and open question (see Theuns, Mo \& Schaye [2001] for some
discussion of the latter).
	
	At lower redshifts, winds even from massive galaxies may
be important, though in our models metal ejection from galaxies of $M \gsim
10^{10.5}\msol$ probably cannot account for all of the metals in
cluster gas unless a rather extreme model is adopted or one of our
methodological assumptions is changed.  Since smaller galaxies eject
metals more efficiently, it is possible that winds could account for
the ICM metallicity if these were included.  Wind ejection also leads
to a mass-metallicity relation comparable to -- but somewhat steeper
than -- that observed.  Using future simulations with a larger range
of galaxy masses, the M-Z relation should be a useful diagnostic of
outflows, as it helps break the degeneracy between our model
parameters in determining the metal ejection efficiency.

The ejection of dust by radiation pressure is another interesting way
that metals may escape galaxies.  We assume that a significant
fraction of galactic dust escapes to the radius where gravitational
and radiation-pressure forces balance.  We find that our basic results
are not strongly sensitive to the assumed IMF or dust correction.  The
results do, however, depend on the grain type and on the very
important assumption that the grains are not confined by the gas or
magnetic fields in galaxies.  The results show that metals removed
from $\gsim 10^{8.5}\msol$ galaxies as dust, then destroyed in the IGM
or galaxy halos by sputtering,
could also account for the mean level of IGM
enrichment observed at $z=3$ -- although again it is unclear whether the
distribution agrees with the observations in detail.  At low redshift,
massive galaxies can enrich the ICM to the observed levels.  Moreover,
since dust ejection does not `avoid' high-pressure regions as winds
do, dust can enrich intragroup gas more uniformly than winds.
Enrichment of the IGM by dust would provide a number of chemical
signatures; in particular, non-depleted elements such at N and Zn
should be underrepresented in the IGM (see Aguirre et al. 2001b).

	Most generally, our simulations support the view that a
significant fraction of cosmic metals lie in the IGM, and our method
provides a useful way to generate predictions of the cosmic
distribution of metals usable in a number of ways.  With
higher-resolution simulations we should be able to more effectively
test the importance of low mass galaxies at all redshifts, as well as
perform more careful resolution tests of our results.  Future
observations, as well as more detailed small-scale simulations, will
help to develop more accurate ejection prescriptions.

\acknowledgements

This work was supported by NASA Astrophysical Theory Grants NAG5-3922,
NAG5-3820, and NAG5-3111, by NASA Long-Term Space Astrophysics Grant
NAG5-3525, and by the NSF under grants ASC93-18185, ACI96-19019, and
AST-9802568.  JG was supported by NASA Grant NGT5-50078 for the
duration of this work, and AA was supported in part by the National
Science Foundation grant no. PHY-9507695 and by a grant in aid from
the W.M. Keck Foundation.  The simulations were performed at the San
Diego Supercomputer Center, and the post-simulation processing used a
computer purchased under the National Science Foundation grant
no. PHY-9507695.

\end{document}